\documentclass[12pt, a4paper]{article}
\usepackage{iem}
\usepackage{mathrsfs,bm,amsmath,amsbsy,amssymb}
\usepackage{microtype}
\usepackage[T1]{fontenc}
\usepackage{listings}
\usepackage{matlab-prettifier}
\usepackage[hidelinks]{hyperref}

\begin{document}
 \lstset{numbers=left, numberstyle=\tiny, stepnumber=2, numbersep=5pt}

\selectlanguage{english}
\title{Ultraspherical/Gegenbauer polynomials to\\[2mm] 
	unify 2D/3D Ambisonic directivity designs}

\author{Franz Zotter}
\authorcontact{zotter@iem.at - http://iem.at/Members/zotter\\
Tel.: +43 389 \{3382\}}
%\pages{\pageref{endmark} pages}

\markboth{}{
Zotter: axisymmetric ultraspherical polynomials and Ambisonic directivities}

%\doctype{Vorlesungsskriptum}

%\date{Juni 2008}

\maketitle
\newpage
\pagestyle{empty}
\hspace{1cm}\vspace{3cm}

\hspace{1cm}\vspace{1cm}

\begin{abstract}
  This report on axisymmetric ultraspherical/Gegenbauer polynomials~\cite{Gegenbauer77} and their use in Ambisonic directivity design in 2D and 3D presents an alternative mathematical formalism to what can be read in, e.g., my and Matthias Frank's book on Ambisonics or J\'{e}r\^{o}me Daniel's thesis, Gary Elko's differential array book chapters, or Boaz Rafaely's spherical microphone array book. \\
  
  My original---but discarded---intention was to already include ultraspherical polynomials in the Ambisonics book to pursue a simple thought: retrieving suitable axisymmetric continuous-direction functions that serve the discrete-direction metrics for experimental psychoacoustic findings in spatial audio uniformly, for any number $\mathrm{D}$ of space dimensions, of course typically $\mathrm{D}=2,3$. But the cost for this natural way to understand spherical/circular polynomials and their directional sampling was too high. This would have required carrying out both the derivations of two entire formalisms from scratch, the ultraspherical/Gegenbauer polynomials and the circular/spherical harmonics, of which only the latter ones define the Ambisonic format. \\
  
  Nevertheless, ultraspherical/Gegenbauer  polynomials are highly valuable when designing axisymmetric beams and understanding spherical $t$ designs so that this report will shed some light on what circular, spherical, and ultraspherical axisymmetric polynomials are. While mathematically interesting by themselves already, they can be useful in spherical beamforming as described in the literature on spherical and differential microphone arrays, e.g.~\cite{Rafaely19,Elko00, Elko04}.
  What are polynomials bases for axisymmetric functions, say axisymmetric harmonics (or ultraspherical or Gegenbauer) polynomials? How can they be utilized to define Ambisonic order weightings or spherical beamformers? \\
  
  In this report, these ultraspherical/Gegenbauer polynomials will be used to uniformly derive for arbitrary dimensions $\mathrm{D}$ the various directivity designs or Ambisonic order weightings known from literature: max-DI/basic \cite{Daniel01}, max-$r_\mathrm{E}$ \cite{Daniel01}, supercardioid \cite{Rafaely19, Elko00, Elko04}, cardioid/inphase \cite{Daniel01}.  Is there a way to relate higher-order cardioids and supercardioids? How could one define directivity patterns with an on-axis flatness constraint?
\end{abstract}

\newpage

\pagestyle{myheadings}
\hspace{1cm}\vspace{2cm}

\tableofcontents
\newpage

\section{Axisymmetric directivity functions}
In the seventies, Cooper~\cite{Cooper72} described the idea of using angular amplitude functions to control the auditory event, which could be thought of as an extension to mid/side decomposition, or as recording with figure-of-eight and omnidirectional directivity patterns.
Stemming from this approach and from this time, the notion and technology of Ambisonics was developed by Felgett~\cite{Felgett}, Gerzon~\cite{Gerzon75}, and Craven~\cite{Craven77}. Since then, the term Ambisonics became common for technology using angular harmonic functions to express surround sound. Read more about all of this in the open Ambisonics book at Springer 
\url{https://doi.org/10.1007/978-3-030-17207-7}.

\autoref{apdx:circle_fourier} shows how 2D Ambisonic panning can be expressed in terms of a finite-order Fourier series using the exponentials $e^{\mathrm{i}m\phi}$ with real coefficients, i.e., a finite cosine $\cos m\phi$ series in the azimuthal distance $\phi=\varphi-\varphi_s$ between array angles $\varphi=\upvarphi_l$ and the given angle $\varphi_s$. For this symmetric finite-resolution function, a regular circular polygon of sufficiently many loudspeaker directions would automatically yield correct aiming of the vector measures $\bm{r}_\mathrm{E}$ and $\bm{r}_\mathrm{V}$ towards the panning direction $\bm\theta_\mathrm{s}$ and panning-invariant loudness and width magnitudes $E$, $P$, $\|\bm{r}_\mathrm{E}\|$, $\|\bm{r}_\mathrm{V}\|$, and a certain directivity factor $Q$.

Instead, we desire defining axisymmetric functions around the panning direction $\bm\theta_\mathrm{s}$ in a space of arbitrarily many dimensions $\mathrm{D}$, and their decomposition into suitable limited degree harmonic functions. The projection of the variable direction vector $\bm\theta$ on given direction $\bm\theta_\mathrm{s}$ yields the cosine of the enclosed angle to the axis $\bm\theta_\mathrm{s}^\mathrm{T}\bm\theta=\cos\phi$, and hereby expresses the only dimension of the axisymmetric function, $x=\cos\phi$. As the $m^\mathrm{th}$ power thereof, $\cos^m\phi$, can be expressed in terms of a cosine series up to the $m^\mathrm{th}$ degree, a power series in $x$ is a reasonable representation of the panning functon on the sphere.

To do this properly, we will first learn about the term harmonics, solve for axisymmetric harmonics, establish the panning function on the circle using these harmonics, and then finally come to the spherical panning function.

\subsection{Axisymmetric harmonics: Gegenbauer polynomials}
The Laplacian is defined in the $\mathrm{D}$-dimensional Cartesian space as
\begin{align}
\bigtriangleup&=\sum_{j=1}^\mathrm{D}\frac{\partial^2}{\partial x_j^2}.
\end{align}
Solving the eigenvalue problem $\bigtriangleup = -m^2\,f=0$ for one dimension $\frac{\partial^2}{\partial x^2}f=-m^2f$ yields with the exponential guess $f=e^{ax}$, inserted and differentiated to $a^2\,f=-m^2\,f$, \emph{harmonic functions} $f=e^{\pm\mathrm{i}mx}$ with $m\in\mathbb{Z}$ if the function should be periodic $f(x)=f(x+2\pi\,k)$, $k\in\mathbb{Z}$. If only a limited number of eigenvalues $m\leq K$ is used, this yields a limited-degree expansion into \emph{harmonics}, a band-limited Fourier series. It is useful to limit the resolution of a represented function.

For 2 dimensions \autoref{apdx:circle_fourier} defines panning function with limited $m$, defined using a series of real part of this complex exponential $e^{\mathrm{i}m\phi}=\cos(m\phi)$ with the polar angle $\phi=\arctan\frac{y}{x}$ enclosed with the $x$ axis. For 3 dimensions, these trigonometric or complex-exponential angular harmonics are not useful anymore, and lose some beneficial properties, for instance, $\cos(m\phi)$ has no uniformly limited resolution and is not analytic around the poles of the sphere.

Generally, homogeneous solutions to the Laplacian $\bigtriangleup g=0$ are called harmonics. In case of a spherical coordinate system with a unit-direction vector $\bm\theta$ and a radius $r$ and a separation approach $g=f(\bm\theta)\,h(r)$, we obtain $\bigtriangleup g=f(\bm\theta)\bigtriangleup h(r)+h(r)\bigtriangleup f(\bm\theta)=0$, for which there needs to be a separation constant $\lambda=\frac{\bigtriangleup h}{h}$ defining Laplacian eigensolutions $\bigtriangleup f=-\lambda\,f$ for the directional part. For suitable Eigenvalues $-\lambda$, these  \emph{directional harmonics} span an orthogonal set of basis functions as solutions.

Chosing the last Cartesian coordinate axis as panning direction $\bm\theta_\mathrm{s}=[\dots,\,0,\,1]^\mathrm{T}$, and writing the direction vector as $\bm\theta=\frac{[x_1,\,\dots,\,x_\mathrm{D}]^\mathrm{T}}{\sqrt{\sum_{j}x_j^2}}=\frac{[x_1,\,\dots,\,x_\mathrm{D}]^\mathrm{T}}{r}$, i.e.\ $\bm{\theta}_\mathrm{s}^\mathrm{T}\bm{\theta}=\frac{x_\mathrm{D}}{r}=\cos\phi$, 
a purely axisymmetric panning function
defined on $x:=\cos\phi$ becomes 
\begin{align}
f&=f(\cos\phi)=f(x).\nonumber
\end{align}
By this choice, the \emph{directional harmonics} become \emph{axisymmetric harmonics} with regard to the enclosed angle $\phi$ to some directional axis. To find their formal solution, we re-write the Laplacian to the axial dimension $-1\leq x\leq1$. Such harmonics are useful to represent limited-resolution patterns of rotational symmetry with regard to their axis. 
As derived in the \autoref{apdx:axiallaplacian}, the Laplacian with regard to the axial coordinate $x=\cos\phi$ becomes:
\begin{align}
r^2\bigtriangleup_x=
(1-x^2)\frac{\mathrm{d}^2}{\mathrm{d}x^2}
-(\mathrm{D}-1)\,x\,\frac{\mathrm{d}}{\mathrm{d} x}.
\end{align}
\paragraph{The Ggegenbauer/ultraspherical differential equation: Laplacian eigenvalue problem for axisymmetric functions.} With the above, we establish the ultraspherical differential equation $(r^2\bigtriangleup_x+\lambda) f=0$ with the eigenvalue $-\lambda$ to the Laplacian
\begin{align}
(1-x^2)f''
-
(\mathrm{D}-1)\,x\,f'
+\lambda\,f=0,
\end{align}
in the hope that if a limited set of discrete eigenvalues $\lambda_n\leq K$ is used, then we obtain harmonics orthogonal on the interval $-1\leq x\leq1$ that contain only limited powers of $x^n$ with $n\leq\mathrm{N}$, given a suitable of $K$ and $\lambda_n$. We know trigonometrically that such functions can only contain oscillations up to $\cos(\mathrm{N}\phi)$, thus have limited angular resolution.

\begin{figure}
	\begin{center}
		\includegraphics[height=4cm]{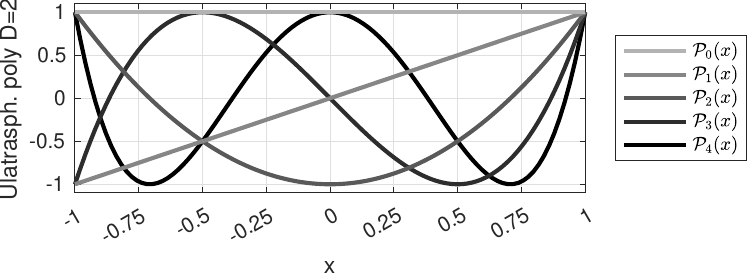}
		\includegraphics[height=4cm]{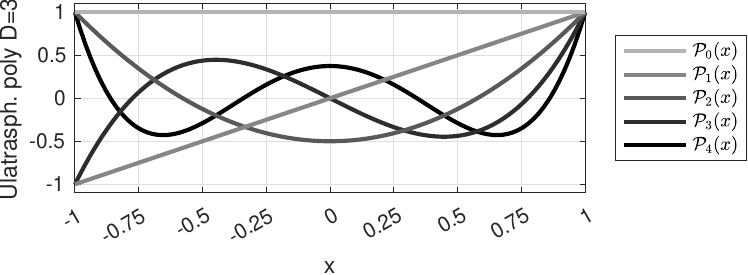}
		\includegraphics[height=4cm]{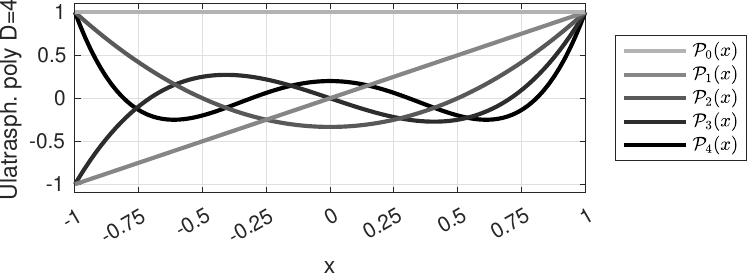}
	\end{center}
\caption{Ultraspherical polynomials $\mathcal{P}_n(x)=\frac{n!}{(2\alpha)^{\overline{n}}}C_n^{(\alpha)}(x)\big|_{1}=1$  for $\mathrm{D}=2,3,4$,  $\alpha=\frac{D-2}{2}$, i.e.\ Chebyshev polynomials $T_n(x)$ for $\mathrm{D}=2$, and Legendre polynomials $P_n(x)$ for $\mathrm{D}=3$\label{fig:Gegenbauer}. They are orthogonal wrt.\ the weight $w(x)=\sqrt{1-x^2}^{\mathrm{D}-3}$.}
\end{figure}

We solve the equation with the power series approach $f=\sum_k c_k\,x^k$, see \autoref{apdx:axisymfrobenius}, and get an interrelation of its coefficients $c_{k+2}=\frac{k^2+k(\mathrm{D}-2)
	-\lambda}{(k+1)(k+2)}\;c_k$.
Choosing $\lambda=\lambda_n=n^2+(\mathrm{D}-2)n$, the power series terminates at $k=n$, so that $c_{k>n}=0$, hence it becomes an $n^\mathrm{th}$-degree polynomial $f=\mathcal{P}_n$, and the relation between the coefficients of $\mathcal{P}_n$ remains
\begin{align}
   c_{k+2}^{(n)}=\frac{(k-n)(k+n+\mathrm{D}-2)}{(k+1)(k+2)}\;c_k^{(n)}.
\end{align}
Also solutions with different $\lambda$ could be acceptable, but their infinite series may diverge, e.g., at $x=1$ or $x=-1$, so we disregard. We re-write the above equation accordingly:
\begin{align}
(1-x^2)\mathcal{P}_n''
-
(\mathrm{D}-1)\,x\,\mathcal{P}_n'
+n(n+\mathrm{D}-2)\,\mathcal{P}_n=0.\label{eq:ultraspherical_ode}
\end{align}
The above equation is called the \emph{ultraspherical} or \emph{Gegenbauer} differential equation~\cite{Gegenbauer77}, and solutions can be constructed from $\mathcal{P}_0(x)=1$, and $\mathcal{P}_1(x)=x$ with the recurrence derived in \autoref{apdx:threeterm}, eq.~\eqref{eq:standard-three-term-recurrence} with the standardization $\mathcal{P}_n(1)=1$ of the \emph{ultraspherical} polynomials, therefore differing from Gegenbauer's typical definition $C_{n}^{(\alpha)}=\frac{(2\alpha)^{\overline{n}}}{n!}\,\mathcal{P}_n(x)$~\cite{Gegenbauer77}, cf.~\autoref{apdx:gegenbauer} eq.~\eqref{eq:gegenbauer_at_1},
\begin{align}
   \mathcal{P}_{n+1}(x)=\frac{2n+\mathrm{D}-2}{n+\mathrm{D}-2}\;x\,\mathcal{P}_{n}(x)-\frac{n}{n+\mathrm{D}-2}\mathcal{P}_{n-1}(x).\label{eq:ultraspherical_rec}
\end{align}
\autoref{fig:Gegenbauer} shows these ultraspherical polynomials for $\mathrm{D}=2,3,4$, $\alpha=\frac{D-2}{2}$. For $\mathrm{D}=2,3$, they become the Chebyshev and the Legendre polynomials.

\newpage
\paragraph{Orthogonality.}
The ultraspherical polynomials $\mathcal{P}_n$ are orthogonal according to the integral $\iiint\,\mathcal{P}_n\,\mathcal{P}_m\,\delta(r-1)\prod_j\mathrm{d} x_j=N_n^2\delta_{nm}$. We re-write the integration elements 
to get the orthogonality relation in $x$, see \autoref{apdx:axialjacobidet}, yielding
\begin{align}
	\int_{-1}^{1}\mathcal{P}_n(x)\mathcal{P}_m(x)\,\sqrt{1-x^2}^{\mathrm{D}-3}\,\mathrm{d}x=N_n^2\,\delta_{nm},
\end{align}
with the squared norm
 \begin{align}
	N_n^2&=\int_{-1}^{1}\mathcal{P}_n^2(x)\;\sqrt{1-x^2}^{\mathrm{D}-3}\,\mathrm{d}x=\frac{n!}{(2n+\mathrm{D}-2)(n+\mathrm{D}-3)!}\frac{S_{\mathrm{D}-1}}{S_{\mathrm{D}-2}}\label{eq:norm_square_maintext}
\end{align}
 of the polynomial, see \autoref{apdx:squarednormgegenbauer} eq.~\eqref{eq:norm}.
Orthogonality can be shown by integrating eq.~\eqref{eq:ultraspherical_ode} over $\mathcal{P}_m$ using the corresponding integration element, and by integration by parts, as shown in \autoref{apdx:orthogonalityproof}.

Orthogonality has the benefit of enabling a transform integral to find the coefficients $\gamma_n$ of a function decomposed into the  polynomials $\mathcal{P}_n$, see \autoref{apdx:orthopoly},
\begin{align}
 g(x)&=\sum_{n=0}^\infty \gamma_n\,\mathcal{P}_n(x), 
 &
 \gamma_n&=\frac{1}{N_n^2}\int_{-1}^{1}g(x)\,\mathcal{P}_n(x)\,\sqrt{1-x^2}^{\mathrm{D}-3}\,\mathrm{d}x.\label{eq:transformintegral_orthopoly}
\end{align}

\subsection{Continuous axisymmetric directivity/panning patterns in D dimensions}
The axisymmetric directivity/panning function is based on the decomposition of a Dirac delta pointing towards the panning direction $\bm\theta_\mathrm{s}$,
\begin{align} 
	\delta(1-\bm\theta_\mathrm{s}^\mathrm{T}\bm\theta)&=
	\begin{cases}
		\infty, & \text{for $\bm\theta=\bm\theta_\mathrm{s}$}\\
		0, & \text{otherwise}
    \end{cases},&
    \text{with }
    \int_{\mathbb{S}^{\mathrm{D}-1}}\delta(1-\bm\theta_\mathrm{s}^\mathrm{T}\bm\theta)\,\mathrm{d}\bm\theta&=1,
\end{align}
into infinitely many axisymmetric harmonics 
\begin{align}
	\delta(1-\bm\theta_\mathrm{s}^\mathrm{T}\bm\theta)&=\frac{1}{S_{\mathrm{D}-2}}\sum_{n=0}^\infty \gamma_n\,\mathcal{P}_n(\bm\theta_\mathrm{s}^\mathrm{T}\bm\theta),
\end{align}
where the factor $\frac{1}{S_{\mathrm{D}-2}}$ accounts for normalization with regard to the integrals over the independent, axisymmetric coordinates, whose value corresponds to the surface of the next lesser dimensional unit sphere $\mathbb{S}^{\mathrm{D}-2}$.

The desired axisymmetric panning function is based on the representation of the Dirac delta, but with limited degree and involving a sidelobe-suppressing weight $a_n$, 
\begin{align}
	g(\bm\theta)&=\frac{1}{S_{\mathrm{D}-2}}\sum_{n=0}^\mathrm{N}a_n\,\gamma_n\,\mathcal{P}_n(\bm\theta_\mathrm{s}^\mathrm{T}\bm\theta).
\end{align}

\paragraph{Simple $\gamma_n$ for axisymmetric Dirac delta at $x=x_0$.}
Given a Dirac delta located at $x=x_0$ as $\frac{\delta(x_0-x)}{\sqrt{1-x^2}^{\mathrm{D}-3}}$ which is correctly normalized to the axial coordinate to fulfill $\int_{-1}^{1}\delta(x_0-x)\;\,\mathrm{d}x=1$, we use the transform integral eq.~\eqref{eq:transformintegral_orthopoly} to find its coefficients $\gamma_m$:
\begin{align}
\frac{\delta(x_0-x)}{\sqrt{1-x^2}^{\mathrm{D}-3}}&=\sum_{n=0}^\infty \gamma_n\,\mathcal{P}_n(x),& \gamma_n=\frac{1}{N_n^2}\;\int_{-1}^1\delta(x_0-x)\,\mathcal{P}_n(x)\;\mathrm{d}x=\frac{\mathcal{P}_n(x_0)}{N_n^2}.\label{eq:axi_delta}
\end{align}
The resulting expression $\frac{\delta(x_0-x)}{w(x)}=\sum_{n=0}^\infty\frac{\mathcal{P}_n(x_0)\mathcal{P}_n(x)}{N_n^2}$ may be simplified by the Christoffel-Darboux identity eq.~\eqref{eq:christoffeldarboux} for summation over a limited degree $n\leq\mathrm{N}$.

When located at the pole $x_0=1$, the ultraspherical/Gegenbauer polynomials normalized to $\mathcal{P}_n(1)=1$ yield a simple expansion coefficient, cf.~eq.~\eqref{eq:norm_square_maintext}, with $a_n=\mathcal{P}_n(x_0)=1$,
\begin{align}
	\gamma_n&=\frac{1}{N_n^2},
	&
	g(x)&=%\frac{1}{S_{\mathrm{D}-1}\,}\sum_{n=0}^\mathrm{N}\frac{(2n+\mathrm{D}-2)\,(n+\mathrm{D}-3)!}{n!}\,\mathcal{P}_n(x)
	\frac{1}{S_{\mathrm{D}-2}\,}\sum_{n=0}^\mathrm{N}\frac{a_n}{N_n^2}\,\mathcal{P}_n(x),\label{eq:dirac}
\end{align}
and $a_n=1$ results from eq.~\eqref{eq:axi_delta}, while $a_n\neq1$ can be chosen to suppress sidelobes.

\begin{figure}
	\begin{center}
		\subfigure[basic\label{fig:basic_patterns}]{\includegraphics[height=5.5cm]{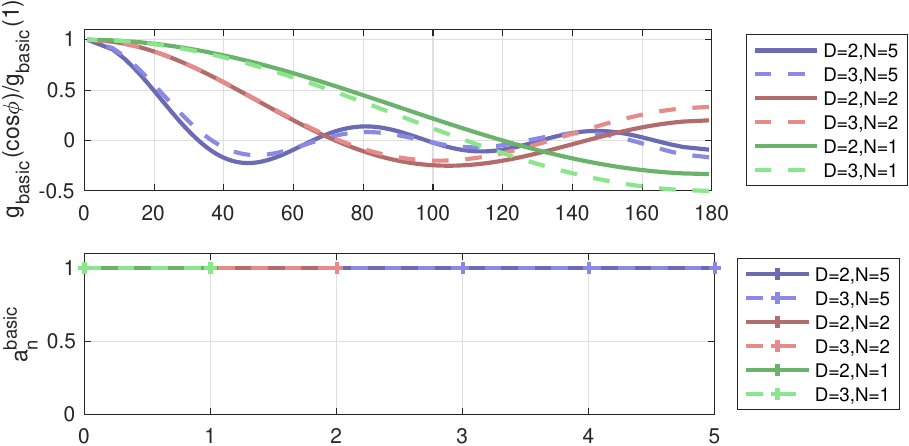}}
		\subfigure[max-$r_\mathrm{E}$\label{fig:maxre_patterns}]{\includegraphics[height=5.5cm]{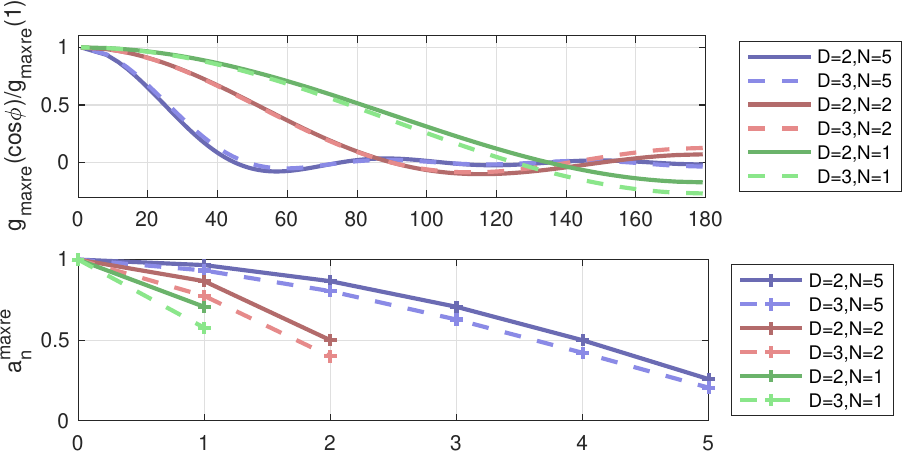}}
	\end{center}
	\caption{Basic and max-$r_\mathrm{E}$, $\mathrm{D}=2,3$ (solid,dashed),  $\mathrm{N}=1,2,5$ (green,red,blue).}
\end{figure}

\paragraph{Metrics $P$, $E$, $Q$, $\bm{r}_\mathrm{V}$, $\bm{r}_\mathrm{V}$, $\mathrm{FBR}$  for finite-degree axisymmetric patterns.}
We still need to find $a_n$, and these criteria were considered helpful. As derived in \autoref{apdx:axisymmetrics}, the metrics of the continuous, limited-degree axisymmetric panning function can be found by solving integrals over the continuous loudspeaker direction $\bm\theta$. The measures only depend on the weight $a_n$; with the integration measure abbreviated as $w(x)=\sqrt{1-x^2}^{\mathrm{D}-3}$,
\begin{align}
	P&=S_{\mathrm{D}-2}\,\int_{-1}^{1}g(x)\,w(x)\mathrm{d}x=a_0,\label{eq:P1}
	\\
	E&=S_{\mathrm{D}-2}\int_{-1}^{1}g^2(x)\,w(x)\mathrm{d}x=\sum_{n=0}^{\mathrm{N}}\frac{|a_n|^2}{S_{\mathrm{D}-2}N_n^2},\label{eq:E1}\\
	Q&=\frac{S_{D-1}\,g^2(1)}{E}=\frac{S_{D-1}\left(\sum_{n=0}^\mathrm{N} \frac{a_n}{S_{D-2}N_n^2}\right)^2}{E},\label{eq:Q1}\\
	\bm r_\mathrm{V}&=\frac{\int_{-1}^{1}g(x)\,x\;w(x)\mathrm{d}x}{P}\,\bm\theta_\mathrm{s}=\frac{a_1}{a_0}\,\bm\theta_\mathrm{s},\label{eq:rV1}\\
	\bm r_\mathrm{E}&=\frac{\int_{-1}^{1}g^2(x)\,x\;w(x)\mathrm{d}x}{E}\,\bm\theta_\mathrm{s}=\frac{\sum_{n=0}^\mathrm{N}\frac{2\beta_{n+1}}{N_n^2}\,a_{n}a_{n+1}}{\sum_{n=0}^\mathrm{N}\frac{1}{N_n^2}|a_n|^2}\,\bm\theta_\mathrm{s}.\label{eq:rE1}\\
	\mathrm{FBR}&=\textstyle\frac{\int_{0}^{1}g^2(x)\,w(x)\,\mathrm{d}x}{\int_{-1}^{0}g^2(x)\,w(x)\,\mathrm{d}x}=\frac{\sum_{n=0}^\mathrm{N}\sum_{m=0}^\mathrm{N}a_ng_{nm}a_{m}}{\sum_{n=0}^\mathrm{N}\sum_{m=0}^\mathrm{N}a_ng_{nm}a_{m}(-1)^{m}}.\label{eq:FBR1}
\end{align}
Obviously, $P$ and $\bm r_\mathrm{V}$ only depend on the weights $a_0$ and $a_1$, while $E$, $Q$, and $\bm r_\mathrm{E}$ involve the weights $a_n$ of all the degrees; integrals of the FBR require the entries of the Gram matrix $g_{nm}$ as derived in \autoref{apdx:gram}. 
\newpage
\paragraph{Basic/maximum-DI weights.}
With the weights $a_n=1$, there is no sidelobe suppression but the narrowest main lobe and highest directivity that can be represented using the limited degrees $n\leq\mathrm{N}$, cf.~\autoref{apdx:maxdi}.
For this choice, the directivity factor is highest $Q=\sum_{n=0}^\mathrm{N}\frac{1}{S_{D-2}N_n^2}$, which is $\sum_{n=0}^\mathrm{N}(2-\delta_m)$ or $\sum_{n=0}^\mathrm{N}(2n+1)$ and so $Q=\{2\mathrm{N}+1,\, (\mathrm{N}+1)^2\}$ for $D=\{2,3\}$. Because there is an equivalent least-squares problem
 $\int_{-1}^1[\delta(1-x)-g(x)]^2w(x)\mathrm{d}x\rightarrow\min$ to the defining eq.~\eqref{eq:dirac} with neutral weights $a_n=1$, the squared pattern $g^2(x)$ gets minimized everywhere except at $x=1$, so that a maximum $\mathrm{DI}=10\,\lg Q$ is reached.
Fig.~\ref{fig:basic_patterns} shows resulting patterns for $\mathrm{D}=2,3$ and $\mathrm{N}=1,2,5$.

Its simple, closed-form expression found in~\cite{Rafaely19} is based on the Christoffel-Darboux identity eq.~\eqref{eq:finite_order_dirac} \autoref{apdx:Chrisfoffel} and for any number of dimensions with $x=\cos\vartheta$, $y=1$
\begin{align}
	g(\cos\vartheta)&=\frac{\mathrm{D}-1}{2\mathrm{N}+\mathrm{D}-1}\frac{\mathcal{P}_{\mathrm{N}+1}(\cos\vartheta)-\mathcal{P}_{\mathrm{N}}(\cos\vartheta)}{\cos\vartheta-1}.
\end{align}

\paragraph{Max$-\bm r_\mathrm{E}$ weights.}
To shape the finite-degree axisymmetric panning function, we choose max-$\bm r_\mathrm{E}$ weights~\cite{Daniel99} $a_n$ that maximize the magnitude $r_{\mathrm{E}}$ of the $\bm r_\mathrm{E}$ vector. In great mathematical detail, \autoref{apdx:maxre_axisym} shows the solution to be an axisymmetric Dirac delta, so $a_n=\mathcal{P}_n$, evaluated at the greatest root $r_{\mathrm{E,max}}$ of $\mathcal{P}_{\mathrm{N}+1}$ in the next higher degree,
\begin{align}
	a_n&=\mathcal{P}_n(r_\mathrm{E,max}), 
		&
    r_\mathrm{E,max}&=\mathrm{arg\,max}_{r_\mathrm{E}}\{\mathcal{P}_{\mathrm{N}+1}(r_\mathrm{E})=0\}.
\end{align}
In general, this requires a root finder or approximation of the greatest root.
For $D=\{2,3\}$, the value is $r_\mathrm{E,max}=\{
\cos\frac{\pi}{2(\mathrm{N}+1)},\,
\cos\frac{\pi}{180^\circ}\frac{137.9^\circ}{\mathrm{N}+1.51}
\}$, where the value for $D=3$ is an approximation~\cite{Zotter12b}.
Fig.~\ref{fig:maxre_patterns} shows max-$r_\mathrm{E}$ patterns for $\mathrm{D}=2,3$ and $\mathrm{N}=1,2,5$. 

The Christoffel-Darboux identity eq.~\eqref{eq:christoffeldarboux} \autoref{apdx:Chrisfoffel} simplifies by $\mathcal{P}_{\mathrm{N}+1}(r_\mathrm{E})=0$ and yields the simple, closed-form expression for the max-$r_\mathrm{E}$ directivity normalized to $g(1)=1$, with removable singularity at $x=r_\mathrm{E,max}$
\begin{align}
	g(\cos\vartheta)&=\frac{(1-r_\mathrm{E,max})\mathcal{P}_{\mathrm{N}+1}(\cos\vartheta)}{\cos\vartheta-r_\mathrm{E,max}}.
\end{align}

\begin{figure}[h]
	\begin{center}
		\subfigure[supercardioid\label{fig:supercard}]{\includegraphics[height=5.5cm]{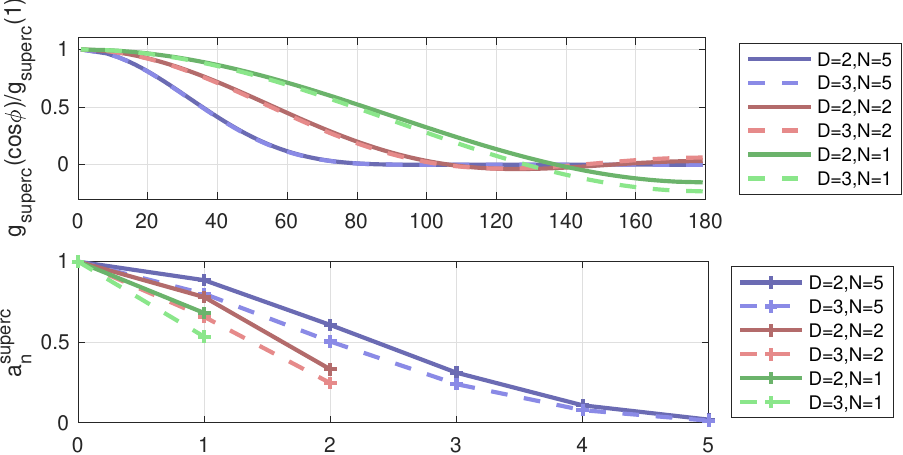}}
		\subfigure[inphase\label{fig:inphase_patterns}]{\includegraphics[height=5.5cm]{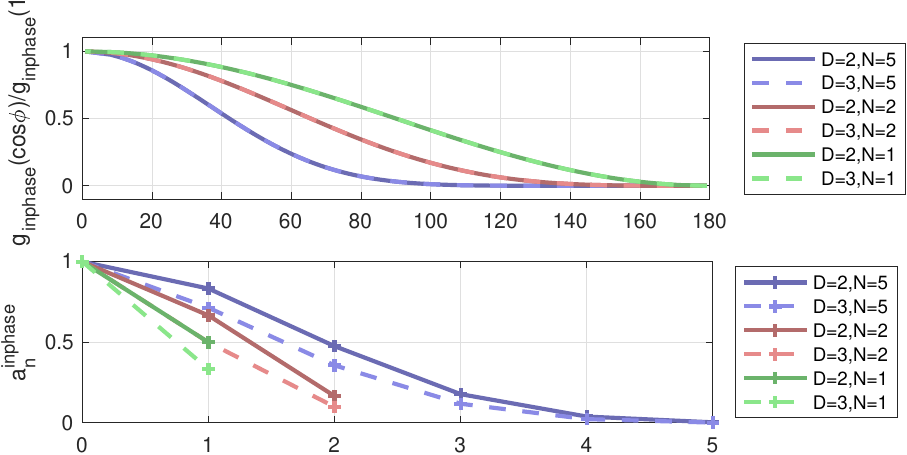}}
	\end{center}
	\caption{Supercardioid and in-phase,  $\mathrm{D}=2,3$ (solid,dashed),  $\mathrm{N}=1,2,5$ (blue,red).}
\end{figure}

\paragraph{Supercardioid/Maximum front-to-back ratio weights.}
For the supercardioid functions, the front-to-back energy ratio (FBR) is maximized. Fig.~\ref{fig:supercard} shows the resulting functions. The resulting problem is an eigenvector problem 
\begin{align}
	\bm a^\top\bm G\bm a&\rightarrow\max\label{eq:supercard}\\
	\text{s.t. } \|\bm a\|&=1.\nonumber
\end{align}
The Gram matrix $\bm G$ contains $g_{nn'}=\int_0^1\frac{\mathcal{P}_{n}(x)\mathcal{P}_{n'}(x)}{N_{n}^2N_{n'}^2}w(x) \mathrm{d}x$ and its entries are derived  in \autoref{apdx:gram}.
The vector $\bm a$ is a normalized vector with the largest-possible eigenvalue. Apparently, the resulting FBR yields an exponential improvement of the optimal front-to-back energy ratios for the circle and sphere, yielding after regression for $N=1,\dots,5$,
\begin{align}
	10\lg \mathrm{FBR_{max}}=13.75\,\mathrm{N}-
	   \begin{cases}
		3.6\thinspace{\mathrm{dB}}, &\mathrm{D=2},\\
		3.0\thinspace{\mathrm{dB}}, &\mathrm{D=3}.
		\end{cases}
\end{align}
As obtaining $a_{n}$ by eigendecomposition can be impractical, the similarity to the weights  $a_{n,\mathrm{inphase}}$ below is helpful: $a_n$ can be approximated by $a_{n,\mathrm{inphase}}^\beta$, modified by an approximate exponent $0.57\leq\beta\leq0.7$ determined numerically for $\mathrm{D}=2,3$ and $\mathrm{N}=1,\dots,10$, 
\begin{align}
	 a_{n}&\approx a_{n,\mathrm{inphase}}^\beta,
	 &
	 \beta&=\frac{0.73\,\mathrm{N}+0.67\mathrm{D}-1.11}{\mathrm{N}+1.11\mathrm{D}-1.5}.
\end{align}
Its absolute approximation error is limited $|a_{n,\mathrm{inphase}}^\beta-a_n|<-44\thinspace\mathrm{dB}$.

\paragraph{In-phase/Higher-order cardioid.}
According to \cite{Daniel99}, it can be helpful to define weights that entirely suppress the side lobes by causing an $\mathrm{N}$-fold zero at $x=-1$, which is often referred to as higher-order cardioid, and which is a max-flat design at $x=-1$,
\begin{align}
	g(x)&\propto(1+x)^{\mathrm{N}},
\end{align}
and
\autoref{apdx:inphase} complements the original derivation of \cite{Daniel01} to formally define the in-phase weighting for arbitrary dimensions
\begin{align}
	a_n=\frac{\mathrm{N}!}{(\mathrm{N}-n)!}\frac{(\mathrm{N}+\mathrm{D}-2)!}{(\mathrm{N}+n+\mathrm{D}-2)!}.\label{eq:inphase1}
\end{align}
Fig.~\ref{fig:inphase_patterns} shows in-phase patterns for $\mathrm{D}=2,3$ and $\mathrm{N}=1,2,5$.

\begin{figure}[h]
	\centering
	\subfigure[$\mathrm{D}=2$]{\includegraphics[height=6cm]{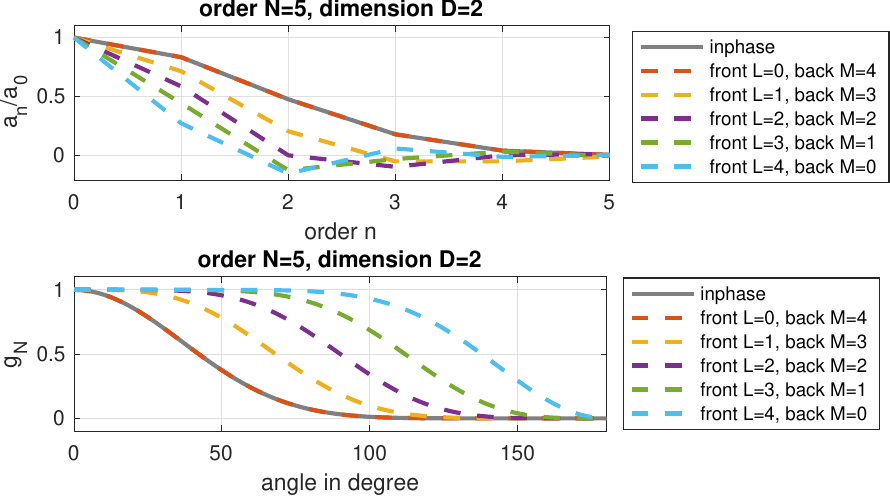}}
	\subfigure[$\mathrm{D}=3$]{\includegraphics[height=6cm]{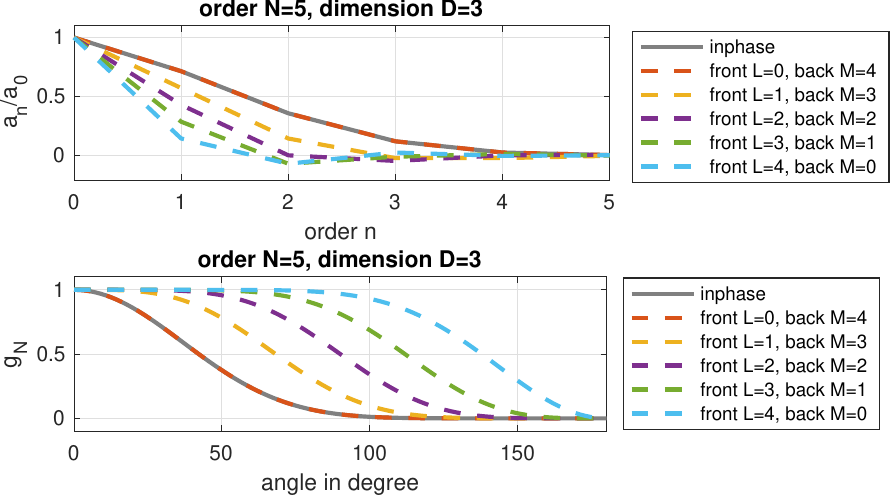}}
	\caption{Examples for maximally flat designs with $\mathrm{L}+1$ flatness degrees on axis, and a zero with $\mathrm{M}+1$ flatness degrees for $\mathrm{N}=\mathrm{L}+\mathrm{M}+1$ and for $\mathrm{D}=2$ and $\mathrm{D}=3$.}\label{fig:maxflat}
\end{figure}
\paragraph{Other max-flat/Butterworth designs} allow to employ some of the polynomial degrees to flatten the directivity on axis, as inspired by the FIR filter design \texttt{firflatLp(N,K,M)} 
from the book~\cite[p.92]{Schuessler}.
The recursive algorithm is defined in \autoref{apdx:inphase} in the equations \eqref{eq:maxflat_iteration} and \eqref{eq:maxflat_a0}. The total oder $\mathrm{N}=\mathrm{L}+\mathrm{M}+1$ is composed of the degree of flatness $\mathrm{M}$ for the zero at $x=-1$ and $\mathrm{L}$ for unity at $x=1$. Examples for $\mathrm{N}=5$, varying $\mathrm{L}$ and $\mathrm{D}=2$ or $\mathrm{D}=3$ are shown in Fig.~\ref{fig:maxflat}; angular flatness at $\phi=0,\pi$ is twice as high.\\
\emph{Inconsistency warning: Butterworth as used by Devaraju~\cite{Devaraju} refers to the weights $a_n=1/{\sqrt{1+(n/n_c)^{2k}}}$ discretizing the continuous spectral roll-off of a Butterworth low pass curve. Despite the inconsistency with Devaraju, I chose to keep the name Butterworth synonymous with the existence of a flatness constraint of a continuous function, which is applicable to the directional pattern across its continuous angle.}

\begin{figure}
	\begin{center}
		\includegraphics[height=5.5cm]{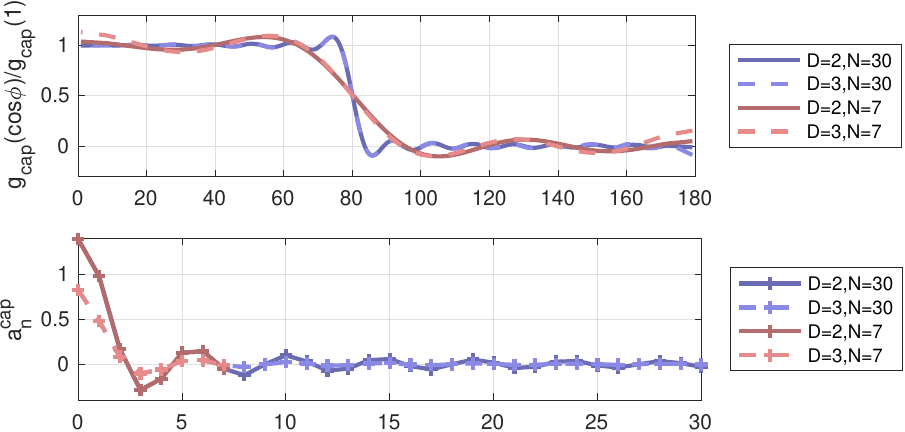}
	\end{center}
	\caption{Cap functions for $\mathrm{D}=2,3$ (solid,dashed), and $\mathrm{N}=7,30$ (blue,red).\label{fig:cap}}
\end{figure}

\paragraph{Cap functions.}
The weights $a_n$ for a spherical cap function from $x_0$ to $1$ are derived in \autoref{apdx:cap} to be
\begin{align}
	a_n=\frac{w(x_0)}{2n+2\alpha}\,\bigg[\mathcal{P}_{n-1}(x_0)-\mathcal{P}_{n+1}(x_0)\bigg].\label{eq:spherecap1}
\end{align}
with the exception at $n=0$, where $a_0=\upvarphi_0$ for $\mathrm{D}=2$ and $a_0=1-x_0$ for $\mathrm{D}=3$. 
Fig.~\ref{fig:cap} shows the cap functions for an $80^\circ$ cap represented with the degrees $\mathrm{N}=\{7,\,30\}$ for $\mathrm{D}=2,3$.

These weights are interesting to model loudspeakers on rigid spheres~\cite{Zotter17}.
In \cite{Zotter22} a trapezoidal function was used to approximate the VBAP kernel, and it consisted of the convolution of two differently sized cap functions, obtained by multiplying $a_n^{(1.375\alpha)}\,a_n^{(0.75\alpha)}$ with an average spacing angle $\alpha$, defining either $x_0=\cos\frac{1.375\alpha}{2}$ or $x_0=\cos\frac{0.75\alpha}{2}$ for the two caps.

\begin{figure}
	\centering
	\includegraphics[height=5.5cm]{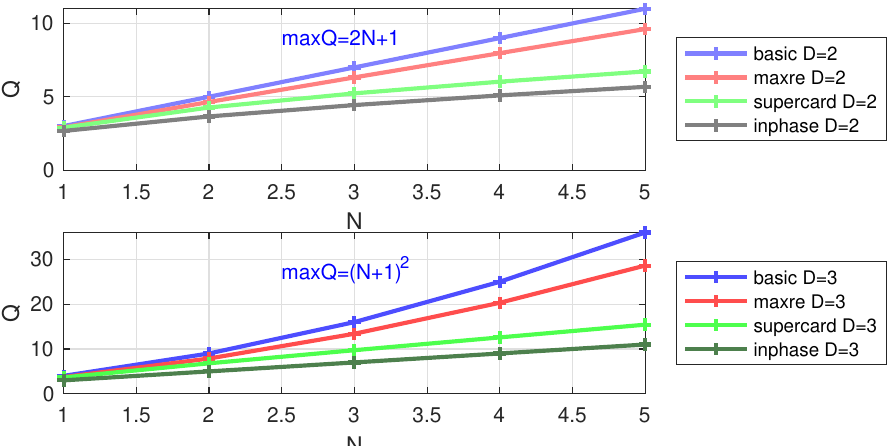}
	\caption{$Q$ values for different pattern designs.}\label{fig:Q}
\end{figure}

\begin{figure}
	\centering
	\includegraphics[height=5.5cm]{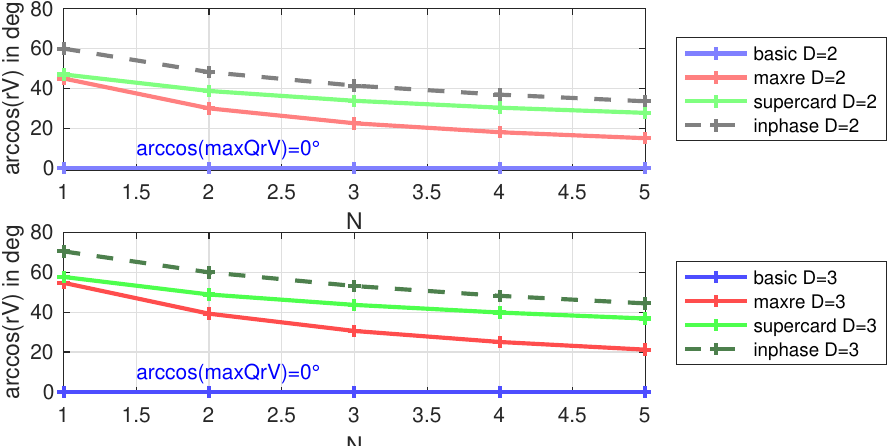}
	\caption{$\arccos r_\mathrm{V}$ values for different pattern designs.}\label{fig:rV}
\end{figure}

\begin{figure}
	\centering
	\includegraphics[height=5.5cm]{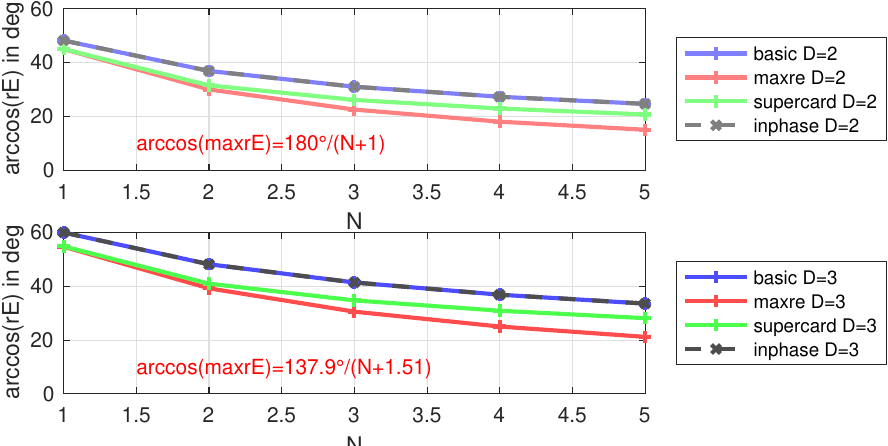}
	\caption{$\arccos r_\mathrm{E}$ values for different pattern designs.}\label{fig:rE}
\end{figure}

\begin{figure}
	\centering
	\includegraphics[height=5.5cm]{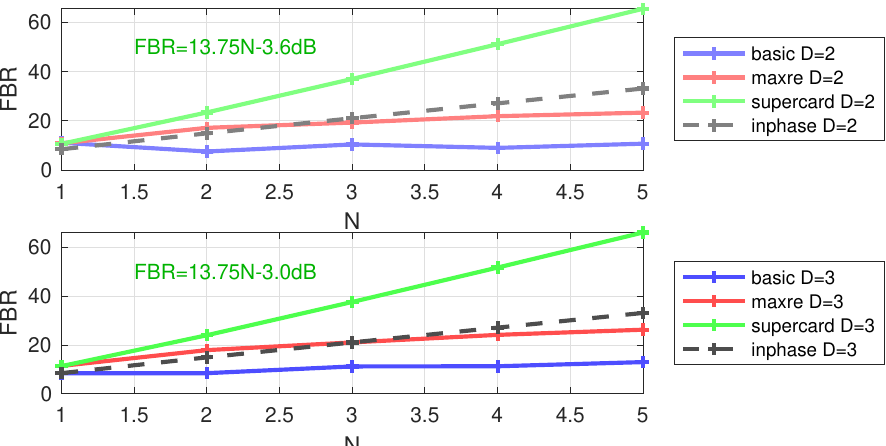}
	\caption{$10\lg\mathrm{FBR}$ values for different pattern designs.}\label{fig:FBR}
\end{figure}

\paragraph{FIGURES WITH METRICS} The Figs.~\ref{fig:Q},\ref{fig:rV},\ref{fig:rE},\ref{fig:FBR} evaluate the patterns presented above in terms of the $Q$, $r_\mathrm{V}$, $r_\mathrm{E}$, $\mathrm{FBR}$ metrics, with implementation example and verification in Appendices~\ref{apdx:metrics} and \ref{apdx:metrics_verify}.

Clearly, in Fig.~\ref{fig:Q}, the basic/max-DI pattern outperforms the others, followed by max-$r_\mathrm{E}$, supercardioid, and inphase as the pattern with the worst directivity factor $Q$.\\
In Fig.~\ref{fig:rV}, the conversion of $\|\bm{r}_\mathrm{V}\|$ to a spread angle by $\arccos\|\bm{r}_\mathrm{V}\|$ clearly indicates that the basic/max-DI is the pattern with zero spread or max-$r_\mathrm{V}$ that is unity, followed by max-$r_\mathrm{E}$, supercardioid, and at last inphase exhibiting the largest spread; similarly as in Fig.~\ref{fig:Q} about the directivity factor Q.\\
This changes in Fig.~\ref{fig:rE}, where the conversion of $\|\bm{r}_\mathrm{E}\|$ to a spread angle by $\arccos\|\bm{r}_\mathrm{E}\|$ clearly leads the smallest spread for max-rE, followed by supercardioid, and inphase en par with basic are worst and exhibit the largest spread.\\
The front-to-back ratio in Fig.~\ref{fig:FBR} is clearly best for supercardioid, followed by inphase at high degrees and max-rE at low degrees, and basic exhibits the worst FBR.\\

This verifies that by definition, basic/max-DI optimally isolates a direction within a diffuse sound field, max-$r_\mathrm{E}$ optimally defines the perceived direction of panned signals in surround sound, inphase is clearly the best in suppressing everything around the opposite-axis direction, and supercardioid is best in isolating the on-axis halfspace from the opposite-axis one.

Dolph-Chebyshev designs were presented by Rafaely~\cite{Rafaely19} and are omitted here for brevity, also some designs mentioned on \url{https://spaudiopy.readthedocs.io} the notable repository of Chris Hold.


\newpage
\subsection{Optimal discretization: spherical $t$ design}\label{sec:tdesign}
Assume that we sample a polynomial in $\left(\bm\theta_\mathrm{s}^\mathrm{T}\bm\theta\right)^n$ limited in degree by $n\leq t$, $t\in\mathbb{N}$. The discretization uses the directions $\{\bm\uptheta_l\}$ at $\mathrm{L}$ points and that this is called $t$ design whenever discrete summation and integral are equivalent, i.e., discrete summation accurately determines the direct component (DC) of the polynomial despite the variable orientation of $\bm\theta_s$, according to the definition in \cite{DeslarteGoethalsSeidel77},
\begin{align}
	\int_{\mathbb{S}_{\mathrm{D}-1}}\mathcal{P}_n\left(\bm\theta_\mathrm{s}^\mathrm{T}\bm\theta\right)\,\mathrm{d}\bm\theta&={\textstyle \frac{S_{\mathrm{D}-1}}{\mathrm{L}}}\sum_{l=1}^\mathrm{L} \mathcal{P}_n\left(\bm\theta_\mathrm{s}^\mathrm{T}\bm\uptheta_l\right).
\end{align}

This simplistic criterion of just estimating the DC correctly is relevant for all models: 
\begin{itemize}
	\item[$\bullet$] In the calculation of $P$, an integral over the $\mathrm{N}^\mathrm{th}$-degree polynomial pattern $g(\bm\theta)$ is evaluated, which is why at least a $t\geq\mathrm{N}$ design should be used to get correct results for $P$ by summing up the discretized pattern $g(\bm\uptheta_l)$ times $\mathrm{d}\Omega=\frac{S_{D-1}}{\mathrm{L}}$.
	\item[$\bullet$] In the calculation of $\bm r_\mathrm{V}$, the $\mathrm{N}^\mathrm{th}$-degree pattern $g(\bm\theta)$ is multiplied by $\bm\theta^\mathrm{T}_\mathrm{s}\bm\theta$, causing an integral over an $(\mathrm{N}+1)^\mathrm{th}$-degree polynomial, therefore discretization requires at least a $t\geq\mathrm{N}+1$ design for the numerator of $\bm{r}_\mathrm{V}$ to sum up $\bm\theta^\mathrm{T}_\mathrm{s}\bm\theta\,g(\bm\uptheta_l)$  correctly. 
	\item[$\bullet$] For calculating $E$, the squared continuous pattern $g(\bm\theta)^2$ is integrated, a $2\mathrm{N}^\mathrm{th}$-degree polynomial, thus discretization should at least consider a $t\geq 2\mathrm{N}$ design so that summing up the discrete pattern $g(\bm\uptheta_l)^2$ times $\mathrm{d}\Omega=\frac{S_{D-1}}{\mathrm{L}}$ correctly estimates ${E}$.
	\item[$\bullet$] In the calculation of $\bm r_\mathrm{E}$, the squared continuous function $g(\bm\theta)^2$ multiplied by $\bm\theta^\mathrm{T}_\mathrm{s}\bm\theta$ is integrated, a $(2\mathrm{N}+1)^\mathrm{th}$-degree polynomial, so discretization should consider a $t\geq2\mathrm{N}+1$ design for correct results for the numerator of $\bm r_\mathrm{E}$ that sums up $\bm\theta^\mathrm{T}_\mathrm{s}\bm\theta\,g(\bm\uptheta_l)^2$.
\end{itemize}
In conclusion, using at least a $t\geq2\mathrm{N}+1$ design ensures that $E$, $P$, $\bm r_\mathrm{V}$, and $\bm r_\mathrm{E}$ obtained by from discrete patterns $g(\bm\uptheta_l)$ complies with results for the continuous function $g(\bm\theta)$.

\emph{Note that the FBR measure is an exception because its lower integration limit may exclude a varying number of discretization nodes, depending on orientation and is therefore more accurate when solved with the analytic values $g_{nm}$ of the Gram matrix, see~\autoref{apdx:gram}.}

\newpage
\paragraph{Solutions in 2 and 3 dimensions.}
For $\mathrm{D}=2$ we get the Chebyshev differential equation and the Chebyshev polynomials $T_n$ as solutions, and for $\mathrm{D}=3$ the Legendre differential equation with Legendre polynomials $P_n$ as solutions
\begin{align}
\mathrm{D}&=2:&(1-x^2)T_n''
-
x\,T_n'
+n^2\,T_n&=0,\\
\mathrm{D}&=3:&(1-x^2)P_n''
-
2\,x\,P_n'
+n(n+1)\,P_n&=0.
\end{align}
These polynomials and their three-term recurrences are listed in Tables~\ref{tab:chebpoly},\ref{tab:legpoly}, yielding polynomials normalized at $x=1$, i.e., $T_n(1)=1$ and $P_n(1)=1$ for all $n$, and they fulfill the orthogonality relations
\begin{align}
   \mathrm{D}&=2:&\int_{-1}^{1}T_n(x)T_m(x)\,\frac{\mathrm{d}x}{\sqrt{1-x^2}}&=\frac{2}{2-\delta_{m,0}}\,\delta_{nm}, &\text{with }N_n^2&=\frac{2-\delta_{m,0}}{2},\\
   \mathrm{D}&=3:&\int_{-1}^{1}P_n(x)P_m(x)\,{\mathrm{d}x}&=\frac{2}{2n+1}\,\delta_{nm},& \text{with }N_n^2&=\frac{2n+1}{2}.
\end{align}
and recurrences
\begin{align}
   \mathrm{D}&=2:& T_{m+1}(x)&=(2-\delta_{m,0})x\,T_{m}(x)-T_{m-1}(x),\label{eq:chebyshev}\\
   \mathrm{D}&=3:& P_{n+1}(x)&=\frac{2n+1}{n+1}x P_n(x)-\frac{n}{n+1}P_{n-1}(x).\label{eq:legendre}
\end{align}

The Dirac delta $\delta(1-\bm\theta_\mathrm{s}^\mathrm{T}\bm\theta)$ pointing to $\bm\theta_\mathrm{s}$ is represented by the coefficients
\begin{align}
	\mathrm{D}=2:\quad\gamma_n&=\frac{2-\delta_{n,0}}{2\pi},&
	\mathrm{D}=3:\quad\gamma_n&=\frac{2n+1}{4\pi}.
\end{align}

\subsection{Continuous circular patterns in Chebyshev polynomials}
%Sec.~\ref{sec:finite-order Fourier series} showed that a cosine series is able to express the 2D panning function, using the enclosed angle $\phi=\varphi-\varphi_\mathrm{s}$. A Dirac delta uses $a_{m}=1$ and $\mathrm{N}=\infty$, i.e. $\delta(\phi)=\frac{1}{2\pi}\sum_{m=-\infty}^\infty e^{\mathrm{i}m\phi}=\frac{1}{2\pi}+\frac{2}{2\pi}\sum_{m=0}^\infty\cos(m\phi)=\sum_{m=0}^\infty\frac{2-\delta_{m,0}}{2\pi}\cos(m\phi)$. The projection of the variable direction on the panning direction is  $x=\bm\theta_\mathrm{s}^\mathrm{T}\bm\theta=\cos\phi$, and the property $T_m(\cos\phi)=\cos(m\phi)$ of the Chebyshev polynomials allows to express the symmetric Fourier series (cosine series) of the Dirac delta in terms of Chebyshev polynomials
On the circle, the equivalence of a Dirac delta  evaluated at the projection of the variable direction $\bm\theta$ on the panning direction $\bm\theta_\mathrm{s}$ is given in terms of an infinite sum of Chebyshev polynomials, see Tab.~\ref{tab:chebpoly},  cf.~\cite{Zotter12b},

\begin{align}
	\sum_{m=0}^\infty\frac{2-\delta_{m,0}}{2\pi}\,
	T_m(\bm\theta_\mathrm{s}^\mathrm{T}\bm\theta)=\delta(1-\bm\theta_\mathrm{s}^\mathrm{T}\bm\theta),\nonumber
\end{align}
with $\int_{\mathbb{S}^1}\delta(1-\bm\theta_\mathrm{s}^\mathrm{T}\bm\theta)\,\mathrm{d}\bm\theta=1$, and $\delta(1-\bm\theta^\mathrm{T}_\mathrm{s}\bm\theta)=\begin{cases}\infty, &\text{for $\bm\theta=\bm\theta_s$},\\ 0, &\text{otherwise}.\end{cases}$\\
The limited-degree panning function should involve the sidelobe-suppressing weights $a_m$
\begin{align}
	g(\bm\theta)=\sum_{m=0}^\mathrm{N}\frac{2-\delta_{m,0}}{2\pi}\,a_m\,
	T_m(\bm\theta_\mathrm{s}^\mathrm{T}\bm\theta),\nonumber
\end{align}
for which we may insert $\mathrm{max}-\bm r_\mathrm{E}$ weights~\cite{Daniel99},
%\begin{align}
	$a_m=T_m\Bigl(\cos\frac{90^\circ}{\mathrm{N}+1}\Bigr)$, as derived in \autoref{apdx:maxre_axisym}.
%\end{align}

\begin{table}[h]
	\caption{\small Coefficients of the Chebyshev polynomials 
		%$T_n(x)=n\sum_{i=0}^n(-2)^i\frac{(n+i-1)!}{(n-i)!(2i)!}(1-x)^i$
		that are orthogonal for integration on the interval $[-1,1]$ with the integration element $\frac{\mathrm{d}x}{\sqrt{1-x^2}}$. The Chebyshev polynomials obey the recurrence $T_{m+1}(x)+T_{m-1}(x)-(2-\delta_{m,0})x\,T_{m}(x)=0$ for $m>1$.    	\label{tab:chebpoly}        }
	\begin{tabular}{crrrrrrrrrr}
		\multicolumn{3}{c}{ $T_m(x)=\sum_{k=0}^m c_k\,x^k$}
		&
		\multicolumn{6}{l}{Coefficients of the Chebyshev polynomials}\\
		\hline
		$m=$ & $c_9$ & $c_8$ & $c_7$ & $c_6$ & $c_5$ & $c_4$ & $c_3$ & $c_2$ & $c_1$ & $c_0$\\
		$0$    & & & & & & & & & & $1$\\
		$1$    & & & & & & & & & $1$ & \\
		$2$    & & & & & & & & $2$ & & $-1$\\
		$3$    & & & & & & & $4$ & & $-3$ & \\
		$4$    & & & & & & $8$ & & $-8$ & & $1$\\
		$5$    & & & & & $16$ & & $-20$ & & $5$ & \\
		$6$    & & & & $32$ & & $-48$ & & $18$& & $-1$\\
		$7$    & & & $64$ & & $-112$ & & $56$ & & $-7$ & \\
		$8$    & & $128$ & & $-256$ & & $160$ & & $-32$& & $1$\\
		$9$    & $256$ & & $-576$ & & $432$ & & $-120$ & & $9$& 
	\end{tabular}
\end{table}

\paragraph{Measures.}
The function is of limited angular resolution $\mathrm{N}$ and allows to be sampled correctly. The $E$, $P$, $Q$, $\bm r_\mathrm{V}$, and $\bm r_\mathrm{E}$ measures of the continuous function can be derived, see \autoref{apdx:axisymmetrics}, and for verification with the Fourier transform \autoref{apdx:circle_fourier},
\begin{align}
	P&=a_0, &&&\bm r_\mathrm{V}&=\frac{a_1}{a_0}\;\bm\theta_\mathrm{s},\\
	E&=\sum_{n=0}^\mathrm{N}\frac{|a_n|^2}{2\pi}, 
	& Q&=\frac{\left|\sum_{n=0}^\mathrm{N}(2-\delta_n)a_n\right|^2}{\sum_{n=0}^\mathrm{N}(2-\delta_n)|a_n|^2},
	&\bm r_\mathrm{E}&=\frac{\sum_{n=0}^\mathrm{N}2\Re\{a_na_{n+1}^*\}}{\sum_{n=0}^\mathrm{N}(2-\delta_n)|a_n|^2}\;\bm\theta_\mathrm{s}.
\end{align}

\subsection{Addition theorem of the circular harmonics}
The 2D Ambisonic format represents source directions at the azimuth angle $\varphi_\mathrm{s}$ along the continuous  azimuth $\varphi$ using the circular harmonics. These are the typical terms of a real-valued Fourier transform of $2\pi$-periodic functions, here in the azimuth angle $\varphi$, 
\begin{align}
	\Phi_m(\bm\theta)=\sqrt{\frac{2-\delta_m}{2\pi}}\begin{cases}
		\cos|m|\varphi, &m\geq0\\
		\sin|m|\varphi, &m<0,
	\end{cases}
\end{align}
The addition theorem $\cos(\alpha-\beta)=\cos\alpha\cos\beta+\sin\alpha\sin\beta$ of the circular harmonics 
\begin{align}
	\Phi_{m}(\bm\theta)\Phi_{m}(\bm\theta_\mathrm{s})+(1-\delta_m)\Phi_{-m}(\varphi)\Phi_{-m}(\varphi_\mathrm{s})={\frac{2-\delta_m}{2\pi}}\,T_m(\bm\theta_\mathrm{s}^\mathrm{T}\bm\theta)
\end{align}
relates them to the polynomials $T_m$, so that we find the equivalence for $g(\bm\theta)$
\begin{align}
	g(\bm\theta)
	=
	\sum_{m=-\mathrm{N}}^\mathrm{N}a_{|m|}\Phi_m(\bm\theta_\mathrm{s})\Phi_m(\bm\theta)
	&=\sum_{n=0}^\mathrm{N}{\frac{2-\delta_m}{2\pi}}\,a_m\,T_m(\bm\theta_\mathrm{s}^\mathrm{T}\bm\theta).
\end{align}
This relation shows for single source directions $\bm\theta_\mathrm{s}$ that axisymmetric functions represented by the 2D Ambisonic format are equivalent to the polynomial representation, here.

\subsection{Optimal circular discretization: circular $t$ design}
Discretization uses the directions $\{\bm\uptheta_l\}$ at $\mathrm{L}$ points that this is called $t$ design equivalently with \autoref{sec:tdesign} whenever discrete summation of the directional function and integral are equivalent. For $\mathrm{D}=2$, this involves the Chebyshev polynomials
\begin{align}
	\int_{-\pi}^{\pi}T_m\left(\bm\theta_\mathrm{s}^\mathrm{T}\bm\theta\right)\,\mathrm{d}\bm\theta&={\textstyle \frac{2\pi}{\mathrm{L}}}\sum_{l=1}^\mathrm{L} T_m\left(\bm\theta_\mathrm{s}^\mathrm{T}\bm\uptheta_l\right).
\end{align}

On the circle, every equiangular sampling $\upvarphi_l=\frac{2\pi}{\mathrm{L}}(l-1)$ is a $t$-design if the number of samples fulfills the sampling criterion $t\leq\mathrm{L}+1$. There should be equally many sampling points than zero crossings of $\cos(m\varphi)$ or $\sin(m\varphi)$, or two samples per oscillation (equivalent to Whittaker/Kotelnikov/Shannon sampling theorem).

Our knowledge about $t$ designs, equiangular sampling on the circle, and the polynomial degrees in the various metrics allow to draw the following conclusion:
We need 
\begin{align}
	\mathrm{L}\geq  2\mathrm{N}+2
\end{align}
equiangularly spaced nodes on the horizon to obtain an optimally placed Ambisonic playback system of the degree $\mathrm{N}$. The $\bm r_\mathrm{E}$ vector is the strictest of the criteria.

\subsection{Continuous spherical patterns in Legendre polynomials} 
On the sphere, the equivalence of a Dirac delta  evaluated at the projection of the variable direction $\bm\theta$ on the panning direction $\bm\theta_\mathrm{s}$ is given in terms of an infinite sum of Legendre polynomials, see Tab.~\ref{tab:legpoly},  cf.~\cite{Zotter12b},
\begin{align}
	 \sum_{n=0}^\infty\frac{2n+1}{4\pi}\,P_n(\bm\theta_\mathrm{s}^\mathrm{T}\bm\theta)=\delta(1-\bm\theta^\mathrm{T}_\mathrm{s}\bm\theta),\nonumber
\end{align}
for which $\int_{\mathbb{S}^2}\delta(1-\bm\theta^\mathrm{T}_\mathrm{s}\bm\theta),\,\mathrm{d}\bm\theta=1$ and $\delta(1-\bm\theta^\mathrm{T}_\mathrm{s}\bm\theta)=\begin{cases}\infty, &\text{for $\bm\theta=\bm\theta_s$},\\ 0, &\text{otherwise}.\end{cases}$\\
The limited-degree summation of the panning function involves the coefficients $a_n$
\begin{align}
	g(\bm\theta)&=\sum_{n=0}^\mathrm{N}\frac{2n+1}{4\pi}\,a_n\,P_n(\bm\theta_\mathrm{s}^\mathrm{T}\bm\theta),
\end{align}
and it may use $\mathrm{max}-\bm r_\mathrm{E}$ weights according to \cite{Zotter12b,Daniel99}, as derived in \autoref{apdx:maxre_axisym},
\begin{align}
	a_n&=P_n\Bigl(\cos\frac{137.9^\circ}{\mathrm{N}+1.51}\Bigr).
\end{align}

\begin{table}[h]
	\caption{\small Coefficients of the Legendre polynomials 
		%$P_n(x)=2^n\sum_{i=0}^nx^i
		%\left(\begin{matrix} n\\
		%i\end{matrix}\right)
		%\left(\begin{matrix} 
		%\frac{n+i-1}{2}\\n
		%\end{matrix}\right)$
		that are orthogonal for the integration on the interval $[-1,1]$ with the integration element $\mathrm{d}x$. The recurrence to obtain these coefficients is $(n+1)P_{n+1}(x)-(2n+1)x P_n(x)+nP_{n-1}(x)=0$.	\label{tab:legpoly}
	}
	\begin{tabular}{crrrrrrrrrr}
		\multicolumn{3}{c}{ $P_n(x)=\sum_{k=0}^nc_k\,x^i$}
		&
		\multicolumn{6}{l}{Coefficients of the Legendre polynomials}\\
		\hline
		$n=$ & $c_9$ & $c_8$ & $c_7$ & $c_6$ & $c_5$ & $c_4$ & $c_3$ & $c_2$ & $c_1$ & $c_0$\\
		$0$    & & & & & & & & & & $1$\\
		$1$    & & & & & & & & & $1$ & \\
		$2$    & & & & & & & & $\frac{3}{2}$ & & $-\frac{1}{2}$\\
		$3$    & & & & & & & $\frac{5}{2}$ & & $-\frac{3}{2}$ & \\
		$4$    & & & & & & $\frac{35}{8}$ & & $-\frac{30}{8}$ & & $\frac{3}{8}$\\
		$5$    & & & & & $\frac{63}{8}$ & & $-\frac{70}{8}$ & & $\frac{15}{8}$ & \\
		$6$    & & & & $\frac{231}{16}$ & & $-\frac{315}{16}$ & & $\frac{105}{16}$& & $-\frac{5}{16}$\\
		$7$    & & & $\frac{429}{16}$ & & $-\frac{693}{16}$ & & $\frac{315}{16}$ & & $-\frac{35}{16}$ & \\
		$8$    & & $\frac{6435}{128}$ & & $-\frac{12012}{128}$ & & $\frac{6930}{128}$ & & $-\frac{1260}{128}$& & $\frac{35}{128}$\\
		$9$    & $\frac{12155}{128}$ & & $-\frac{25740}{128}$ & & $\frac{18018}{128}$ & & $-\frac{4620}{128}$ & & $\frac{315}{128}$& 
	\end{tabular}
\end{table}

\paragraph{Measures.}
The function is of limited angular resolution $\mathrm{N}$ and allows to be sampled correctly. The $E$, $Q$, $P$, $\bm r_\mathrm{V}$, and $\bm r_\mathrm{E}$ measures of the continuous function can be derived, see \autoref{apdx:axisymmetrics}, 
\begin{align}
	P&=a_0, &&&\bm r_\mathrm{V}&=\frac{a_1}{a_0}\;\bm\theta_\mathrm{s},\\
		E&=\sum_{n=0}^\mathrm{N}\frac{2n+1}{4\pi}|a_n|^2,
		&Q&=\frac{\left|\sum_{n=0}^\mathrm{N}(2n+1)a_n\right|^2}{\sum_{n=0}^\mathrm{N}(2n+1)|a_n|^2},
		 &\bm r_\mathrm{E}&=\frac{\sum_{n=0}^\mathrm{N}(n+1)2\Re\{a_na_{n+1}^*\}}{\sum_{n=0}^\mathrm{N}(2n+1)|a_n|^2}\;\bm\theta_\mathrm{s}.
\end{align}

\subsection{Addition theorem of the spherical harmonics}
The 3D Ambisonic format is based on representing source directions using the spherical harmonics. These involve the azimuth angle $\varphi$ and the zenith angle $\vartheta$ of the direction vector $\bm\theta$, and the associated Legendre functions $P_n^m(x)$, which are equivalent to the Legendre polynomials for $m=0$,
\begin{align}
	Y_n^m(\bm\theta)=(-1)^{|m|}\sqrt{\frac{(2-\delta_m)}{4\pi}\frac{(2n+1)\,(n-|m|)!}{(n+|m|)!}}\,P_n^{|m|}(\cos\vartheta)\,
   \begin{cases}
	\cos|m|\varphi, &m\geq0\\
	\sin|m|\varphi, &m<0.
   \end{cases}
\end{align}
Their addition theorem relates them to the Legendre polynomials $P_n(x)$
\begin{align}
	\sum_{m=-n}^nY_n^{m}(\bm\theta)Y_n^{m}(\bm\theta_\mathrm{s})=
	{\frac{2n+1}{4\pi}}\,P_n(\bm\theta_\mathrm{s}^\mathrm{T}\bm\theta),
\end{align}
so that we find the equivalence for $g(\bm\theta)$
\begin{align}
	g(\bm\theta)
	=
	\sum_{n=0}^\mathrm{N}\sum_{m=-n}^na_nY_n^{m}(\bm\theta)Y_n^{m}(\bm\theta_\mathrm{s})
	&=\sum_{n=0}^\mathrm{N}{\frac{2n+1}{4\pi}}\,a_n\,P_n(\bm\theta_\mathrm{s}^\mathrm{T}\bm\theta).
\end{align}
Single source directions $\bm\theta_\mathrm{s}$ represented within 3D Ambisonics are therefore equivalent to the Legendre-polynomial representation.

\subsection{Optimal spherical discretization: spherical $t$-design}
Spherical $t$-designs are $\mathrm{L}$ sampling nodes $\{\bm\uptheta_l\}$ on the sphere that allow to replace the integration over any spherical polynomial of limited degree $n\leq t$ to be evaluated correctly using a summation over the sampled polynomials. Note that the direction $\bm\theta_\mathrm{s}$ stays arbitrary, and a $t$-design establishes the exact equivalence
\begin{align}
	\int_{\mathbb{S}^2} P_n(\bm\theta_\mathrm{s}^\mathrm{T}\bm\theta)\,\mathrm{d}\bm\theta
	=
	{\textstyle \frac{4\pi}{\mathrm{L}}}\sum_{l=1}^\mathrm{L}P_n(\bm\theta_\mathrm{s}^\mathrm{T}\bm\uptheta_l).
\end{align}
The measures $P$, $E$, $\bm r_\mathrm{V}$, and $\bm r_\mathrm{E}$ of the continuous functions, as shown in the \autoref{apdx:axisymmetrics}, require integration over $\mathrm{N}^\mathrm{th}$, $(\mathrm{N}+1)^\mathrm{th}$, $2\mathrm{N}^\mathrm{th}$, and $(2\mathrm{N}+1)^\mathrm{th}$ degree polynomials, respectively. Exact equivalence between continuous and discretized functions therefore require the discrete sampling nodes $\{\bm\uptheta_l\}$ to be at least a
\begin{align}
	t\geq 2\mathrm{N}+1 \text{ design.}
\end{align}

%On the sphere, there are only the following geometrically regular polyhedra
%\begin{itemize}
%	\item[$\bullet$] Tetrahedron,  (2-design),
%	\item[$\bullet$] Octahedron, $\mathrm{L}=6$, (3-design),
%	\item[$\bullet$] Hexahedron/Cube, $\mathrm{L}=8$, (3-design),
%	\item[$\bullet$] Icosahedron, $\mathrm{L}=12$, (5-design),
%	\item[$\bullet$] Dodecahedron, $\mathrm{L}=20$, (5-design).
%\end{itemize}
%For instance, for $\mathrm{N}=1$, the octahedron is a suitable spherical design, for $\mathrm{N}=2$, the icosahedral or octahedral layouts are suitable.

Apart from the exception of 5 regular polyhedra (tetrahedron with $\mathrm{L}=4$ and $t=2$, octahedron with $\mathrm{L}=6$ and $t=3$, cube with $\mathrm{L}=8$ and $t=3$, icosahedron with $\mathrm{L}=12$ and $t=5$, dodecahedron with $\mathrm{L}=20$ and $t=5$), spherical designs are otherwise generally found by optimization, as, e.g., by Hardin and Sloane~\cite{Hardin96} and Gr\"af and Potts~\cite{Graef11}, as available on the following websites\\ \emph{http://neilsloane.com/sphdesigns/dim3/} and\\ \emph{http://homepage.univie.ac.at/manuel.graef/quadrature.php}\\ (Chebyshev-type Quadratures on $\mathbb{S}^2$).

\clearpage
\begin{appendices}
    
\section{Laplacian in axisymmetric coordinates}\label{apdx:axiallaplacian}
A function in the last, normalized Cartesian coordinate $x:=\frac{x_\mathrm{D}}{r}=\cos\vartheta$ is axisymmetric with regard to $\mathbf{i}_\mathrm{D}=[0,\dots,0,1]^\top$, and the generalized chain rule yields for the Laplacian
\begin{align}
	\bigtriangleup=\sum_j\frac{\partial^2}{\partial x_j^2}=
	\sum_j\frac{\partial}{\partial x_j}
	\left(
	\frac{\partial x}{\partial x_j}\frac{\mathrm{d}}{\mathrm{d}x}
	\right)
	=
	\sum_j\frac{\partial^2 x}{\partial x_j^2}\frac{\mathrm{d}}{\mathrm{d}x}
	+
	\sum_j\left(\frac{\partial x}{\partial x_j}\right)^2\frac{\mathrm{d}^2}{\mathrm{d} x^2}.
\end{align}
The derivative of $\frac{1}{r}$ keeps re-appearing. It is with regard to any of the coordinates
\begin{align}
	\frac{\partial}{\partial x_j}\frac{1}{r}=-\frac{1}{r^2}\frac{\partial r}{\partial x_j}=-\frac{1}{r^2}\frac{\partial[\sum_jx_j^2]^\frac{1}{2}}{\partial x_j}=
	-\frac{1}{r^2}\frac{[\sum_jx_j^2]^{-\frac{1}{2}}}{2}\frac{\partial \sum_jx_j^2}{\partial x_j}=-\frac{1}{r^3}\frac{1}{2}2x_j=-\frac{x_j}{r^3}.
\end{align}
The derivatives of $x:=\frac{x_\mathrm{D}}{r}$ are for $j\neq 1$:
\begin{align}
	\frac{\partial x}{\partial x_j}&=\frac{\partial }{\partial x_j}\frac{x_\mathrm{D}}{r}=-\frac{x_\mathrm{D}\,x_j}{r^3},
	\\
	\frac{\partial^2x}{\partial^2x_j}&=
	-\frac{\partial}{\partial x_j}\frac{x_\mathrm{D}\,x_j}{r^3}=-\frac{x_\mathrm{D}}{r^3}+3\frac{x_\mathrm{D}x_j^2}{r^5}=\frac{3x_\mathrm{D}x_j^2-r^2x_\mathrm{D}}{r^5}.
\end{align}
The first derivative of $x$ to $x_\mathrm{D}$ involves one more term depending on $x_\mathrm{D}$ that we could skip in the product rule before:
\begin{align}
	\frac{\partial x}{\partial x_\mathrm{D}}&=\frac{\partial}{\partial x_\mathrm{D}}\frac{x_\mathrm{D}}{r}=-\frac{x_\mathrm{D}^2}{r^3}+\frac{1}{r}=\frac{r^2-x_\mathrm{D}^2}{r^3}=\frac{\sum_{j\backslash\mathrm{D}}x_j^2}{r^3}=\frac{r_{\mathrm{D}-1}^2}{r^3},
	\\
	\frac{\partial^2 x}{\partial x_\mathrm{D}^2}&=\frac{\partial}{\partial x_\mathrm{D}}\frac{r_{\mathrm{D}-1}^2}{r^3}=\frac{-3\,x_\mathrm{D}\,r_{\mathrm{D}-1}}{r^5}.
\end{align}
We obtain for the factors involved in front of the second and first-order derivatives of the axisymmetric Laplacian:
\begin{align}
	\sum_j\left(\frac{\partial x}{\partial  x_j}\right)^2&=\frac{r_{\mathrm{D}-1}^4}{r^6}+\frac{x_\mathrm{D}^2\sum_{j\backslash\mathrm{D}}x_j^2}{r^6}=
	\frac{(r^2-x_\mathrm{D}^2)r_{\mathrm{D}-1}^2}{r^6}+\frac{x_\mathrm{D}^2r_{\mathrm{D}-1}^2}{r^6}=\frac{r^2r_{\mathrm{D}-1}^2}{r^6}=\frac{r_{\mathrm{D}-1}^2}{r^4},   \\
	\sum_j\frac{\partial^2\phi}{\partial  x_j^2}&=
	\frac{-3x_\mathrm{D}\,r_{\mathrm{D}-1}}{r^5}+
	\frac{3x_\mathrm{D}\sum_{j\backslash\mathrm{D}}x_j^2-\sum_{j\backslash\mathrm{D}}r^2x_\mathrm{D}}{r^5}=-(\mathrm{D}-1)\frac{x_\mathrm{D}}{r^3}.
\end{align}
Considering $\frac{x_\mathrm{D}}{r}=x$ and $\frac{r_{\mathrm{D}-1}}{r}=\sqrt{1-x^2}$, i.e., $\frac{x_\mathrm{D}}{r^3}=\frac{x}{r^2}$ and $\frac{r_{\mathrm{D}-1}^2}{r^4}=\frac{1-x^2}{r^2}$, the axisymmetric Laplacian $\bigtriangleup_x$ becomes
\begin{align}
	r^2\bigtriangleup_x=
	(1-x^2)\frac{\mathrm{d}^2}{\mathrm{d}x^2}
	-(\mathrm{D}-1)\,x\,\frac{\mathrm{d}}{\mathrm{d} x}.
\end{align}
We can re-write the axisymmetric Laplacian (which can be proven using the chain rule), involving the definition $\alpha=\frac{\mathrm{D}}{2}-1$, and $w(x)=(1-x^2)^{\alpha-\frac{1}{2}}$, see \autoref{apdx:axialjacobidet}
\begin{align}
	r^2\bigtriangleup_x\,f&=
	(1-x^2)f''
	-(2\alpha+1)\,x\,f'\\
	&=\frac{1}{(1-x^2)^{\alpha-\frac{1}{2}}}\left[(1-x^2)^{\alpha+\frac{1}{2}}f'\right]'=
	\frac{1}{w(x)}\left[(1-x^2)\,w(x)\;f'\right]'.
\end{align}

\section{Jacobi determinant of axisymmetric coordinates}\label{apdx:axialjacobidet}
For the axisymmetric coordinate $x$, we re-write the integration constants $\prod_{j=1}^\mathrm{D}\mathrm{d} x_j=|\bm{J}|\,\mathrm{d}x\,\prod_{j=1}^{\mathrm{D}-1}\mathrm{d}\tilde{x}_{j}$ with the determinant of the \emph{Jacobian} matrix that describes how the surface element is stretched. The Jacobi matrix is filled with derivatives of the old coordinates with regard to the new ones. One of the new coordinates is $x$ that defines $x_\mathrm{D}=r\,x$. The other $\mathrm{D}-1$ coordinates are just copies of $x_j$ but normalized to preserve the information about the radius as the new radius $r=\sqrt{\sum_{i}\tilde{x}_i^2}$ of the $(\mathrm{D}-1)$-dimensional subspace. This defines 
\begin{align}
	&x_\mathrm{D}=x\,r,
	&x_j&=\tilde{x}_j\,\sqrt{1-x^2},
\end{align}
and herewith the derivatives for $\bm J$ are
\begin{align}
	\bm J&=
	\begin{bmatrix}
		[\frac{\partial x_j}{\partial \tilde{x}_i}]_{i,j<\mathrm{D}},&
		[\frac{\partial x_\mathrm{D}}{\partial \tilde{x}_i}]_{i<\mathrm{D}}\\
		[\frac{\partial x_j}{\partial x}]_{j<\mathrm{D}}&	
		\frac{\partial x_\mathrm{D}}{\partial x}
	\end{bmatrix}
	=
	\begin{bmatrix}
		\mathbf{I}\sqrt{1-x^2},&
		\bigl[\frac{x\,\tilde{x}_i}{r}\bigr]_{i<\mathrm{D}}\\
		\Bigl[-\frac{\tilde x_j\,x}{\sqrt{1-x^2}}\Bigr]_{j<\mathrm{D}}&	
		r
	\end{bmatrix},    
\end{align}
which makes the determinant particularly simple to calculate: (i) the determinant of a product of matrices is the product of the determinants, (2) for a block-triangular matrix, the product of the diagonal blocks would be the determinant. A block-matrix $\bm E$ can be made block-triangular by LU decomposition
\begin{align}
	\bm J&=\begin{bmatrix}
		\bm A&\bm B\\\bm C&\bm D
	\end{bmatrix}=
	\begin{bmatrix}
		\bm A&\bm 0\\\bm C&\bm I
	\end{bmatrix}
	\begin{bmatrix}
		\bm I&\bm A^{-1}\bm B\\\bm 0&\bm D-\bm C\bm A^{-1}\bm B
	\end{bmatrix},\nonumber\\
	\text{so that }
	\det\bm J&=\det\bm A\,\det(\bm D-\bm C\bm A^{-1}\bm B).
\end{align}
We get the weight of the integration constant $|\bm J|\,\mathrm{d}x$:
\begin{align}
	|\bm J|&=\sqrt{1-x^2}^{\mathrm{D}-1}\;\left(r+\frac{1}{\sqrt{1-x^2}}\,\sum_{j}\frac{x\,\tilde{x}_j}{r}\frac{x\,\tilde{x}_j}{\sqrt{1-x^2}}\right)\\
	&=\frac{\sqrt{1-x^2}^{\mathrm{D}-1}}{r}\left(1+\frac{x^2 r^2}{r^2(1-x^2)}\right)=\frac{\sqrt{1-x^2}^{\mathrm{D}-3}}{r}=\frac{w(x)}{r}.
\end{align}

\section{Solving the ultraspherical differential equation:\\Gegenbauer polynomials}\label{apdx:axisymfrobenius}
With $\alpha=\frac{\mathrm{D}}{2}-1$, the Laplacian eigenfunction problem $r^2\bigtriangleup y(x)=-\lambda\,y(x)$ is the ultraspherical or Gegenbauer differential equation for the purely axisymmetric function $y$
\begin{align}
(1-x^2)y''
-
(2\alpha+1)\,x\,y'
+\lambda\,y=0.\nonumber
\end{align}
It is solved by a power series (Frobenius Ansatz)
\begin{align}
 y=\sum_{k=0}^\infty c_k\;x^k.\nonumber
\end{align}
Inserted into the differential equation, we can simplify, shift the indices $k$ to fit the powers $x^k$, and find by comparison of coefficients.
\begin{align}
(1-x^2)\,\sum_{k=2}^\infty c_k\,k\,(k-1)\,x^{k-2}
-(2\alpha+1)\,x\,\sum_{k=1}^\infty c_k\,k\,x^{k-1}
+\lambda\sum_{k=0}^\infty c_k\,x^{k}&=0
\nonumber
\\
\sum_{k=2}^\infty c_k\,k\,(k-1)\,x^{k-2}
-\sum_{k=2}^\infty c_k\,k\,(k-1)\,x^k
-(2\alpha+1)\,\sum_{k=1}^\infty c_k\,k\,x^{k}
+\lambda\sum_{k=0}^\infty c_k\,x^{k}&=0
\nonumber
\\
\sum_{k=0}^\infty c_{k+2}\,(k+1)\,(k+1)\,x^{k}
-\sum_{k=2}^\infty c_k\,k\,(k-1)\,x^k
-(2\alpha+1)\,\sum_{k=1}^\infty c_k\,k\,x^{k}
+\lambda\sum_{k=0}^\infty c_k\,x^{k}&=0
\nonumber
\\
\sum_{k=0}^\infty\Bigl\{
\bigl[
\lambda-k\,(k-1)\,u[k-2]
-(2\alpha+1)\,k\,u[k-1]
\bigr]\,c_k
+(k+1)\,(k+2)\,c_{k+2}
\Bigr\}x^{k}&=0\nonumber\\
c_{k+2}=\frac{(2\alpha+1)\,k+k\,(k-1)
	-\lambda}{(k+1)(k+2)}\;c_k.\quad\text{for $k\geq2$}\nonumber\\
\Longrightarrow
c_{k+2}=\frac{k(k+2\alpha)
	-\lambda}{(k+1)(k+2)}\;c_k.\quad\text{for $k\geq2$}\nonumber
\end{align}
Choosing $\lambda=\lambda_n=n(n+2\alpha)$, the series terminates at $k=n$.
The recurrence of the coefficients for a certain $n$ can be re-written to
\begin{align}
c_{k+2}^{(n)}=\frac{k^2+2\alpha\,k
	-n^2-2\alpha\,n}{(k+1)(k+2)}\;c_k=-\frac{(n-k)(n+k+2\alpha)}{(k+1)(k+2)}\;c_k^{(n)}.\nonumber
\end{align}
For even $n=2m$, odd-degree coefficients vanish $c_{2l+1}^{(2m)}=0$ and for odd $n=2m+1$, coefficients of even powers vanish $c_{2l}^{(2m+1)}$.
The non-zero coefficient series are composed factorials and double factorials (containing every second factor) when applying the recurrence repeatedly, and they are either based on a starting value $c_0^{(n)}$ or $c_1^{(n)}$.
%\begin{align}
%\text{even $n$:}&&    c_{k}^{(n)}&=\cdots
%    \frac{(4-n)(n+\mathrm{D}+2)}{5\cdot 6}
%    \overbrace{\frac{(2-n)(n+\mathrm{D}+0)}{3\cdot 4}
%    \underbrace{\frac{(0-n)(n+\mathrm{D}-2)}{1\cdot 2}\,c_0^{(n)}}_{c_2^{(n)}}}^{c_4^{(n)}},\nonumber\\
%\text{odd $n$:}&&     c_k^{(n)}&=\cdots
%    \frac{(5-n)(n+\mathrm{D}+3)}{6\cdot 7}
%    \overbrace{\frac{(3-n)(n+\mathrm{D}+1)}{4\cdot 5}
%    \underbrace{\frac{(1-n)(n+\mathrm{D}-1)}{2\cdot 3}\,c_1^{(n)}}_{c_3^{(n)}}}^{c_5^{(n)}}.\nonumber
%\end{align}
%or in short, using the double factorial symbol $(l)!!$, which consists of every second product term only, e.g. $7!!=1\cdot3\cdot5\cdot7$, $8!!=2\cdot4\cdot6\cdot8$, and with the modulo operator $\mathrm{mod}_{a,b}$
%\begin{align} 
%	c_k^{(n)}&=
%	\delta_{0,\mathrm{mod}_{k+n,2}}\mathrm{i}^{k+\mathrm{mod}_{n,2}}
%	\frac{(n+\mathrm{mod}_{n,2})!!}{(n+\mathrm{mod}_{n,2}+\mathrm{D}-4)!!}
%	\frac{(n+\mathrm{D}+k-4)!!}{(n-k)!!}\frac{1}{k!}\,c^{(n)}_{\mathrm{mod}_{n,2}}.
%\end{align}

We may replace the constants $k+2=n-2l$ and arrive at
\begin{align}
	c_{n-2l}^{(n)}&=-4(l+1)(n-l-1+\alpha)\frac{c_{n-2l-2}^{(n)}}{(n-2l)(n-2l-1)}
\end{align}
Starting from $l=0$, re-iterate the recurrence until arriving at $c_{n-2\lfloor\frac{n}{2}\rfloor}^{(n)}$ after $2l\leq n$ iterations, or with round-off operator $l\leq\lfloor\frac{n}{2}\rfloor$. For $l=\lfloor\frac{n}{2}\rfloor$, the term $n-l$ becomes $n-\lfloor\frac{n}{2}\rfloor=\lceil\frac{n}{2}\rceil$. We use rising factorials or Gamma functions $(\lceil\frac{n}{2}\rceil+\alpha-1)^{\overline{\lfloor\frac{n}{2}\rfloor-l}}=\frac{\Gamma(n-l+\alpha)}{\Gamma(\lceil\frac{n}{2}\rceil+\alpha)}=(\lceil\frac{n}{2}\rceil+\alpha-1)\dots(n-l+\alpha-2)(n-l+\alpha-1)$ to generalize the ratio of factorials to real arguments
\begin{align}
	c_{n-2l}^{(n)}&=\frac{(-1)^l\,2^{n-2l}\lfloor\frac{n}{2}\rfloor!(\lceil\frac{n}{2}\rceil+\alpha-1)^{\overline{\lfloor\frac{n}{2}\rfloor-l}}}{l!(n-2l)!}c_{n-2\lfloor\frac{n}{2}\rfloor}^{(n)}\nonumber\\
	&=
	\frac{(-1)^l\,2^{n-2l}\lfloor\frac{n}{2}\rfloor!}{l!(n-2l)!} \frac{\Gamma(n-l+\alpha)}{\Gamma(\lceil\frac{n}{2}\rceil+\alpha)}c_{n-2\lfloor\frac{n}{2}\rfloor}^{(n)}.\nonumber
\end{align}
The typical choice for the coefficient $c_{n-2\lfloor\frac{n}{2}\rfloor}^{(n)}=\frac{\Gamma(\lceil\frac{n}{2}\rceil+\alpha)}{\lfloor\frac{n}{2}\rfloor!\Gamma(\alpha)}$ removes clutter depending on integer division of $n$ and keeps the values of the Gamma functions $\frac{\Gamma(n-l+\alpha)}{\Gamma(\alpha)}$ finite, or more slim with the factorial $(\alpha)^{\overline{n-l}}$ rising from $(\alpha)$ to $(\alpha+n-l-1)$
\begin{align}
	c_{n-2l}^{(n)}&=\frac{(-1)^l\,2^{n-2l}}{l!(n-2l)!}
	 \frac{\Gamma(n-l+\alpha)}{\Gamma(\alpha)}
	 =\frac{(-1)^l\,2^{n-2l}(\alpha)^{\overline{n-l}}}{l!(n-2l)!},
\end{align}
defining the Gegenbauer polynomials \cite{Gegenbauer77}
\begin{align}
	C_n^{(\alpha)}(x)&=\sum_{l=0}^{\lfloor\frac{n}{2}\rfloor}\frac{(-1)^l\,2^{n-2l}(\alpha)^{\overline{n-l}}}{l!(n-2l)!}
	\,x^{n-2l},\label{eq:gegenbauer}
\end{align}
and for $\alpha=0$ the rising factorial simplifies to $(\alpha)^{\overline{n-l}}=(n-l-1)!$.

\section{Orthogonality of axisymmetric harmonics}\label{apdx:orthogonalityproof}
The axisymetric Laplacian differential equation 
$(1-x^2)\mathcal{P}''_n-(\mathrm{D}-1)\,x\,\mathcal{P}'_n
+\lambda_n\,\mathcal{P}_n=0$ with $\lambda_n=n(n+2\alpha)$ can be re-written compactly as
\begin{align}
   \frac{1}{w(x)}\left[(1-x^2)\,w(x)\,\mathcal{P}'_n\right]'+
  \lambda_n\mathcal{P}_n=0.
\end{align}
Multiplied by another solution $\mathcal{P}_m$ and integrated over $w(x)\,\mathrm{d}x$ it becomes
\begin{align}
\int_{-1}^1\left[(1-x^2)\,w(x)\,
\mathcal{P}'_n\right]'\mathcal{P}_m\,\mathrm{d}x
+\lambda_n\,
\int_{-1}^1
\mathcal{P}_n\,
\mathcal{P}_m\,w(x)\,\mathrm{d}x
=0.\nonumber
\end{align}
The first expression can be dealt with using partial integration 
%(fg)'=f'g+fg'
% int(fg)'=int f'g + int fg'
% fg = int f'g + int fg'
% int f'g = fg - int fg'
% int fg = (int f)g - int (int f)g'
$\int f'g = fg - \int fg':$
\begin{align}
\int_{-1}^1\left[(1-x^2)\,w(x)\,
\mathcal{P}'_n\right]'\mathcal{P}_m\,\mathrm{d}x
&=
(1-x^2)\,w(x)\,
\mathcal{P}'_n\,\mathcal{P}_m\Big|_{-1}^1-
\int_{-1}^{1}(1-x^2)\,w(x)\,
\mathcal{P}'_n\,\mathcal{P}'_m\,\mathrm{d}x\nonumber.
\end{align}
%In our case $\mathrm{D}>1$, i.e.\ $\alpha>-\frac{1}{2}$, the first expression $(1-x^2)\,w(x)=(1-x^2)^{\alpha-\frac{1}{2}}=0$ vanishes for the boundaries $x=\pm1$ while the polynomial $\mathcal{P}_n'\mathcal{P}_m$ stays finite.
%We get
%\begin{align}
%-\int_{-1}^{1}(1-x^2)\,w(x)\,
%\mathcal{P}'_n\,\mathcal{P}'_m\,\mathrm{d}x
%+
%\lambda_n\,
%\int_{-1}^1
%\mathcal{P}_n\,
%\mathcal{P}_m\,w(x)\,\mathrm{d}x
%=0. \nonumber
%\end{align}
To show that the solutions $\mathcal{P}_n$ and $\mathcal{P}_m$ with $n\neq m$ are orthogonal, we subtract a similar equation, the one with the roles of $\mathcal{P}_n$ and $\mathcal{P}_m$ interchanged, and get:
\begin{align}
(1-x^2)\,w(x)[\mathcal{P}_n'\mathcal{P}_m-\mathcal{P}_m'\mathcal{P}_n]\bigg|_{-1}^{1}+(\lambda_n-\lambda_m)\,
\int_{-1}^1
\mathcal{P}_n\,
\mathcal{P}_m\,w(x)\,\mathrm{d}x&=0,\label{eq:orthogonalityproof_intermediate}
\end{align}
because the two integrals $\int(1-x^2)\,w(x)\,\mathcal{P}_n'\mathcal{P}_m'\mathrm{d}x$ are identical. With the integration boundaries $-1\leq x\leq 1$, the factor  $(1-x^2)\,w(x)|_{\pm1}=0$ in front of the mixed derivatives vanishes, which happens for any $-\frac{1}{2}\geq\alpha$, i.e., $\mathrm{D}\geq1$. We get:
\begin{align}
(\lambda_n-\lambda_m)\,
\int_{-1}^1
\mathcal{P}_n\,
\mathcal{P}_m\,w(x)\,\mathrm{d}x=0.
\end{align}
If the indices match $n=m$, the first term vanishes and the integral needs not vanish. For different indices $n\neq m$ our eigenvalues are unique as $\lambda_n=n(n+\mathrm{D}-2)$, hence $\lambda_n\neq\lambda_m$, so the first term won't vanish, consequently the integral must vanish: 

Any pair of solutions $\mathcal{P}_n$ and $\mathcal{P}_m$ with different index $n\neq m$ is orthogonal under the integration weight $w(x)=(1-x^2)^{\alpha-\frac{1}{2}}$.

\section{Orthogonal polynomials}\label{apdx:orthopoly}
A set of orthogonal polynomials $\mathcal{P}_n(x)$ sorted by their degree $n$, up to which they contain powers of the variable $\mathcal{P}_n=\sum_{k=0}^n c_{n,k}\,x^k$ can naturally be used to re-expand any other limited-degree polynomial $\mathscr{P}_n=\sum_{k=0}^n c_k\,x^k$:
\begin{align}
\mathscr{P}_n(x)=\sum_{m=0}^n \gamma_m\,\mathcal{P}_m(x).
\end{align}
This is not surprising. What makes orthogonal polynomials useful is their orthogonality on the interval $[a,b]$:
\begin{align}
\int_a^b \mathcal{P}_n(x) \mathcal{P}_m(x)\; w(x)\,\mathrm{d}x = 
\begin{cases}
0\qquad\text{for $n\neq m$},\\
N_n^2\qquad\text{for $n=m$}.
\end{cases}
\end{align}
Here, $N_n^2=\int_a^b \mathcal{P}_n^2(x)\;w(x)\,\mathrm{d}x$ is the squared magnitude of each function integrated over the entire interval.

Note that by the ability of limited-degree polynomials $\mathcal{P}_{m<n}$ to express any $m^\mathrm{th}$-degree polynomial with $m<n$, the orthogonality automatically extendeds to \emph{any} lower-degree polynomial $\mathscr{P}_{m<n}=\sum_{l=0}^m\gamma_l\,\mathcal{P}_l$
\begin{align}
	\int_a^b \mathcal{P}_n(x) \mathscr{P}_{m<n}(x)\; w(x)\,\mathrm{d}x = 0.\label{eq:extended_orthogonality}
\end{align}

\subsection{The transform integral: a benefit of orthogonality}
The orthogonality allows to decompose any given function $g(x)$ into coefficients $\gamma_m$ of the orthogonal polynomials rather easily by an integral transformation. We write down the decomposition that we desire and integrate over a test function $\mathcal{P}_n(x)w(x)\mathrm{d}x\int$. Orthogonality simplifies the expression that we get
\begin{align}
g(x)&=\sum_{m=0}^\infty \gamma_m\,\mathcal{P}_m(x)\qquad\qquad\Big|\; \mathcal{P}_n(x)w(x)\mathrm{d}x\int_a^b\nonumber\\
\int_a^bg(x)\,\mathcal{P}_n(x)\;w(x)\,\mathrm{d}x&=\sum_{m=0}^\infty \gamma_m\,\underbrace{\int_a^b\mathcal{P}_m(x)\mathcal{P}_n(x)\;w(x)\,\mathrm{d}x}_{N_n^2\delta_{nm}}=N_n^2\;\gamma_n,\nonumber
\end{align}
We hereby found the expansion coefficients by the integral without matrix inversion. The transform pair is:
\begin{align}
g(x)&=\sum_{m=0}^\infty \gamma_m\,\mathcal{P}_m(x)\qquad\qquad \gamma_n=\frac{1}{N_n^2}\;\int_a^bg(x)\,\mathcal{P}_n(x)\;w(x)\,\mathrm{d}x.
\end{align}
As infinitely many orthogonal polynomials consist of any power $x^k$ with $k=0,1,2,\dots,\infty$, the representation is able to represent any function that can be described by an infinite power series.

\subsection{Symetries and interval $-1\leq x\leq 1$.} 
If $-a=b=1$ and $w(x)=w(-x)$, the integration range and integration weight is symmetrical. 
We can re-write the orthogonality condition
\begin{align}
\int_{-1}^{1}\mathcal{P}_n\,\mathcal{P}_m\;w\,\mathrm{d}x&=  \phantom{-}\int_{-1}^{0}\mathcal{P}_n(x)\,\mathcal{P}_m(x)\;w(x)\,\mathrm{d}x+
\int_{0}^{1}g(x)\,\mathcal{P}_n(x)\;w(x)\,\mathrm{d}x\nonumber\\
&=
-\int_{0}^{-1}\mathcal{P}_n(x)\,\mathcal{P}_m(x)\;w(x)\,\mathrm{d}x+
\int_{0}^{1}\mathcal{P}_n(x)\,\mathcal{P}_m(x)\;w(x)\,\mathrm{d}x\nonumber\\
&=
-\int_{0}^{1}\mathcal{P}_n(-x)\,\mathcal{P}_m(-x)\;w(-x)\,\mathrm{d}(-x)+
\int_{0}^{1}\mathcal{P}_n(x)\,\mathcal{P}_m(x)\;w(x)\,\mathrm{d}x\nonumber\\
&=
\phantom{-}\int_{0}^{1}\mathcal{P}_n(-x)\,\mathcal{P}_m(-x)\;w(x)\,\mathrm{d}x+
\int_{0}^{1}\mathcal{P}_n(x)\,\mathcal{P}_m(x)\;w(x)\,\mathrm{d}x\nonumber\\
&=
\phantom{-}\int_{0}^{1}\underbrace{\Bigl[\mathcal{P}_n(-x)\,\mathcal{P}_m(-x)+\mathcal{P}_n(x)\,\mathcal{P}_m(x)\Bigr]}_{=0\text{ for $n\neq m$}}\;w(x)\,\mathrm{d}x = N_n^2\,\delta_{nm}.
\end{align}
The expression in brackets $[\mathcal{P}_n(-x)\,\mathcal{P}_m(-x)+\mathcal{P}_n(x)\,\mathcal{P}_m(x)]$ vanishes whenever exclusively one of the two functions is odd, either $\mathcal{P}_n(-x)=-\mathcal{P}_n(x)$ or $\mathcal{P}_m(-x)=-\mathcal{P}_m(x)$, while the other one is even. The highest power in a polynomial corresponds to its degree, odd powers of $x^{2k+1}$ to odd function, even powers of $x^{2k}$ to even functions, so consequently orthogonal polynomials for the interval $-1\leq x\leq 1$ and weight $w(-x)=w(x)$ alternate between even and odd in $m$ to obtain orthogonality: 
Odd-degree orthogonal polynomials contain only odd powers, even-degree orthogonal polynomials only even powers.

\subsection{Three-term recurrence of orthogonal polynomials}\label{apdx:threeterm}
Normally, we define $\mathcal{P}_0(x)=1$ and any higher-degree polynomial by a simple recurrence relation, hence based on the lower degree polynomials. Finding the recurrence relation requires some consideration.

\noindent Orthogonality yields that we don't need an orthogonal polynomial of the $n^\mathrm{th}$ degree $\mathcal{P}_{n}(x)$ to represent an arbitrary lower-degree polynomial $\mathscr{P}_{m<n}(x)$. This becomes clear when expanding the polynomial $\mathscr{P}_{m<n}(x)=\sum_{k=0}^{m<n}\beta_k\,\mathcal{P}_k(x)$ and considering the transform integral with $\mathcal{P}_n$ vanishing $\int \mathcal{P}_{k<n}\mathcal{P}_n=0$ for every term because of orthogonality. We may write
\begin{align}
\int_a^b\mathscr{P}_{m<n}(x)\,\mathcal{P}_n(x)\;w(x)\,\mathrm{d}x &= 0.
\end{align}
After multiplication with $x$, an orthogonal polynomial of degree $n-1$ becomes an $n^\mathrm{th}$-degree polynomial 
\begin{align}
x\,\mathcal{P}_{n-1}(x)=\mathscr{P}_{n}(x).
\end{align}
We can find the expansion coefficients, $\beta_m$, thereof into orthogonal polynomials by integrating $x\,\mathcal{P}_{n-1}(x)$ over $\mathcal{P}_m(x)$. By applying the knowledge from above, we see which coefficients must vanish:
\begin{align}
\beta_{m<n-2}&=\frac{1}{N_{m<n-2}^2}\int_a^b\overbrace{\mathcal{P}_{m<n-2}(x)\,x}^{\mathscr{P}_{m<n-1}(x)}\,\mathcal{P}_{n-1}(x)\;w(x)\,\mathrm{d}x=0,\nonumber\\
\beta_{m>n}&=\frac{1}{N_{m>n}^2}\int_a^b\mathcal{P}_{m>n}(x)\,\underbrace{x\,\mathcal{P}_{n-1}(x)}_{\mathscr{P}_{n}(x)}\;w(x)\,\mathrm{d}x=0.\nonumber
\end{align}
Obviously, $x\,\mathcal{P}_{n-1}(x)$ only consists of $\mathcal{P}_{m}(x)$ with $m=n-2,n-1,n$: \emph{Any set of orthogonal polynomials obeys a three-term recurrence:}
\begin{align}
	x\,\mathcal{P}_{n-1}(x)&=\beta_n^{(n)}\,\mathcal{P}_{n}(x)+\beta_{n-1}^{(n)}\,\mathcal{P}_{n-1}(x)+\beta_{n-2}^{(n)}\,\mathcal{P}_{n-2}(x)
\end{align} 
that can be used as $\mathcal{P}_{n}(x)=\frac{1}{\beta_n^{(n)}}\Bigl[\bigl(x-\beta_{n-1}^{(n)}\bigr)\,\mathcal{P}_{n-1}(x)-\beta_{n-2}^{(n)}\,\mathcal{P}_{n-2}(x)\Bigr]$ to calculate higher-degree polynomials based on the lower-degree ones.
%The scalar $\frac{1}{\beta_n}$ may be removed as orthogonal polynomials need not be normalized.
The recurrence reduces for $-a=b=1$ and $w(-x)=w(x)$ as this requires $\mathcal{P}_{n}$ to only contain even or odd powers of $x$, 
\begin{align}
  x\,\mathcal{P}_{n-1}(x)=\beta_n^{(n)}\,\mathcal{P}_{n}(x)+\beta_{n-2}^{(n)}\,\mathcal{P}_{n-2}(x).\label{eq:threeterm_symm}
\end{align}

\subsection{Explicit three-term recurrence}
For the Gegenbauer polynomials  $C_n^{(\alpha)}(x)=\sum_{l=0}^{\lfloor\frac{n}{2}\rfloor}\frac{(-1)^l\,2^{n-2l}(\alpha)^{\overline{n-l}}}{l!(n-2l)!}
\,x^{n-2l}$ of eq.~\eqref{eq:gegenbauer}, we can obtain an explicit three-term recurrence by comparing the highest powers; divided by common factors $\frac{2^{n}(\alpha)^{\overline{n}}}{0!n!}$ and $\frac{2^{n-1}(\alpha)^{\overline{n-1}}}{0!(n-2)!}$ in the columns on the right
\begin{align}
	C_{n+1}^{(\alpha)}&=\frac{2^{n+1}(\alpha)^{\overline{n+1}}}{0!(n+1)!}x^{n+1}&-\frac{2^{n-1}(\alpha)^{\overline{n}}}{1!(n-1)!}x^{n-1}&\dots\bigg|
	&\frac{2(\alpha+n)}{n+1}&x^{n+1}\bigg| &-(\alpha+n-1)&x^{n-1}
	\nonumber\\
	xC_{n}^{(\alpha)}&=\frac{2^{n}(\alpha)^{\overline{n}}}{0!n!}x^{n+1}&-\frac{2^{n-2}(\alpha)^{\overline{n-1}}}{1!(n-2)!}x^{n-1}&\dots\bigg|
	& &x^{n+1}\bigg| &-\frac{n-1}{2}&x^{n-1}
	\nonumber\\
	C_{n-1}^{(\alpha)}&=&\frac{2^{n-1}(\alpha)^{\overline{n-1}}}{0!(n-2)!}x^{n-1}&\dots\bigg|
	&&\phantom{x^{n+1}}\bigg|&&x^{n-1}
	\nonumber
\end{align}
Combining $(n+1)C_{n+1}^{(\alpha)}-2(\alpha+n)xC_{n}^{(\alpha)}$ obviously cancels the power $x^{n+1}$ and leaves $-(n+1)(\alpha+n-1)+(n-1)(\alpha+n)=-2\alpha-n+1$ to cancel by $C_{n-1}^{(\alpha)}$. We get 
\begin{align}
	2(n+\alpha)\,x\,C_n^{(\alpha)}=(n+1)C_{n+1}^{(\alpha)}+(n+2\alpha-1)C_{n-1}^{(\alpha)}.\label{eq:gegenbauer_rec}
\end{align}

\paragraph{For standardization to $\mathcal{P}_n(1)=1$}
If we standardize the Gegenbauer polynomials to ultraspherical ones $\mathcal{P}_n(1)=1$ for all $n$, eq.~\eqref{eq:threeterm_symm} imposes the condition that $\beta_{n}^{(n)}+\beta_{n-2}^{(n)}=1$, so that $\beta_{n-2}^{(n)}=1-\beta_{n}^{(n)}$. We skip the superscript and write $\beta_n$ for the recurrence coefficient,
\begin{align}
	x\,\mathcal{P}_{n-1}(x)=\beta_n\,\mathcal{P}_{n}(x)+(1-\beta_{n})\,\mathcal{P}_{n-2}(x).\label{eq:three-term-recurrence}
\end{align}
We obtain the recurrence coefficient $\beta_n$ from comparing coefficients of the highest powers in $\mathcal{P}_n$ and $\mathcal{P}_{n-1}$. This is done based on the coefficient recurrence for $c_n^{(n)}$, 
\begin{align}
	\beta_n\,c_{n}^{(n)}&=c_{n-1}^{(n-1)}\Longrightarrow \beta_n=\frac{c_{n-1}^{(n-1)}}{c_n^{(n)}}.\nonumber
\end{align}

From above we know about the existence and nature of three-term recurrences. 
For the polynomials with $\mathcal{P}_n(1)=1$ in eqs.\eqref{eq:gegenbauer} \eqref{eq:gegenbauer1}, we get a ratio of the $l=0$ coefficients for $n$ and $n+1$ as
$\beta_{n+1}=\frac{c_{n}^{(n)}}{c_{n+1}^{(n+1)}}=\frac{2^{n}\,\Gamma(n+\alpha)}{\Gamma(n+2\alpha)}\frac{\Gamma(n+2\alpha+1)}{2^{n+1}\,\Gamma(n+\alpha+1)}=\frac{n+2\alpha}{2(n+\alpha)}$ and $1-\beta_{n+1}=\frac{n}{2(n+\alpha)}$ for a recurrence as in eq.~\eqref{eq:three-term-recurrence}
\begin{align}
	2(n+\alpha)\,x\,\mathcal{P}_n&=(n+2\alpha)\mathcal{P}_{n+1}+n\mathcal{P}_{n-1}\label{eq:standard-three-term-recurrence}.
\end{align}

\subsection{Squared norm of orthogonal polynomials}
According to Law and Sledd~\cite{LawSledd75}, the recurrence coefficient $\beta_n$ can also be used to evaluate the normalization $N_n^2=\int_{-1}^{1}[\mathcal{P}_n(x)]^2\,w(x)\,\mathrm{d}x$ of the polynomial recursively.
This is done by re-formulating the three-term recurrence to 
\begin{align}
	\mathcal{P}_{n}&=(A_nx+B_n)\mathcal{P}_{n-1}-C_n\mathcal{P}_{n-2}.\label{eq:lawsledd_rec}
\end{align}
Law and Sledd's derivation is based on explicitly evaluating only the first term of the recurrence until $\mathcal{P}_0=1$, while the rest must stay some polynomial of the degree $n-1$
\[
\mathcal{P}_{n}=A_nx\mathcal{P}_{n-1}+\mathscr{A}_{n-1}=A_nA_{n-1}x^2\mathcal{P}_{n-2}+\mathscr{B}_{n-1}=A_nA_{n-1}\dots A_{1}x^n+\mathscr{C}_{n-1},
\]
or more neatly
\begin{align}
	x^n=(A_nA_{n-1}\dots A_{1})^{-1}\mathcal{P}_{n}+\mathscr{P}_{n-1}.\label{eq:lawsledd_mu_power}
\end{align}
Integration over $\mathcal{P}_n\,w\,\mathrm{d}x\,\int_a^b$ yields
\begin{align}
	\int_a^bx^n\mathcal{P}_n\,w\,\mathrm{d}x
	&=(A_nA_{n-1}\dots A_{1})^{-1}\underbrace{\int_a^b(\mathcal{P}_n)^2\,w\,\mathrm{d}x}_{=N_n^2}+
	\underbrace{\int_a^b\mathscr{P}_{n-1}\,\mathcal{P}_n\,w\,\mathrm{d}x}_{=0}\nonumber\\
	\int_a^bx^n\mathcal{P}_n\,w\,\mathrm{d}x&=(A_nA_{n-1}\dots A_{1})^{-1}\,N_n^2.\label{eq:lawsledd_3}
\end{align}
$x^n\mathcal{P}_n$ is re-expressed by re-arranging eq.~\eqref{eq:lawsledd_rec} to
$x\mathcal{P}_{n}=
A_{n+1}^{-1}[\mathcal{P}_{n+1}-B_{n+1}\mathcal{P}_{n}+C_{n+1}\mathcal{P}_{n-1}]$
and multiplication with $x^{n-1}=(A_{n-1}\dots A_1)^{-1}[\mathcal{P}_{n-1}+\mathscr{P}_{n-2}]$ from eq.~\eqref{eq:lawsledd_mu_power}, 
\begin{align}
	\int_a^bx^n\mathcal{P}_n\,w\,\mathrm{d}x
	&=\frac{\int_a^b
		[\mathcal{P}_{n+1}-B_{n+1}\mathcal{P}_{n}+C_{n+1}\mathcal{P}_{n-1}][\mathcal{P}_{n-1}+\mathscr{P}_{n-2}]
		\,w\,\mathrm{d}x}{A_{n+1}(A_{n-1}\dots A_1)}
	\nonumber\\
	&=\frac{C_{n+1}}{A_{n+1}(A_{n-1}\dots A_1)}\underbrace{\int_a^b
		(\mathcal{P}_{n-1})^2
		\,w\,\mathrm{d}x}_{=N_{n-1}^2}.
\end{align}
Together with eq.~\eqref{eq:lawsledd_3}, we get for 
\begin{align}
	N_{n}^2=\frac{C_{n+1}A_n}{A_{n+1}}N_{n-1}^2,\label{eq:normalizationrecurrence}
\end{align}
which is in our $\mathcal{P}_n(1)=1$ standardized case with
$\beta_{n}\mathcal{P}_{n}(x)=x\mathcal{P}_{n-1}-(1-\beta_{n})\mathcal{P}_{n-2}$
hence $A_n=\frac{1}{\beta_{n}}$, and $C_n=\frac{1-\beta_{n}}{\beta_{n}}$, as well as $\mathcal{P}_0=1$ and $\int_{-1}^1(\mathcal{P}_0)^2\,w(x)\,\mathrm{d}x=\frac{S_{\mathrm{D}-1}}{S_{\mathrm{D}-2}}$, and so
\begin{align}
	N_{n}^2&=\frac{1-\beta_{n+1}}{\beta_{n}}\,N_{n-1}^2=\frac{S_{\mathrm{D}-1}}{S_{\mathrm{D}-2}}\prod_{k=1}^n\frac{1-\beta_{k+1}}{\beta_k}\label{eq:normalizationrecurrence1};
\end{align}
see Appdx.~\ref{apdx:squarednormgegenbauer} for more details.

\subsection{Christoffel-Darboux formula}\label{apdx:Chrisfoffel}
From the general three-term recurrence
\begin{align}
	A_{n+1}x\mathcal{P}_{n}(x)&=\mathcal{P}_{n+1}(x)-B_{n+1}\mathcal{P}_{n}(x)+C_{n+1}\mathcal{P}_{n-1}(x)\nonumber
\end{align}
one finds the Cristoffel-Darboux formula by expanding with $\mathcal{P}_n(y)$ and $(x-y)$
\begin{align}
	A_{n+1}(x-y)\mathcal{P}_{n}(x)\mathcal{P}_{n}(y)
	&=[\mathcal{P}_{n+1}(x)\mathcal{P}_{n}(y)-\mathcal{P}_{n}(x)\mathcal{P}_{n+1}(y)]\nonumber\\&+C_{n+1}[\mathcal{P}_{n-1}(x)\mathcal{P}_{n}(y)-\mathcal{P}_{n}(x)\mathcal{P}_{n-1}(y)].\nonumber
\end{align}
Multiplying with $\frac{1}{A_{n+1}N_n^2}$ yields with the normalization recurrence $\frac{A_{n+1}}{C_{n+1}}N_n^2=A_n\,N_{n-1}^2$ from eq.~\eqref{eq:normalizationrecurrence}
\begin{align}
	\frac{\mathcal{P}_{n}(x)\mathcal{P}_{n}(y)}{N_n^2}
	&=\underbrace{\frac{\mathcal{P}_{n+1}(x)\mathcal{P}_{n}(y)-\mathcal{P}_{n}(x)\mathcal{P}_{n+1}(y)}{(x-y)\,A_{n+1}N_n^2}}_{\mathcal{F}_n(x,y)}+\underbrace{\frac{\mathcal{P}_{n-1}(x)\mathcal{P}_{n}(y)-\mathcal{P}_{n}(x)\mathcal{P}_{n-1}(y)}{(x-y)\,A_{n}N_{n-1}^2}}_{-\mathcal{F}_{n-1}(x,y)}.\nonumber
\end{align}
Adding all expressions of smaller $n$ until $n=0$, where $\mathcal{F}_{-1}(x,y)=0$, yields the Christoffel-Darboux formula decomposing an axisymmetric Dirac delta $\delta(x-y)$ 
\begin{align}
	\sum_{n=0}^\mathrm{N}\frac{\mathcal{P}_{n}(x)\mathcal{P}_{n}(y)}{N_n^2}
	&=\frac{\mathcal{P}_{\mathrm{N}+1}(x)\mathcal{P}_{\mathrm{N}}(y)-\mathcal{P}_{\mathrm{N}}(x)\mathcal{P}_{\mathrm{N}+1}(y)}{(x-y)\,A_{\mathrm{N}+1}N_\mathrm{N}^2}.\label{eq:christoffeldarboux}
\end{align}
If $y=x$, the fraction exhibits a pole and the confluent Christoffel-Darboux identity is obtained by $y=x+\Delta x$ and taking the limit $\lim_{\Delta x\rightarrow0}$
\begin{align}
	\sum_{n=0}^\mathrm{N}\frac{[\mathcal{P}_{n}(x)]^2}{N_n^2}
	&=\lim_{\Delta x\rightarrow0}\frac{\mathcal{P}_{\mathrm{N}+1}(x)\mathcal{P}_{\mathrm{N}}(x+\Delta x)-\mathcal{P}_{\mathrm{N}}(x)\mathcal{P}_{\mathrm{N}+1}(x+\Delta x)}{-\Delta x\,A_{\mathrm{N}+1}N_\mathrm{N}^2}\nonumber\\
	&=\frac{\mathcal{P}_{\mathrm{N}}(x)\mathcal{P}_{\mathrm{N}+1}'(x)-\mathcal{P}_{\mathrm{N}+1}(x)\mathcal{P}_{\mathrm{N}}'(x)}{A_{\mathrm{N}+1}N_\mathrm{N}^2},
\end{align}
which is obviously finite.
This equation is useful to simplify the finite-degree representation of a Dirac delta at $y=1$, as, e.g., found in Rafaely's book~\cite{Rafaely19}. For normalization at $x=1$, all factors can be disregarded and using the derivative at $x=1$ in eq.~\eqref{eq:derivatives_at_1}, we find $\mathcal{P}_\mathrm{N+1}'(1)-\mathcal{P}_\mathrm{N}'(1)=\frac{(\mathrm{N}+1)(\mathrm{N}+\mathrm{D}-1)-\mathrm{N}(\mathrm{N}+\mathrm{D}-2)}{\mathrm{D}-1}=\frac{2\mathrm{N}+\mathrm{D}-1}{\mathrm{D}-1}$ so that $g(1)=1$ and elsewhere eq.~\eqref{eq:christoffeldarboux} holds with the newly obtained factor
\begin{align}
	g(x)&=\frac{\mathrm{D}-1}{2\mathrm{N}+\mathrm{D}-1}\frac{\mathcal{P}_{\mathrm{N}+1}(x)-\mathcal{P}_{\mathrm{N}}(x)}{x-1}.\label{eq:finite_order_dirac}
\end{align}

\subsection{Polynomials $C_n^{(\alpha)}(x)$ and $\mathcal{P}_n(x)$ at $x=1$}\label{apdx:gegenbauer}
To find the values $C_n^{(\alpha)}(1)$, we regard the three-term recurrence eq.~\eqref{eq:gegenbauer_rec}  for $x=1$\begin{align}
	(n+1)C_{n+1}^{(\alpha)}-2(\alpha+n)\,C_{n}^{(\alpha)}+(2\alpha+n-1)C_{n-1}^{(\alpha)}=0.
	\nonumber
\end{align}
As there is multiplication of $C_{n+1}^{(\alpha)}$ with $(n+1)$, which is a hint for $\frac{1}{n!}$, and multiplication of $C_{n-2}^{(\alpha)}$ with $(2\alpha+n-1)$, hinting towards $(2\alpha)^{\overline{n}}$, we test a solution $C_n^{(\alpha)}=\frac{(2\alpha)^{\overline{n}}}{n!}$ 
\begin{align}
	\frac{(2\alpha)^{\overline{n}}}{(n-1)!}-2(\alpha+n-1)\,\frac{(2\alpha)^{\overline{n-1}}}{(n-1)!}+\frac{(2\alpha)^{\overline{n-1}}}{(n-2)!}=0
	\nonumber\\
	(2\alpha+n-1)(2\alpha)^{\overline{n-1}}
	-(2\alpha+2n-2)\,(2\alpha)^{\overline{n-1}}+(n-1)(2\alpha)^{\overline{n-1}}=0.\qquad\blacksquare
	\nonumber
\end{align}
Therefore the value of the Gegenbauer polynomials at $x=1$ become
\begin{align}
	C_n^{(\alpha)}(1)&=\frac{(2\alpha)^{\overline{n}}}{n!}, &
	\text{and } 
	C_n^{(\alpha)}(x)&=\frac{(2\alpha)^{\overline{n}}}{n!}\mathcal{P}_n^{(\alpha)}(x).\label{eq:gegenbauer_at_1}
\end{align}
Division of eq.~\eqref{eq:gegenbauer} by $\frac{(2\alpha)^{\overline{n}}}{n!}$ lets us also get an explicit power-series expansion of $\mathcal{P}_n$
\begin{align}
	\mathcal{P}_n(x)&=\sum_{l=0}^{\lfloor\frac{n}{2}\rfloor}\frac{(-1)^l2^{n-2l}\,n!}{l!(n-2l)!}
	\frac{(\alpha)^{\overline{n-l}}}{(2\alpha)^{\overline{n}}}\,x^{n-2l},\label{eq:gegenbauer1}
\end{align}
which yields numerical difficulties at $\alpha\rightarrow0$. We  re-formulate the rising factorials as $(\alpha)^{\overline{n-l}}=\Gamma(\alpha+n-l)|_0/\Gamma(\alpha)=(n-l-1)!/\Gamma(\alpha)$ and $(2\alpha)^{\overline{n}}=\Gamma(2\alpha+n)|_0/\Gamma(2\alpha)=(n-1)!/\Gamma(2\alpha)$,  use  duplication~\cite[5.5.E5]{nist} $\Gamma(2z)/\Gamma(z)=2^{2z-1/2}\Gamma(z+1/2)/\sqrt{2\pi}$, and get with $\Gamma(1/2)=\sqrt{\pi}$ the limit 
$\lim_{\alpha\rightarrow 0}\frac{\Gamma(2\alpha)}{\Gamma(\alpha)}=1/2$, and so  $\lim_{\alpha\rightarrow0}
\frac{(\alpha)^{\overline{n-l}}}{(2\alpha)^{\overline{n}}}=\frac{(n-l-1)!}{(n-1)!}$,
\begin{align}
	\mathcal{P}_n(x)&=\lim_{\alpha\rightarrow 0}\frac{\Gamma(2\alpha)}{\Gamma(\alpha)}\sum_{l=0}^{\lfloor\frac{n}{2}\rfloor}\frac{(-1)^l2^{n-2l}\,n!}{l!(n-2l)!}
	\frac{(n-l-1)!}{(n-1)!}\,x^{n-2l},\nonumber\\
	&=\sum_{l=0}^{\lfloor\frac{n}{2}\rfloor}
	\frac{(-1)^l\,2^{n-2l-1}n!\;(n-l-1)!}{l!(n-2l)!}\,x^{n-2l}.\nonumber
\end{align}

\subsection{Derivative of the polynomials $\mathcal{P}_n(x)$ at $x=1$}
The ultraspherical differential equation $(1-x^2)\,\mathcal{P}_n''-(2\alpha+1)\,x\,\mathcal{P}_n'+n(n+2\alpha)\,\mathcal{P}_n=0$ becomes at $x=1$ with $(1-x^2)=0$ and $\mathcal{P}_n(1)=1$ 
\begin{align}
	\mathcal{P}_n'(1)&=\frac{n(n+2\alpha)}{2\alpha+1}=\frac{n(n+\mathrm{D}-2)}{\mathrm{D}-1}.
	\label{eq:derivatives_at_1}
\end{align} 

\subsection{Value of the polynomials $C^{(\alpha)}_n(x)$ and $\mathcal{P}_n(x)$ at $x=0$}
By the explicit series expansions, we can evaluate the Gegenbauer polynomials at $x=0$. Only such with even degree have a non-zero value there:
\begin{align}
	C_{2m}^{(\alpha)}(0)&=\sum_{l=0}^{m}\frac{(-1)^{l}(\alpha)^{\overline{2m-l}}}{l!(2m-2l)!}\underbrace{(2x)^{n-2l}\bigg|_{0}}_{\neq0:l=m}=\frac{(-1)^m\,(\alpha)^{\overline{m}}}{m!}.
\end{align}
With the ultraspherical polynomials $\mathcal{P}_n$, this becomes by division with $\frac{(2\alpha)^{\overline{n}}}{n!}$
\begin{align}
	\mathcal{P}_{2m}^{(\alpha)}(0)&=\frac{(-1)^m\,(2m)!(\alpha)^{\overline{m}}}{m!(2\alpha)^{\overline{2m}}}.\label{eq:gegenbauer_at_0}
\end{align}

%\subsection{Generating function}
%In the inhomogeneous potential equation $\bigtriangleup G=-\delta(\bm x-\bm x_0)$, if the source coordinate $\bm x_0$ is aligned with the last coordinate axis $x_\mathrm{D}$, and $x=\cos\vartheta$ measures the inclination of the other position, and $r$ measures the radius of $\bm x$ from the origin, while the source is at $r_0=\|\bm x_0\|=1$, the so called generating function constitutes the solution 
%\begin{align}
%	\frac{1}{(1-2rx+r^2)^\alpha}&=\sum_{n=0}^\infty r^n\,C_n^{(\alpha)}(x),
%\end{align}
%and for $\alpha=0$ it will be exceptionally $-\ln(1-2rx+r^2)$.
%Deriving with regard to $r$ and collecting the coefficients of $r^n$ also yields the multiplication recurrence relation.

\subsection{Rodrigues' formula for Gegenbauer polynomials.}
The French mathematician Rodrigues provided a formula for the Legendre polynomials that are orthogonal using the integral measure $w(x)=1$. A general Rodrigues formula for a more arbitrary measure $w(x)=(1-x^2)^{\alpha-\frac{1}{2}}$ can be given to specify the corresponding orthogonal polynomials $\mathcal{P}_n$. 

The ultraspherical (Gegenbauer) differential equation for axisymmetric harmonic solutions in $\alpha=\frac{\mathrm{D}}{2}-1$ dimensions is 
\begin{align}
(1-x^2)\mathcal{P}_n''(x)
-(2\alpha+1)x\mathcal{P}_n'(x)+n(n+2\alpha)\mathcal{P}_n(x)&=0,\label{eq:gegenbauerdgl}
\end{align}
and it is solved by the generalized Rodrigues formula
\begin{align}
\mathcal{P}_n(x)=e_n^{-1}(1-x^2)^{-\alpha+{\textstyle\frac{1}{2}}}\frac{\mathrm{d}^n}{\mathrm{d}x^n}(1-x^2)^{\alpha-{\textstyle\frac{1}{2}}+n}.\nonumber
\end{align}

\paragraph{Proof.}
To show this, we start with the inner-most function 
\[
f_n=(1-x^2)^{\alpha-{\textstyle\frac{1}{2}}+n}
\]
that the solution is constructed of and derive it
\[
f_n'=(n+\alpha-{\textstyle\frac{1}{2}})(1-x^2)^{\alpha-\frac{1}{2}+n-1}(-2x)=-2(n+\alpha-{\textstyle\frac{1}{2}})x(1-x^2)^{-1}f_n,
\]
yielding a differential equation after multiplication with $(1-x^2)$:
\[
(1-x^2)f_n'+(2n+2\alpha-1)x f_n=0.
\]
Deriving again yields a second-order differential equation
\begin{align}
(1-x^2)f_n''-2x f_n'+(2n+2\alpha-1)x f_n'+(2n+2\alpha-1)f_n&=0\nonumber\\
(1-x^2)f_n''+(2n+2\alpha-3)x f_n'+(2n+2\alpha-1)f_n&=0.\nonumber
\end{align}
Now this second-order differential equation in $f_n=(1-x^2)^{\alpha-\frac{1}{2}+n}$ is derived $n$ times requiring for the first two expressions the Leibniz rule for repeated differentiation of products $\frac{\mathrm{d}^n}{\mathrm{d}x^n}(f\,g)=\sum_{k=0}^n\frac{n!}{k!(n-k)!}f^{(k)}\,g^{(n-k)}$, 
\begin{align}
(1-x^2)f_n^{(n+2)}
-2nx f_n^{(n+1)}
-n(n-1)f_n^{(n)} \nonumber\\%erste runde
+(2n+2\alpha-3)x f_n^{(n+1)}
+n(2n+2\alpha-3)f_n^{(n)} \nonumber\\%zweite runde
+(2n+2\alpha-1)f_n^{(n)}&=0\nonumber\\
(1-x^2)f_n^{(n+2)} 
+(2\alpha-3)x f_n^{(n+1)}
+\left[n(n+2\alpha)
+(2\alpha-1)\right]
f_n^{(n)} 
&=0.\nonumber
\end{align}
The above differential equation is second-order for $f_n^{(n)}=\frac{\mathrm{d}^n}{\mathrm{d}x^n}(1-x^2)^n$, which is related to the Rodrigues solution by $f_n^{(n)}=e_n(1-x^2)^{\alpha-\frac{1}{2}}\,\mathcal{P}_n$, which upon insertion and omission of $e_n$ becomes
\begin{align}
(1-x^2)[(1-x^2)^{\alpha-\frac{1}{2}}\,\mathcal{P}_n]''
+(2\alpha-3)x[(1-x^2)^{\alpha-\frac{1}{2}}\,\mathcal{P}_n]'\nonumber\\
+\left[n(n+2\alpha)
+(2\alpha-1)\right](1-x^2)^{\alpha-\frac{1}{2}}\,\mathcal{P}_n
&=0\nonumber\\
(1-x^2)[
(1-x^2)^{\alpha-\frac{1}{2}}\,\mathcal{P}_n''
-2(2\alpha-1) x(1-x^2)^{\alpha-\frac{3}{2}}\mathcal{P}_n'\quad\nonumber\\
-(2\alpha-1) (1-x^2)^{\alpha-\frac{3}{2}}\mathcal{P}_n
+(2\alpha-1)(2\alpha-3) x^2(1-x^2)^{\alpha-\frac{5}{2}}\mathcal{P}_n
]\nonumber\\%erste runde
+(2\alpha-3)x[
(1-x^2)^{\alpha-\frac{1}{2}}\,\mathcal{P}_n'
-(2\alpha-1) x(1-x^2)^{\alpha-\frac{3}{2}}\,\mathcal{P}_n
]\nonumber\\%erste runde
+\left[n(n+2\alpha)
+(2\alpha-1)\right](1-x^2)^{\alpha-\frac{1}{2}}\,\mathcal{P}_n(x)
&=0\quad\Big|(1-x^2)^{-\alpha+\frac{1}{2}}\nonumber\\
(1-x^2)[\mathcal{P}_n''
-2(2\alpha-1) x(1-x^2)^{-1}\mathcal{P}_n'\quad\nonumber\\
-\cancel{(2\alpha-1) (1-x^2)^{-1}\mathcal{P}_n}
+\cancel{(2\alpha-1)(2\alpha-3) x^2(1-x^2)^{-2}\mathcal{P}_n}
]\nonumber\\%erste runde
+(2\alpha-3)x[
\mathcal{P}_n'
-\cancel{(2\alpha-1) x(1-x^2)^{-1}\,\mathcal{P}_n}
]\nonumber\\%erste runde
+\left[
n(n+2\alpha)+
\cancel{(2\alpha-1)}\right]\,\mathcal{P}_n(x) 
&=0\nonumber\\
(1-x^2)\mathcal{P}_n''(x)
-(2\alpha+1)x\mathcal{P}_n'(x)+n(n+2\alpha)\mathcal{P}_n(x)&=0.\nonumber
\end{align}
As this is the ultraspherical (Gegenbauer) differential equation, and we obtained it by the $(n+2)^\mathrm{th}$ derivative of $f_n=(1-x^2)^{\alpha-\frac{1}{2}+n}$ and its substitution $f_n^{(n)}=e_n(1-x^2)^{\alpha-\frac{1}{2}}\mathcal{P}_n$, we hereby proved Rodrigues' solution 
$\mathcal{P}_n=e_n^{-1}(1-x^2)^{-\alpha+\frac{1}{2}}\frac{\mathrm{d}^n}{\mathrm{d}x^n}(1-x^2)^{\alpha-\frac{1}{2}+n}$.

To find the normalization factor $e_n$ for standardization of $\mathcal{P}_n(1)$, here for $\mathcal{P}_n(1)=1$, we apply Leibniz's rule for repeated differentiation
\begin{align}
	\mathcal{P}_n(x)&=e_n^{-1}(1-x^2)^{-\alpha+\frac{1}{2}}\frac{\mathrm{d}^n}{\mathrm{d}x^n}[(1-x)^{\alpha-\frac{1}{2}+n}(1+x)^{\alpha-\frac{1}{2}+n}]\nonumber\\
	&=e_n^{-1}\sum_{k=0}^n\frac{n!}{k!(n-k)!}\frac{(1+x)^{-\alpha+\frac{1}{2}}\mathrm{d}^{k}(1+x)^{\alpha-\frac{1}{2}+n}}{\mathrm{d}x^{k}}\frac{(1-x)^{-\alpha+\frac{1}{2}}\mathrm{d}^{n-k}(1-x)^{\alpha-\frac{1}{2}+n}}{\mathrm{d}x^{n-k}}\nonumber\\
	&=e_n^{-1}\sum_{k=0}^n\frac{n!}{k!(n-k)!}\frac{\Gamma(\alpha+n+\frac{1}{2})(1+x)^{n-k}}{\Gamma(\alpha+n-k+\frac{1}{2})}\underbrace{\frac{\Gamma(\alpha+n+\frac{1}{2})(-1)^{n-k}(1-x)^{k}}{\Gamma(\alpha+k+\frac{1}{2})}}_{\neq0:x=1:k=0}
	\nonumber\\
	\mathcal{P}_n(x)\Big|_{1}&=e_n^{-1}\,\frac{\Gamma(\alpha+n+\frac{1}{2})(2)^{n}(-1)^{n}}{\Gamma(\alpha+\frac{1}{2})}=1\nonumber\\
	e_n&=\frac{(2)^n\Gamma(\alpha+n+\frac{1}{2})}{(-1)^n\Gamma(\alpha+\frac{1}{2})}=\frac{2^n(\alpha+\frac{1}{2})^{\overline{n}}}{(-1)^n}.
\end{align}
Hereby, Rodrigues' formula for Gegenbauer polynomials standardized to ultraspherical ones $\mathcal{P}_n(1)=1$ becomes:
\begin{align}
	\mathcal{P}_n(x)=\frac{(-1)^n}{2^n(\alpha+\frac{1}{2})^{\overline{n}}}
	(1-x^2)^{-\alpha+\frac{1}{2}}\frac{\mathrm{d}^n}{\mathrm{d}x^n}(1-x^2)^{\alpha-\frac{1}{2}+n},\label{eq:rodrigues1}
\end{align}
and alternatively we could write with the integration weight $w(x)=(1-x^2)^{\alpha-\frac{1}{2}}$ inserted 
$\mathcal{P}_n(x)=\frac{(-1)^n}{(2\alpha+1)^{\overline{n}}\,w(x)}
\frac{\mathrm{d}^n}{\mathrm{d}x^n}\left[(1-x^2)^{n}\,w(x)\right]$.

The typical Gegenbauer standardization is $C_n^{(\alpha)}(1)=\frac{\Gamma(n+2\alpha)}{n!\Gamma(2\alpha)}=\frac{(2\alpha)^{\overline{n}}}{n!}$ as in eq.~\eqref{eq:gegenbauer_at_1}, and its Rodrigues' formula  is obtained by multiplying the above with $\frac{(2\alpha)^{\overline{n}}}{n!}$, cf.~eq.~\eqref{eq:gegenbauer_at_1},
\begin{align}
	C_n^{(\alpha)}(x)=
	\frac{(-1)^n(2\alpha)^{\overline{n}}}{n!2^n(\alpha+\frac{1}{2})^{\overline{n}}}
	(1-x^2)^{-\alpha+\frac{1}{2}}\frac{\mathrm{d}^n}{\mathrm{d}x^n}(1-x^2)^{\alpha-\frac{1}{2}+n}.
\end{align}

\subsection{Norm of Gegenbauer polynomials}\label{apdx:squarednormgegenbauer}
For the transform integral, the squared norm $N_n^2=\int_{-1}^{1}|\mathcal{P}_n(x)|^2\,w(x)\,\mathrm{d}x$ plays an important role for normalization.
As the zeroth-degree polynomial is $C_0^{(\alpha)}(x)=\mathcal{P}_0(x)=1$, its norm $N_0^2=\int_{-1}^{1}w(x)\,\mathrm{d}x$ should be easiest to find. The others follow from eq.~\eqref{eq:normalizationrecurrence}.

\paragraph{Finding $N_0^2$.}
Using all coordinates of the $\mathrm{D}$-dimensional space, the integral over $x$ the last one of multiple integrals.
In the sequence of the $(\mathrm{D}-1)$-fold integration over $1$ to determine the surface $S_{\mathrm{D}-1}$ of the $\mathbb{S}^{\mathrm{D}-1}$ unit sphere in a $\mathrm{D}$-dimensional space, the integral over $x$ delivers the distinguishing factor between the surfaces  $S_{\mathrm{D}-1}$ and $S_{\mathrm{D}-2}$, the one of the next lower-dimensional unit sphere, cf.~\cite[vol.2, p.387]{HilbertCourant},
\begin{align}
	S_{\mathrm{D}-1}&=\underbrace{\iiint\,\mathrm{d} S_{\mathrm{D}-2}}_{=S_{\mathrm{D}-2}}\cdot \,\underbrace{\int_{-1}^{1}\,\sqrt{1-x^2}^{\mathrm{D}-3}\,\mathrm{d}x}_{=N_0^2}.
\end{align}
With $S_{\mathrm{D}-1}=\frac{2\,\sqrt{\pi}^{\mathrm{D}}}{\Gamma(\frac{\mathrm{D}}{2})}$~\cite[vol.2, p.387]{HilbertCourant}, we conclude that
\begin{align}
	N_0^2&=\frac{S_{\mathrm{D}-1}}{S_{\mathrm{D}-2}}=\frac{\sqrt{\pi}\,\Gamma(\frac{\mathrm{D}-1}{2})}{\Gamma(\frac{\mathrm{D}}{2})}.
\end{align}
For 2 dimensions $\mathrm{D}=2$ we get $N_0^2=\frac{2\pi}{2}=\pi$, and for 3 dimensions $\mathrm{D}=3$ we get $N_0^2=\frac{4\pi}{2\pi}=2$, for both kinds $\mathcal{P}_n$ and $C_n^{(\alpha)}$.

\paragraph{Squared norm for any degree $N_n^2$ for $C_n^{(\alpha)}$.} 
eq.~\eqref{eq:normalizationrecurrence} obtains the norm $	N_{n}^2=\frac{C_{n+1}A_n}{A_{n+1}}N_{n-1}^2$ of higher-degree polynomials recursively from $N_0^2$.
We use the recurrence $C_{n}^{(\alpha)}=\underbrace{\textstyle\frac{2(n+\alpha-1)}{n}}_{A_n}x\,C_{n-1}^{(\alpha)}-\underbrace{\textstyle\frac{n+2\alpha-2}{n}}_{C_n}\,C_{n-2}^{(\alpha)}$
to find the squared norm, and $\alpha=\frac{\mathrm{D}}{2}-1$,
\begin{align}
	N_n^2&=\frac{C_{n+1}A_n}{A_{n+1}}N_{n-1}^2=\frac{\frac{n+2\alpha-1}{n+1}\frac{2(n+\alpha-1)}{n}}{\frac{2(n+\alpha)}{n+1}}N_{n-1}^2=\frac{(n+2\alpha-1)(2n+2\alpha-2)}{n(2n+2\alpha)}N_{n-1}^2\nonumber\\
	&=\frac{(n+2\alpha-1)\cancel{(2n+2\alpha-2)}}{n(2n+2\alpha)}\frac{(n+2\alpha-2)(2n+2\alpha-4)}{(n-1)\cancel{(2n+2\alpha-2)}}N_{n-2}^2\nonumber\\
	&=\frac{(n+2\alpha-1)!}{n!(2n+2\alpha)}N_0^2=
	\frac{(n+\mathrm{D}-3)!}{n!(2n+\mathrm{D}-2)}
	\frac{S_{\mathrm{D}-1}}{S_{\mathrm{D}-2}},
\end{align}
what yields for $C_n^{(0)}$ the norm $N_n^2=\pi(2-\delta_{n,0})$ when $\mathrm{D}=2$ and $N_n^2=\frac{2}{2n+1}$ for $\mathrm{D}=3$.
\paragraph{Squared norm for any degree $N_n^2$ for $\mathcal{P}_n$.} 
For the $\mathcal{P}_n(1)=1$ ultraspherical polynomials with $\mathcal{P}_{n}=\underbrace{\textstyle\frac{2(n+\alpha-1)}{n+2\alpha-1}}_{A_n}x\,\mathcal{P}_{n-1}-\underbrace{\textstyle\frac{n-1}{n+2\alpha-1}}_{C_n}\,\mathcal{P}_{n-2}$ this is for $n>0$
\begin{align}
     N_n^2&=\frac{C_{n+1}A_n}{A_{n+1}}N_{n-1}^2=\frac{\frac{n}{n+2\alpha}\frac{2(n+\alpha-1)}{n+2\alpha-1}}{\frac{2(n+\alpha)}{n+2\alpha}}N_{n-1}^2=
     \frac{n(n+\alpha-1)}{(n+\alpha)(n+2\alpha-1)}N_{n-1}^2\nonumber\\
     &=\frac{n!(\alpha)^{\overline{n}}N_0^2}{(\alpha+1)^{\overline{n}}(2\alpha)^{\overline{n}}}=\frac{n!}{(2n+\mathrm{D}-2)\;(n+\mathrm{D}-3)!}\,\frac{S_{\mathrm{D}-1}}{S_{\mathrm{D}-2}},\label{eq:norm}
%	N_n^2&=\frac{n(2n+\mathrm{D}-4)}{(2n+\mathrm{D}-2)(n+\mathrm{D}-3)}\,N_{n-1}^2.\nonumber
\end{align}
%Using $N_0^2=\frac{S_{\mathrm{D}-1}}{S_{\mathrm{D}-2}}$ as a starting value, we get from the entire sequence
%\begin{align}
%	N_{1}^2&=\frac{1}{\mathrm{D}}\,N_0^2, \quad
%	N_{2}^2=\frac{2}{(\mathrm{D}+2)(\mathrm{D}-1)}\,N_0^2, \quad
%	N_{3}^2=\frac{6}{(\mathrm{D}+4)(\mathrm{D}-1)\mathrm{D}}\,N_0^2,\nonumber\\   N_{n}^2&=\frac{n!}{(2n+\mathrm{D}-2)(n+\mathrm{D}-3)!}\frac{S_{\mathrm{D}-1}}{S_{\mathrm{D}-2}}\quad\text{for $n>1$},
%\end{align}
yielding $N_n^2=\frac{\pi}{2-\delta_{n,0}}$ for $\mathrm{D}=2$ and $N_n^2=\frac{2}{2n+1}$ for $\mathrm{D}=3$; $N_0^2=\frac{S_{D-1}}{S_{D-2}}$ can be omitted.

\subsection{Integral recurrence of the Gegenbauer polynomials}
With the power-series representation of the Gegenbauer polynomials $C_n^{(\alpha)}$ in eq.~\eqref{eq:gegenbauer}, we find derivative recurrences after deriving $C_{n+1}^{(\alpha)}$ and $C_{n-1}^{(\alpha)}$. We separating the factors distinguishing the series to the coefficients of $C_n^{(\alpha)}$, after an index shift in $l\rightarrow l-1$ for the derivative of $C_{n-1}^{(\alpha)}$:
\begin{align}
	C_{n+1}^{(\alpha)'}&=\sum_{l=0}^{\lfloor\frac{n+1}{2}\rfloor}
\frac{(-1)^l2^{n-2l+1}(\alpha)^{\overline{n-l+1}}\cancel{(n-2l+1)}x^{n-2l}}{l!(n-2l\cancel{+1})!}\nonumber\\
&=\sum_{l=0}^{\lfloor\frac{n+1}{2}\rfloor}2(\alpha+n-l)\,
\frac{(-1)^l2^{n-2l}(\alpha)^{\overline{n-l}}}{l!(n-2l)!}\,x^{n-2l}\nonumber\\
		C_{n-1}^{(\alpha)'}&=\sum_{l=0}^{\lfloor\frac{n-1}{2}\rfloor}
	\frac{(-1)^l2^{n-2l-1}(\alpha)^{\overline{n-l-1}}\cancel{(n-2l-1)}x^{n-2l-2}}{l!(n-2l-\cancel{1}^{2})!}\nonumber\\
	&=\sum_{l=\cancel{1}^0}^{\lfloor\frac{n+1}{2}\rfloor}(-1)2l
	\frac{(-1)^{l}2^{n-2l}(\alpha)^{\overline{n-l}}x^{n-2l}}{l!(n-2l)!}.\nonumber
\end{align}
Combining $C_{n+1}^{(\alpha)'}-C_{n-1}^{(\alpha)'}$ obviously leaves 
\begin{align}
	2(\alpha+n)\,\sum_{l=0}^{\lfloor\frac{n+1}{2}\rfloor}
	\frac{(-1)^l2^{n-2l}(\alpha)^{\overline{n-l}}}{l!(n-2l)!}\,x^{n-2l}=2(\alpha+n)\,C_{n}^{(\alpha)}, \nonumber
\end{align}
and hence
\begin{align}
	2(\alpha+n)C_{n}^{(\alpha)}&=[C_{n+1}^{(\alpha)}-C_{n-1}^{(\alpha)}]'.
\end{align}
Similarly, we find for the Gegenbauer polynomials standardized to $\mathcal{P}_n(1)=1$ eq.~\eqref{eq:gegenbauer1}
\begin{align}
	\mathcal{P}'_{n+1}
	&=\sum_{l=0}^{\lfloor\frac{n+1}{2}\rfloor}\frac{2(n+1)(\alpha+n-l)}{2\alpha+n}\,
	\frac{(-1)^l2^{n-2l}\,n!}{l!(n-2l)!}
	\frac{(\alpha)^{\overline{n-l}}}{(2\alpha)^{\overline{n}}}\,x^{n-2l}\label{eq:derivative_P}\\
	\mathcal{P}'_{n-1}&=\sum_{l=\cancel{1}^0}^{\lfloor\frac{n+1}{2}\rfloor}\frac{(-1)2l(2\alpha+n-1)}{n}\,
	\frac{(-1)^l2^{n-2l}\,n!}{l!(n-2l)!}
	\frac{(\alpha)^{\overline{n-l}}}{(2\alpha)^{\overline{n}}}\,x^{n-2l},\nonumber
\end{align}
leaving
\begin{align}
  2(\alpha+n)\mathcal{P}_n=\Big[\frac{2\alpha+n}{n+1}\mathcal{P}_{n+1}-\frac{n}{2\alpha+n-1}\mathcal{P}_{n-1}\Big]',
\end{align}
which leaves for $\alpha=\frac{1}{2}$ and Legendre polynomials $(2n+1)P_n=[P_{n+1}-P_{n-1}]'$, and $\alpha=0$ with Chebyshev polynomials 
$2T_n=[\frac{1}{n+1}T_{n+1}-\frac{1}{n-1}T_{n-1}]'$.
By this, we get the recurrence for integration
\begin{align}
	2(n+\alpha)\int\mathcal{P}_n\mathrm{d}x=\frac{n+2\alpha}{n+1}\mathcal{P}_{n+1}-\frac{n}{n+2\alpha-1}\mathcal{P}_{n-1}.\label{eq:int_recurrence}
\end{align}

\subsection{Additional recurrence relations}
From the definition of the differential equation~\eqref{eq:gegenbauerdgl}, we directly get
\begin{align}
	(1-x^2)\,\mathcal{P}_n''&=(1+2\alpha)\,x\,\mathcal{P}_n'-n(n+2\alpha)\,\mathcal{P}_n.
\end{align}
From partial integration of $x\,\mathcal{P}_n'$, we get via the multiplication recurrence \eqref{eq:standard-three-term-recurrence} and integration recurrence \eqref{eq:int_recurrence}:
\begin{align}
	\int x\,\mathcal{P}_n'\,\mathrm{d}x&=x\,\mathcal{P}_n-\int\mathcal{P}_n\,\mathrm{d}x\nonumber\\
	&=\big[{\textstyle\frac{n+2\alpha}{2(n+\alpha)}}-{\textstyle\frac{n+2\alpha}{2(n+\alpha)(n+1)}}\big]\mathcal{P}_{n+1}+
	\big[{\textstyle\frac{n}{2(n+\alpha)}}+{\textstyle\frac{n}{2(n+\alpha)(n+2\alpha-1)}}\big]\mathcal{P}_{n-1}\nonumber\\
	2(n+\alpha)\int x\,\mathcal{P}_n'\,\mathrm{d}x&=
	n(n+2\alpha)\bigg[\frac{1}{n+1}\mathcal{P}_{n+1}+
	\frac{1}{n+2\alpha-1}\mathcal{P}_{n-1}\bigg].
\end{align}
From partial integration of the differential equation, we get
\begin{align}
	\int(1-x^2)\,\mathcal{P}_n''\,\mathrm{d}x-(2\alpha+1)\,\int x\,\mathcal{P}_n'\,\mathrm{d}x+n(n+2\alpha)\,\int\mathcal{P}_n\,\mathrm{d}x&=0\nonumber\\
\textstyle	\big[(1-x^2)\mathcal{P}_n'+2\int x\,\mathcal{P}_n'\,\mathrm{d}x\big]	-(2\alpha+1)\int x\,\mathcal{P}_n'\,\mathrm{d}x
	+n(n+2\alpha)\int\mathcal{P}_n\,\mathrm{d}x&=0\nonumber\\
\textstyle	(1-x^2)\mathcal{P}_n'-(2\alpha-1)\int x\,\mathcal{P}_n'\,\mathrm{d}x
	+n(n+2\alpha)\int\mathcal{P}_n\,\mathrm{d}x&=0\nonumber\\
\textstyle	(1-x^2)\mathcal{P}_n'-(2\alpha-1)\big[x\,\mathcal{P}_n-\int\mathcal{P}_n\,\mathrm{d}x\big]
	+n(n+2\alpha)\int\mathcal{P}_n\,\mathrm{d}x&=0\nonumber\\
\textstyle	(1-x^2)\mathcal{P}_n'-(2\alpha-1)\,x\,\mathcal{P}_n
	+[n(n+2\alpha)+(2\alpha-1)]\int\mathcal{P}_n\,\mathrm{d}x&=0\nonumber\\
\textstyle	(1-x^2)\mathcal{P}_n'-(2\alpha-1)\,x\,\mathcal{P}_n
+(n+1)(n+2\alpha-1)\int\mathcal{P}_n\,\mathrm{d}x&=0\nonumber	
\end{align}
\begin{align}
		(1-x^2)\mathcal{P}_n'&=(2\alpha-1)\,x\,\mathcal{P}_n
	-(n+1)(n+2\alpha-1)\int\mathcal{P}_n\,\mathrm{d}x\nonumber	\\
	&=\big[{\textstyle\frac{(2\alpha-1)(n+2\alpha)}{2(n+\alpha)}}-{\textstyle\frac{\cancel{(n+1)}(n+2\alpha-1)(n+2\alpha)}{2(n+\alpha)\cancel{(n+1)}}}\big]\mathcal{P}_{n+1}+
	\big[{\textstyle\frac{n(2\alpha-1)}{2(n+\alpha)}}+{\textstyle\frac{n(n+1)\cancel{(n+2\alpha-1)}}{2(n+\alpha)\cancel{(n+2\alpha-1)}}}\big]\mathcal{P}_{n-1}\nonumber
\end{align}
\begin{align}
	2(n+\alpha)(1-x^2)\mathcal{P}_n'&=n(n+2\alpha)\big[-\mathcal{P}_{n+1}+
	\mathcal{P}_{n-1}\big]
\end{align}

\newpage
\section{Designs and metrics \texorpdfstring{$P$, $E$, $Q$, $FBR$, $\bm r_\mathrm{V}$, and $\bm r_\mathrm{E}$}{P, E, Q, FBR, rV, and rE} for axisymmetric directivity/panning patterns}\label{apdx:axisymmetrics}
%\section{P, E, Q, FBR, rV, and rE metrics for axisymmetric panning function}\label{apdx:axisymmetrics}
The panning function or directivity pattern
\begin{align}
	g(x)&=\sum_{n=0}^\mathrm{N}\frac{a_n}{S_{\mathrm{D}-2}N_n^2}\,\mathcal{P}_n(x)
\end{align}
is rotationally symmetric around the $x$ axis of a $\mathrm{D}$-dimensional space. For the scalar measures, we integrate over $x$ with the surface weight $w(x)=\sqrt{1-x^2}^{\mathrm{D}-3}$, and we use the surface $S_{D-2}$ of the unit sphere in the $(\mathrm{D}-1)$-dimensional space to denote the result of the integrals independent of $x$:
\begin{align}
	P&=S_{\mathrm{D}-2}\,\int_{-1}^{1}\sum_{n=0}^\mathrm{N}\frac{a_n}{S_{\mathrm{D}-2}\,N_n^2}\,\mathcal{P}_n(x)\,w(x)\mathrm{d}x=\sum_{n=0}^\mathrm{N}\frac{a_n}{N_n^2}\overbrace{\int_{-1}^{1}\mathcal{P}_n(x)\,w(x)\mathrm{d}x}^{N_n^2\delta_{n0}}=a_0\label{eq:P}\\
	E&=S_{\mathrm{D}-2}\sum_{n,n'}\frac{a_{n'}^*}{S_{\mathrm{D}-2}N_{n'}^2}\frac{a_n}{S_{\mathrm{D}-2}N_n^2}\overbrace{\int_{-1}^{1}\mathcal{P}_n(x)\mathcal{P}_{n'}(x)\,w(x)\mathrm{d}x}^{N_n^2\delta_{nn'}}=\sum_{n=0}^{\mathrm{N}}\frac{|a_n|^2}{S_{\mathrm{D}-2}N_n^2},\label{eq:E}\\
	Q&=\frac{S_{D-1}\left(\sum_{n=0}^\mathrm{N}\frac{a_n}{S_{D-2}\,N_n^2}\right)^2}{E},\qquad\quad
	FBR=\frac{\sum_{n'=0}^\mathrm{N}\sum_{n=0}^\mathrm{N}g_{nn'}a_na_{n'}^*}{\sum_{n'=0}^\mathrm{N}\sum_{n=0}^\mathrm{N}g_{nn'}a_na_{n'}^*(-1)^{n+n'}},\label{eq:QFBR}
\end{align}
with the Gram  matrix $g_{nn'}$ of the polynomials integrated over the frontal half-space, only
\begin{align}
	g_{nn'}&=\frac{1}{N_n N_{n'}}\int_0^1\mathcal{P}_n(x)\mathcal{P}_{n'}(x)\,w(x)\,\mathrm{d}x,
\end{align}
see \autoref{apdx:gram}, also~\cite{Rafaely19, Elko00, Elko04} for more information.

For rotationally symmetric shapes in space, the vector models always indicate the direction of axial symmetry. The remaining integral for $\bm{r}_\mathrm{V}$ and $\bm{r}_\mathrm{E}$ is a one-dimensional calculation along $x$ and determines the length.
% $\beta_0^{(2)}=1-\beta_{2}^{(2)}=1-\frac{\mathrm{D}-1}{\mathrm{D}}=$)
\begin{align}
	r_\mathrm{V}&=\frac{S_{\mathrm{D}-2}\sum_{n=0}^\mathrm{N}\frac{a_n}{N_n^2S_{\mathrm{D}-2}}\overbrace{\int_{-1}^{1}\mathcal{P}_n(x)\,x\,w(x)\mathrm{d}x}^{N_1^2\delta_{n,1}}
	}{P}=\frac{a_1}{a_0}.\label{eq:rV}
\end{align}
For the $\bm r_\mathrm{V}$, the expression below the brace was evaluated by inserting the recurrence of the Gegenbauer polynomials $x \mathcal{P}_{n}(x)=\dots$, 
\begin{align}
	\int_{-1}^1[(1-\beta_{n+1})P_{n-1}(x)+\beta_{n+1}P_{n+1}(x)]w(x)\mathrm{d}x=(1-\beta_{2})N_0^2\delta_{n,1}=N_1^2\delta_{n,1},
\end{align}
and by noticing that one can re-write $(1-\beta_2)N_0^2$ to $\beta_1N_1^2$ because of the recurrence of the normalization in eq.~\eqref{eq:normalizationrecurrence}, $N_1^2=\frac{1-\beta_2}{\beta_1}N_0^2$, and further to $N_1^2$ given that $\beta_1=1$.

For $\bm r_\mathrm{E}$, the expression with the integral yields
\begin{align}
	r_\mathrm{E}&=\frac{S_{\mathrm{D}-2}\,\sum_{n,n'}\frac{a_n}{S_{\mathrm{D}-2}N_n^2}\frac{a_{n'}^*}{S_{\mathrm{D}-2}N_{n'}^2}\overbrace{\int_{-1}^{1}\mathcal{P}_{n}(x)\mathcal{P}_{n'}(x)\,x\,w(x)\,\mathrm{d}x}}{E}=\frac{
		\sum_{n}\frac{\beta_{n+1}}{N_{n}^2}2\Re\{a_na_{n+1}^*\}
	}{\sum_{n=0}^{\mathrm{N}-1}\frac{1}{N_n^2}|a_n|^2}.\label{eq:rE}
\end{align}
To evaluate the expression below the brace, the recurrence was used
\begin{align}
	\textstyle	\int_{-1}^1 \mathcal{P}_{n}(x)
	\bigl[(1-\beta_{n'+1})\mathcal{P}_{n'-1}(x)&+\beta_{n'+1}P_{n'+1}(x)\bigr]w(x)\mathrm{d}x\nonumber\\
	%w(x)\mathrm{d}x=N_n^2\left[\beta_{n'-1}^{(n'+1)}\delta_{n,n'-1}+\beta_{n'+1}^{(n'+1)}\delta_{n,n'+1}}{2n'+1}\nonumber\\
&=N_{n}^2(1-\beta_{n'+1})\delta_{n+1,n'}+N_n^2\beta_{n'+1}\delta_{n,n'+1},\nonumber
\end{align}
and re-insertion gave
\begin{align}
r_\mathrm{E}&=\frac{
	\sum_{n=0}^{\mathrm{N}-1}\frac{a_na_{n+1}^*}{S_{\mathrm{D}-2}N_{n+1}^2}(1-\beta_{n+2})+
	\sum_{n'=0}^{\mathrm{N}-1}\frac{a_{n'+1}a_{n'}^*}{S_{\mathrm{D}-2}N_{n'+1}^2}\beta_{n'+1}
}{E}=\frac{
	\sum_{n'=0}^{\mathrm{N}-1}\frac{a_{n}a_{n+1}^*+a_n^*a_{n+1}}{S_{\mathrm{D}-2}N_{n'}^2}\beta_{n+1}\label{eq:rED_recurrence}
}{E},
\end{align}
for which eq.~\eqref{eq:normalizationrecurrence} was used to re-write $\frac{1-\beta_{n+2}}{N_{n+1}^2}=\frac{\beta_{n+1}}{N_{n}^2}$ in the first numerator term, and a shift of $n$ by one in the second one.

\subsection{Supercardioid: Gram matrix for half interval}\label{apdx:gram}
For the design of a supercardioid pattern, the entries of the Gram matrix are needed
\begin{align}
	g_{nm}&=\int_{x_0}^{1}\frac{\mathcal{P}_n(x)}{N_n^2}\frac{\mathcal{P}_m(x)}{N_m^2}\,w(x)\,\mathrm{d}x.\label{eq:Gram}
\end{align}
As $\mathcal{P}_n(x)\mathcal{P}_m(x)=(-1)^{n+m}\mathcal{P}_n(-x)\mathcal{P}_m(-x)$ uses $(-1)^{2(l+k)}$ or $(-1)^{2(l+k)+2}$ there is no sign change between the half intervals for matching parity. In these cases, eq.~\eqref{eq:Gram} multiplied by $\frac{1}{2}$ and integrated over the full range $\int_{-1}^1$ is equivalent and orthogonality preserved, therefore and the entries $g_{2l,2k}=g_{2l+1,2k+1}=0$ for $l\neq k$.
The entries of the diagonal $n=m$ is the half-interval integral of the squared function, therefore half the squared normalization $\frac{N_n^2}{2}$ divided by $N_n^4$. Omitting the unit-sphere ratio  $\frac{S_{D-1}}{S_{D-2}}$ by setting $N_0^2=1$ in eq.~\eqref{eq:norm} for simplicity, we get
\begin{align}
	g_{nn}&=\frac{1}{2\,N_n^2}=\frac{(\alpha+1)^{\overline{n}}(2\alpha)^{\overline{n}}}{2\;n!(\alpha)^{\overline{n}}}.%\frac{(n+\alpha)\,(2\alpha)^{\overline{n}}}{n!}.
\end{align}
eq.~\eqref{eq:orthogonalityproof_intermediate} gives a hint for $n\neq m$, using $w(x)=(1-x^2)^{\alpha-\frac{1}{2}}$, so $(1-x^2)\,w(x)=(1-x^2)^{\alpha+\frac{1}{2}}$, which vanishes at the upper bound $x=1$:
\begin{align}
	g_{nm}=\int_{x_0}^1\frac{\mathcal{P}_n(x)}{N_n^2}\frac{\mathcal{P}_m(x)}{N_m^2}\,w(x)\,\mathrm{d}x&=\frac{(1-x^2)\,w(x)}{N_n^2N_m^2}\,\frac{\mathcal{P}_n'\mathcal{P}_m-\mathcal{P}_m'\mathcal{P}_n}{n(n+2\alpha)-m(m+2\alpha)}\bigg|_{x_0}^1\\
	&=\frac{-(1-x_0^2)^{\alpha+\frac{1}{2}}}{N_n^2N_m^2}\,\frac{\mathcal{P}_n'(x_0)\mathcal{P}_m(x_0)-\mathcal{P}_m'(x_0)\mathcal{P}_n(x_0)}{n(n+2\alpha)-m(m+2\alpha)}.\nonumber
\end{align}
%With the known derivatives at $x=1$ that are $\mathcal{P}_n(1)'=\frac{n(n+2\alpha)}{2\alpha+1}$ from eq.~\eqref{eq:derivatives_at_1} and $\mathcal{P}_n(1)=1$, 
At the lower bound $x_0=0$, the factor $(1-x^2)^{\alpha+\frac{1}{2}}=1$ is unity, and 
we get  $\mathcal{P}_{2l+1}(0)=0$ because of the odd parity and $\mathcal{P}_{2k}'(0)=0$ due to the even parity
\begin{align}
	g_{2l+1,2k}&=\frac{-1}{N_n^2N_m^2}\,\frac{\mathcal{P}_{2l+1}'(0)\mathcal{P}_{2k}(0)-\mathcal{P}_{2k}'(0)\mathcal{P}_{2l+1}(0)}{(2l+1)(2l+2\alpha+1)-(2k)(2k+2\alpha)}\nonumber\\
	&=\frac{1}{N_{2l+1}^2N_{2k}^2}
	\,\frac{-\mathcal{P}_{2l+1}'(0)\mathcal{P}_{2k}(0)}{(2l+1)(2l+2\alpha+1)-(2k)(2k+2\alpha)}.
\end{align}
With 
$\mathcal{P}_{2k}(0)=\frac{(-1)^k(2k)!(\alpha)^{\overline{k}}}{k!\,(2\alpha)^{\overline{2k}}}$  and $\mathcal{P}_{n+1}'(x)=\sum_l\frac{(-1)^l2^{n-2l+1}(n+1)!(\alpha)^{\overline{n-l+1}}}{l!(n-2l)!\,(2\alpha)^{\overline{n+1}}}x^{n-2l}$ from eqs.\ \eqref{eq:gegenbauer_at_0} \eqref{eq:derivative_P}, with $n=2l$ and $x=0$, so 
$\mathcal{P}_{2l+1}'(0)=\frac{2\,(-1)^l(2l+1)!(\alpha)^{\overline{l+1}}}{l!\,(2\alpha)^{\overline{2l+1}}}$ we have with $\frac{1}{N_n^2}$,$\frac{1}{N_m^2}$
\begin{align}
	g_{2l+1,2k}&=\textstyle\frac{1}{N_0^2}
	\frac{(\alpha+1)^{\overline{2l+1}}\cancel{(2\alpha)^{\overline{2l+1}}}}{\cancel{(2l+1)!}(\alpha)^{\overline{2l+1}}}
	\frac{2\,(-1)^l\cancel{(2l+1)!}(\alpha)^{\overline{l+1}}}{l!\,\cancel{(2\alpha)^{\overline{2l+1}}}}
	\frac{(\alpha+1)^{\overline{2k}}\cancel{(2\alpha)^{\overline{2k}}}}{\cancel{(2k)!}(\alpha)^{\overline{2k}}}
	\frac{(-1)^k\cancel{(2k)!}(\alpha)^{\overline{k}}}{k!\,\cancel{(2\alpha)^{\overline{2k}}}}\cdots
	\,\nonumber\\
	&\qquad\textstyle\cdots 
	\frac{-1
	}{(2l+1)(2l+2\alpha+1)-(2k)(2k+2\alpha)}\nonumber\\
	&=\textstyle
	\frac{\Gamma(\alpha+1)}{\sqrt{\pi}\Gamma(\alpha+\frac{1}{2})}\frac{(-1)^{l+1}(\alpha+1)^{\overline{2l+1}}}{l!\,(\alpha+1+l)^{\overline{l}}}
	\frac{(-1)^{k}(\alpha+1)^{\overline{2k}}}{k!\,(\alpha+k)^{\overline{k}}}
	\,
	\frac{2}{(2l+1)(2l+2\alpha+1)-(2k)(2k+2\alpha)}.
\end{align}
Interchanging $n\leftrightarrow m$ in eq.~\eqref{eq:Gram} yields the values $g_{2l+1,2k}= g_{2k,2l+1}$ on the other side of the diagonal. \emph{(No idea from where the $N_0^2=\frac{S_\mathrm{D-1}}{S_\mathrm{D-2}}$ sneaked in, but it was numerically tested to be necessary for $\mathrm{D}=\{2,2.1,2.5,3,3.5,4,4.5,5\}$) to restore balance between diagonal/offdiagonal).}

\subsection{Basic pattern with maximum DI/maximum \texorpdfstring{$\bm r_\mathrm{V}$}{rV} vector}\label{apdx:maxdi}
With $a_n=1$, the value $r_\mathrm{V}=\frac{a_1}{a_0}=1$. The maximum directivity factor $Q$ is obtained, because $a_n=1$ decomposes a Dirac delta, which practically minimizes the squares of the directivity everywhere except at $x=1$.
%One could also zero the derivative of $Q$ wrt.\ $a_{n'}$ 
%\begin{align}
%	\frac{1}{S_{\mathrm{D}-1}}\frac{\partial Q}{\partial a_{n'}}&=
%	\frac{\frac{\partial g(1)^2}{\partial a_{n'}}}{E}-\frac{g(1)^2\frac{\partial E}{\partial a_{n'}}}{E^2}
%	=
%	2\frac{g(1)\,\frac{1}{S_{\mathrm{D}-2}{N_{n'}}^2}}{E}-2\frac{g(1)^2\,\frac{a_{n'}}{S_{\mathrm{D}-2}N_{n'}^2}}{E^2}=0\nonumber\\
%	\Rightarrow a_n&=\frac{E}{g(1)}=\text{const}.
%\end{align}
The resulting value of the directivity factor is 
\begin{align}
	\mathrm{max} Q&=S_{\mathrm{D}-2}\frac{(\sum_{n=0}^\mathrm{N}\frac{1}{N_n^2})^2}{\sum_{n=0}^\mathrm{N}\frac{1}{N_n^2}}=\sum_{n=0}^{\mathrm{N}}\frac{S_{\mathrm{D}-2}}{N_n^2}=
	\begin{cases}
		\sum_{n=0}^\mathrm{N}(2-\delta_n)=2\mathrm{N}+1, & \mathrm{D}=2\\
		\sum_{n=0}^\mathrm{N}(2n+1)=(\mathrm{N}+1)^2, & \mathrm{D}=3.
	\end{cases}
\end{align}

\subsection{Pattern with maximum \texorpdfstring{$\bm r_\mathrm{E}$}{rE} vector}
%\section{Maximizing the rE vector}
\label{apdx:maxre_axisym}
The energy vector length is determined by the real-valued coefficients $a_n$ in terms of
$r_\mathrm{E}=\frac{	\sum_{n=0}^{\mathrm{N}-1}\frac{2\beta_{n+1}}{N_{n}^2}a_na_{n+1}
}{\sum_{n=0}^{\mathrm{N}}\frac{1}{N_n^2}a_n^2}=\frac{A}{E}$. We may zero its derive it with regard to $a_n$ to maximize its length, yielding $r_\mathrm{E}'=\frac{A'}{E}-\frac{A}{E^2}E'=\frac{A'-r_\mathrm{E}\,E'}{E}=0$, where $A/E=r_\mathrm{E}$ was re-inserted as described by Daniel \cite{Daniel99,Daniel01}. It is sufficient to zero $A'-r_\mathrm{E}\,E'=0$ and insert the derivatives of the numerator and denominator, $A'$, $E'$, respectively. For the numerator $A'$  
\begin{align}
\frac{\partial}{\partial a_n}\sum_{n'}\frac{2\beta_{n'+1}a_{n'}a_{n'+1}}{N_{n'}^2}=\frac{2\beta_{n+1}a_{n+1}}{N_n^2}+\frac{2\beta_{n}a_{n-1}}{N_{n-1}^2}=
\frac{2}{N_n^2}\left[\beta_{n+1}a_{n+1}+(1-\beta_{n+1})a_{n-1}\right]\nonumber
\end{align} 
the equivalence  $\frac{\beta_{n}}{N_{n-1}^2}=\frac{1-\beta_{n+1}}{N_{n}^2}$ from eq.~\eqref{eq:normalizationrecurrence} helped to unify expressions. 
For the denominator $E'$ we get $\frac{\partial}{\partial a_n}\sum_{n'}\frac{a_{n'}^2}{N_{n'}^2}=\frac{2}{N_n^2}a_n$, and we insert $A'$ and $E'$ into $A'-r_\mathrm{E}\,E'=0$
\begin{align}
\beta_{n+1}a_{n+1}+(1-\beta_{n+1})a_{n-1}-r_\mathrm{E}\,a_n=0.
\end{align}
We recognize the recurrence relation  $\beta_{n+1}\mathcal{P}_{n+1}+(1-\beta_{n+1})\mathcal{P}_{n-1}-x\,\mathcal{P}_n(x)=0$ of the orthogonal polynomials and hereby find $a_n=\mathcal{P}_n(x)$ with $r_\mathrm{E}=x$ as solution scheme
\begin{align}
a_n=\mathcal{P}_n(r_\mathrm{E}).
\end{align} 
The difference equation does not consider the finite-degree summation boundaries of $r_\mathrm{E}$ in eq.~\eqref{eq:rED_recurrence}, yet. Only for an infinite sum, $r_\mathrm{E}=1$ is feasible. Otherwise, feasible choices  for finite-degree $r_\mathrm{E}$ are found after inserting the scheme into $r_\mathrm{E}=A/E$ in the form $A=r_\mathrm{E}E$ of eq.~\eqref{eq:rED_recurrence} considering the finite range of $n\in[0;\mathrm{N}-1]$ for $A$ and $n\in[0;\mathrm{N}]$ for $E$,
\begin{align}
\bigg[\sum_{n=0}^{\mathrm{N}-1}\frac{\mathcal{P}_n(r_\mathrm{E})\mathcal{P}_{n+1}(r_\mathrm{E})}{S_{\mathrm{D}-2}N_{n+1}^2}(1-\beta_{n+2})+
\sum_{n'=0}^{\mathrm{N}-1}\frac{\mathcal{P}_{n'+1}(r_\mathrm{E})\mathcal{P}_{n'}(r_\mathrm{E})}{S_{\mathrm{D}-2}N_{n'+1}^2}\beta_{n'}\bigg]&=
r_\mathrm{E}\bigg[\sum_{n=0}^\mathrm{N}\frac{\mathcal{P}_n(r_\mathrm{E})\mathcal{P}_n(r_\mathrm{E})}{S_{\mathrm{D}-2}N_{n}^2}\bigg]\nonumber
\end{align}
We'd nearly precisely get the same right-hand-side term after expanding one of the $\mathcal{P}_n(r_\mathrm{E})$ terms in the sum by the factor $r_\mathrm{E}$ from outside the sum, using the recurrence $r_\mathrm{E}\,\mathcal{P}_n(r_\mathrm{E})=\beta_{n+1}\mathcal{P}_{n+1}(r_\mathrm{E})+(1-\beta_{n+1})\mathcal{P}_{n-1}(r_\mathrm{E})$. And yet, there is a decisive difference: no sum cuts away the degree $\mathrm{N}+1$ expression, which wasn't part of the sums on the left. Therefore, a feasible finite-degree solution for $r_\mathrm{E}$ requires a vanishing term
\begin{align}
\beta_{\mathrm{N}+1}\,\mathcal{P}_{\mathrm{N}+1}(r_\mathrm{E})\mathcal{P}_{\mathrm{N}}(r_\mathrm{E})=0.
\end{align}
Because $a_\mathrm{N}\neq0$ should not vanish and $\beta_{\mathrm{N}+1}\neq0$, the feasibility constraint is that
\begin{align}
\arg\max \mathcal{P}_{\mathrm{N}+1}(r_\mathrm{E})=0.\label{eq:maxre}
\end{align}
Therefore, as defined in \cite{Daniel99,Daniel01}, the largest root of $\mathcal{P}_{\mathrm{N}+1}$ defines the coefficients $a_n$ for a maximum-length $\bm r_\mathrm{E}$ vector.

Interestingly, this corresponds to decomposing an axisymmetric Dirac delta $\delta(x-\max r_\mathrm{E})$: a concentric ring (3D) or a pair of points (2D) around the panning direction, which could be expressed in short by the Christoffel-Darboux identity eq.~\eqref{eq:christoffeldarboux} with $\mathcal{P}_{\mathrm{N}+1}(\max r_\mathrm{E})=0$
\begin{align}
g(\cos\vartheta)&=\frac{\mathcal{P}_{\mathrm{N}+1}(\cos\vartheta)\mathcal{P}_{\mathrm{N}}(\max r_\mathrm{E})}{\cos\vartheta-\max r_\mathrm{E}}.
\end{align}

\subsection{Particular max-\texorpdfstring{$r_\mathrm{E}$}{rE} solutions for $\mathrm{D}=3$}

A fairly good approximation for practical use has been given in~\cite{Zotter12b} by
\begin{align}
r_\mathrm{E}=\cos\frac{137.9^\circ}{\mathrm{N}+1.51},
\end{align}
and it saves us the time for polynomial root finding.

\newpage
\subsection{(Ultra-)Spherical cap function}\label{apdx:cap}
For the coefficients of a spherical cap function, one needs to integrate the polynomial over $w(x)$ and the corresponding segment of the sphere $x_0\leq x\leq 1$.

The coefficient needed for expanding a spherical cap function is obtained by integrating $a_n=\int_{x_0}^1\mathcal{P}_n(x)\,w(x)\,\mathrm{d}x$, with the weight function $w(x)=(1-x^2)^{\alpha-\frac{1}{2}}$. 

We can manipulate Rodrigues' formula eq.~\eqref{eq:rodrigues1} in the form with $n+1$
\begin{align}
	\frac{2^{n+1}(\alpha+\frac{1}{2})^{\overline{n+1}}}{(-1)^{n+1}}\;w(x)\,\mathcal{P}_{n+1}(x)&=
	\frac{\mathrm{d}^{n+1}}{\mathrm{d}x^{n+1}}(1-x^2)^{\alpha+n+\frac{1}{2}},\nonumber
\end{align}
by carrying out one of the derivatives on the right to get
\begin{align}
	\frac{2^{n+1}(\alpha+\frac{1}{2})^{\overline{n+1}}}{(-1)^{n+1}}\,&w(x)\,\mathcal{P}_{n+1}(x)=
	(\alpha+n+{\textstyle\frac{1}{2}})\frac{\mathrm{d}^{n}}{\mathrm{d}x^{n}}(-2x)(1-x^2)^{\alpha+n-\frac{1}{2}},\nonumber\\
	&=-	(2\alpha+2n+1)\bigg[\underbrace{x\frac{\mathrm{d}^{n}(1-x^2)^{\alpha+n-\frac{1}{2}}}{\mathrm{d}x^{n}}}_{	\frac{2^n(\alpha+\frac{1}{2})^{\overline{n}}}{(-1)^n}x\,w(x)\,\mathcal{P}_{n}(x)}
	+n\,\underbrace{\frac{\mathrm{d}^{n-1}(1-x^2)^{\alpha+n-\frac{1}{2}}}{\mathrm{d}x^{n-1}}}_{	\frac{2^n(\alpha+\frac{1}{2})^{\overline{n}}}{(-1)^n}\int\mathcal{P}_{n}(x)\,w(x)\,\mathrm{d}x}\bigg]\nonumber\\
	\frac{2\alpha+2n+1}{2\alpha+2n+1}\;w(x)\,&\mathcal{P}_{n+1}(x)=
	x\,w(x)\,\mathcal{P}_{n}(x)
	+n\,\int\mathcal{P}_{n}(x)\,w(x)\,\mathrm{d}x\nonumber\\
	&\mathcal{P}_{n+1}(x)=
	x\,\mathcal{P}_{n}(x)
	+\frac{n}{w(x)}\,\int\mathcal{P}_{n}(x)\,w(x)\,\mathrm{d}x
\end{align}
\begin{align}
	\Longrightarrow \int\mathcal{P}_{n}(x)\,w(x)\,\mathrm{d}x&=\frac{w(x)}{n}\,\bigg[\mathcal{P}_{n+1}(x)-x\,\mathcal{P}_n(x)\bigg].
\end{align}
The recurrence $x\mathcal{P}_n(x)=\frac{n+2\alpha}{2n+2\alpha}\mathcal{P}_{n+1}(x)+\frac{n}{2n+2\alpha}\mathcal{P}_{n-1}(x)$ yields the coefficients
\begin{align}
	a_n=\int_{x_0}^1\mathcal{P}_{n}(x)\,w(x)\,\mathrm{d}x&=
	\frac{w(x)}{2n+2\alpha}\,\bigg[\mathcal{P}_{n+1}(x)-\mathcal{P}_{n-1}(x)\bigg]\Bigg|_{x_0}^1\label{eq:spherecap}\\
	&=\frac{w(x_0)}{2n+2\alpha}\,\bigg[\mathcal{P}_{n-1}(x_0)-\mathcal{P}_{n+1}(x_0)\bigg].\nonumber
\end{align}
For the exception $n=0$, the integral to solve is $a_0=\int_{x_0}^1w(x)\,\mathrm{d}x$, which is
\begin{align}
	a_0=\int_{x_0}^1\sqrt{1-x^2}^{\mathrm{D}-3}\,\mathrm{d}x=
	\begin{cases}
		\arccos(x_0)=\upvarphi_0, &\mathrm{D}=2\\
		1-x_0, &\mathrm{D}=3\\
		F(\frac{1}{2},\frac{2-\mathrm{D}}{2}; \frac{3}{2}, 1) - x_0\,F(\frac{1}{2},\frac{2-\mathrm{D}}{2}; \frac{3}{2}, x_0^2), &\text{any $\mathrm{D}$}
	\end{cases}\label{eq:spherecap0}
\end{align}
using Gau{\ss}' hypergeometric function $F(a,b;c,z)$.

\newpage
\subsection{Inphase/higher-order cardioid pattern}\label{apdx:inphase}
A higher-order cardioid pattern has an $\mathrm{N}$-fold zero at $x=-1$ and is defined as
\begin{align}
	g_\mathrm{N}&={\textstyle\frac{1}{2^{\mathrm{N}}}}\,(1+x)^\mathrm{N},
\end{align}
and we want to find the coefficients $a_n$ for its expansion
\begin{align}
	g_\mathrm{N}&=\sum_{n=0}^\infty \frac{a_n}{N_n^2}\,\mathcal{P}_n(x).
\end{align}
Deriving $\frac{\mathrm{d}}{\mathrm{d}x}(1+x)^{\mathrm{N}+1}=(\mathrm{N}+1)(1+x)^\mathrm{N}$, and then integrating it from its zero at $x=-1$ to $x$, we se that $g_\mathrm{N}$ can be obtained from the integral equation $(1+x)\,y=(\mathrm{N}+1)\int_{-1}^xy\;\mathrm{d}x$. 
We use the integration and multiplication recurrences:\\
$2(n+\alpha)\int_{-1}^x\mathcal{P}_n\mathrm{d}x=\frac{n+2\alpha}{n+1}\mathcal{P}_{n+1}-\frac{n}{n+2\alpha-1}\mathcal{P}_{n-1}$, and\\
$2(n+\alpha)(1+x)\,\mathcal{P}_n=(n+2\alpha)\mathcal{P}_{n+1}+2(n+\alpha)\mathcal{P}_{n}+n\mathcal{P}_{n-1}$ \\ to write $(\mathrm{N}+1)\int y\,\mathrm{d}x-(1+x)\,y=0$:
\begin{align}
	\sum_{n=0}^\mathrm{N} {\textstyle\frac{a_n}{N_n^22(n+\alpha)}}\,[
	{\textstyle\frac{(\mathrm{N}+1)(n+2\alpha)-(n+1)(n+2\alpha)}{n+1}}\mathcal{P}_{n+1}
	-2(n+\alpha)\mathcal{P}_n
	-{\textstyle\frac{(\mathrm{N}+1)n+(n+2\alpha-1)n}{n+2\alpha-1}}\mathcal{P}_{n-1}]&=0,\nonumber\\
	\sum_{n=0}^\mathrm{N} {\textstyle\frac{a_n}{N_n^2}}\,[
	{\textstyle\frac{(\mathrm{N}-n)(n+2\alpha)}{2(n+\alpha)(n+1)}}\mathcal{P}_{n+1}
	-\mathcal{P}_n
	-{\textstyle\frac{(\mathrm{N}+n+2\alpha)n}{2(n+\alpha)(n+2\alpha-1)}}\mathcal{P}_{n-1}]&=0,\nonumber
\end{align}
and by applying orthogonality when integrating over $\int_{-1}^1\mathcal{P}_{n'}\mathcal{P}_{n}\mathrm{d}x=N_n^2\delta_{nn'}$ we find
after using 
$N_{n}^2N_{n-1}^{-2}=\frac{(n+\alpha-1)n}{(n+\alpha)(n+2\alpha-1)}$ and
$N_{n}^2N_{n+1}^{-2}=\frac{(n+\alpha+1)(n+2\alpha)}{(n+\alpha)(n+1)}$ the difference equation
\begin{align}		
	{\textstyle\frac{(\mathrm{N}-n+1)(n+2\alpha-1)}{2(n+\alpha-1)n}}{\textstyle\frac{N_n^2\,a_{n-1}}{N_{n-1}^2}}
	-{\textstyle\frac{N_n^2\,a_{n}}{N_{n}^2}}
	-{\textstyle\frac{(\mathrm{N}+n+2\alpha+1)(n+1)}{2(n+\alpha+1)(n+2\alpha)}}{\textstyle\frac{N_n^2\,a_{n+1}}{N_{n+1}^2}}&=0,
\nonumber \\
\textstyle(\mathrm{N}-n+1)\,a_{n-1}
-2(n+\alpha)\,a_{n}
-(\mathrm{N}+n+2\alpha+1)\,a_{n+1}&=0,
\nonumber
\end{align}
and we can guess that $a_n$ must contain $(\mathrm{N}-n)^{-1}$ to cancel the factor $(\mathrm{N}-n+1)$ with an index shift to $a_{n-1}$ and $(\mathrm{N}+n+2\alpha)^{-1}$ to cancel the factor $(\mathrm{N}+n+2\alpha+1)$ with an index shift to $a_{n+1}$. We guess a decaying solution of factorials ending at these numbers as denominator, we insert and divide by the solution to test
\begin{align}
	1/a_n&=(\mathrm{N}-n)!(\mathrm{N}+n+2\alpha)!\nonumber\\
     	(\mathrm{N}+n+2\alpha)
			-2(n+\alpha)
			-(\mathrm{N}-n)
		&=0,\nonumber\qquad\blacksquare
\end{align}
We express the solution normalized to its first weight $a_0=1$ and with the dimensionality constant written as
$2\alpha=\mathrm{D}-2$ as
\begin{align}
	a_n=\frac{\mathrm{N}!(\mathrm{N}+\mathrm{D}-2)!}{(\mathrm{N}-n)!(\mathrm{N}+n+\mathrm{D}-2)!}.\label{eq:inphase}
\end{align}
The $\mathrm{N}$-fold zero at $x=-1$ can be modulated to $x=1$ by sign alteration for the odd polynomials, i.e.\ $a_n=(-1)^n\frac{\mathrm{N}!(\mathrm{N}+\mathrm{D}-2)!}{(\mathrm{N}-n)!(\mathrm{N}+n+\mathrm{D}-2)!}$.

\subsection{Max-flat pattern: generalized higher-order cardioid}\label{apdx:maxflat}
This was inspired by the maxflat FIR filter \texttt{firflatLp(N,K,M)} 
in Sch\"u{\ss}ler's DSP2 book~\cite[p.92]{Schuessler}.
A polynomial with $\mathrm{L}$-fold zero at $x=1$ and $\mathrm{M}$-fold zero at $x=-1$ integrated yields a pattern that is zero with $\mathrm{L}$-degree flatness at $x=-1$ and a positive constant with $\mathrm{M}$-degree flatness at $x=1$, with $\mathrm{N}=\mathrm{L}+\mathrm{M}+1$, a Butterworth design
\begin{align}
	g_\mathrm{N}&=\frac{(\mathrm{L+M}+1)!}{2^{\mathrm{L+M}+1}\,\mathrm{L}!\,\mathrm{M}!}\int_{-1}^x(1-x)^\mathrm{L}(1+x)^\mathrm{M}\,\mathrm{d}x.
\end{align}
The factor provides $g_\mathrm{N}(1)=1$ and is obtained by applying partial integration recursively,
\begin{align}
	(m+1)\int_{-1}^1(1-x)^l(1+x)^m\,\mathrm{d}x
	&=(1-x)^l(1+x)^{m+1}\big|_{-1}^1+l\,\int_{-1}^1(1-x)^{l-1}(1+x)^{m+1}\,\mathrm{d}x\nonumber\\
   (m+1)\,g_{lm}&=g_{l,m+1}'+l\,g_{l-1,m+1}\nonumber\\
	(m+1)(m+2)\,g_{lm}&=(m+2)g_{l,m+1}'+l\,g_{l-1,m+2}'+l\,(l-1)\,g_{l-2,m+2}\nonumber\\
	%(m+1)(m+2)(m+3)\,g_{lm}&=(m+2)(m+3)g_{l,m+1}'+(m+3)l\,g_{l-1,m+2}'\nonumber\\
	%&+l\,(l-1)\,g_{l-2,m+3}'+l(l-1)(l-2)g_{l-3,m+3}\nonumber\\
	\frac{(m+l+1)!}{m!}g_{lm}&=\underbrace{\sum_{k=0}^{l}\frac{(m+l+1)!\,l!}{(m+k+1)!(l-k)!}\,g_{l-k,m+k+1}'}_{g_{l-k,m+k+1}'|_1=0\forall k\neq l, \quad g_{l-k,m+k+1}'|_{-1}=0\forall k}+0\,g_{-1,m+l+1}\nonumber\\
	\frac{(m+l+1)!}{m!}\,g_{lm}\big|_1&=l!\,g_{0,m+l+1}'\big|_1=l!\,2^{m+l+1}\nonumber\\
	g_{lm}\big|_1&=\frac{2^{l+m+1}\,l!\,m!}{(l+m+1)!}.\nonumber
\end{align}

We can call $g_\mathrm{N}=y$ and derive twice to get 
$(1-x^2)\;y''=\mathrm{M}(1-x)\;y'-\mathrm{L}(1+x)\;y'$, i.e.\
$(1-x^2)\;y''+(\mathrm{M}+\mathrm{L})\,x\,y'-(\mathrm{M}-\mathrm{L})\,y'=0$.
 We can insert $y=\sum_{n=0}^\mathrm{N}\frac{a_n}{N_n^2}\mathcal{P}_n$ 
and get:
\begin{align}
	(1-x^2)\sum_{n=0}^\mathrm{N}\frac{a_n}{N_n^2}\mathcal{P}_n''+
	(\mathrm{M}+\mathrm{L})\, x\sum_{n=0}^\mathrm{N}\frac{a_n}{N_n^2}\mathcal{P}_n'+(\mathrm{M}-\mathrm{L})\sum_{n=0}^\mathrm{N}\frac{a_n}{N_n^2}\mathcal{P}_n'&=0,
\end{align}
and make use of 
$(1-x^2)\mathcal{P}_n''=
(1+2\alpha)\,x\,\mathcal{P}_n'-n(n+2\alpha)\,\mathcal{P}_n$, 
\begin{align}
	\sum_{n=0}^\mathrm{N}\frac{a_n}{N_n^2}\,\left[(\mathrm{M}+\mathrm{L}+2\alpha+1)x\mathcal{P}_n'
	-n(n+2\alpha)\,\mathcal{P}_n
	+(\mathrm{M}-\mathrm{L})\,\mathcal{P}_n'
	\right]
	=0,
\end{align}
We integrate to get 
\begin{multline}
	\sum_{n=0}^\mathrm{N}\frac{a_n}{N_n^2}\,
	\bigg\{(\mathrm{M}+\mathrm{L}+2\alpha+1)\int x\,\mathcal{P}_n'\,\mathrm{d}x
	-n(n+2\alpha)\,\int\mathcal{P}_n\,\mathrm{d}x
	+(\mathrm{M}-\mathrm{L})\,\mathcal{P}_n
	\bigg\}
	=C,
\end{multline}
and make use of\\
$2(n+\alpha)\int_{-1}^x x\,\mathcal{P}_n'\,\mathrm{d}x=
n(n+2\alpha)[\frac{1}{n+1}\mathcal{P}_{n+1}+
\frac{1}{n+2\alpha-1}\mathcal{P}_{n-1}]$\\
$2(n+\alpha)\int_{-1}^x\mathcal{P}_n\mathrm{d}x=\frac{(n+2\alpha)}{n+1}\mathcal{P}_{n+1}-\frac{n}{n+2\alpha-1}\mathcal{P}_{n-1}$
\begin{multline}\textstyle
	\sum_{n=0}^\mathrm{N}\frac{a_n}{N_n^2}\,\bigg\{
	\frac{n(n+2\alpha)(\mathrm{M}+\mathrm{L}-n+1)}{(n+1)2(n+\alpha)}\,\mathcal{P}_{n+1}+\frac{n(n+2\alpha)(\mathrm{M}+\mathrm{L}+n+2\alpha+1)}{(n+2\alpha-1)2(n+\alpha)\,}\,\mathcal{P}_{n-1}
	+(\mathrm{M}-\mathrm{L})\,\mathcal{P}_n
	\bigg\}
	=C.\nonumber
\end{multline}
With the orthogonality when integrating over $\mathcal{P}_n$
\begin{multline}
	\textstyle
	\frac{N_{n}^2a_{n-1}}{N_{n-1}^2}\,
	\frac{(n-1)(n+2\alpha-1)(\mathrm{M}+\mathrm{L}-n+2)}{2n(n+\alpha-1)}
	+\frac{N_{n}^2a_{n+1}}{N_{n+1}^2}\,
	\frac{(n+1)(n+2\alpha+1)(\mathrm{M}+\mathrm{L}+n+2\alpha+2)}{2(n+2\alpha)(n+\alpha+1)}
	+a_n\,(\mathrm{M}-\mathrm{L})
	=E\,\delta_n.\nonumber
\end{multline}
$N_{n}^2N_{n-1}^{-2}=\frac{2(n+\alpha-1)n}{2(n+\alpha)(n+2\alpha-1)}$ and
$N_{n}^2N_{n+1}^{-2}=\frac{2(n+\alpha+1)(n+2\alpha)}{2(n+\alpha)(n+1)}$ yield
the difference equation
\begin{multline}
	\textstyle
	a_{n-1}\,
	(n-1)(\mathrm{M}+\mathrm{L}-n+2)
	+a_{n+1}\,
	(n+2\alpha+1)(\mathrm{M}+\mathrm{L}+n+2\alpha+2)\\
	+a_n\,(\mathrm{M}-\mathrm{L})2(n+\alpha)
	=E\,\delta_n.\nonumber
\end{multline}
As the resulting sequence $a_n$ should be limited to the degree $\mathrm{N}=\mathrm{M}+\mathrm{L}+1$, we can abbreviate the recurrence by $\mathrm{N}$ and $\Delta\mathrm{N}$ as 
\begin{align}
	\textstyle
	(\mathrm{N}-n+1)(n-1)\,a_{n-1}
	+2\Delta\mathrm{N}(n+\alpha)\,a_n
	+
	(\mathrm{N}+n+2\alpha+1)(n+2\alpha+1)\,a_{n+1}
	=E\,\delta_n.\nonumber
\end{align}
We can focus on using the recurrence for $n>0$ so that we can omit $E\,\delta_n$ and write
\begin{align}
	a_{n+1}=-\frac{(\mathrm{N}-n+1)(n-1)\,a_{n-1}
	+2\Delta\mathrm{N}(n+\alpha)\,a_n}{(\mathrm{N}+n+2\alpha+1)(n+2\alpha+1)}, \quad n\geq1.\label{eq:maxflat_iteration}
\end{align}
It is useful to start from assuming $a_1=1$ and obtaining $a_{n>1}$ from the above iteration with $n=1,\dots,\mathrm{N}-1$.
The missing value for $a_0$ can be obtained as the offset at $x=-1$, where $g_\mathrm{N}(-1)=0$ should have a zero, and where $\mathcal{P}_n(-1)=(-1)^n$ consequently yields
\begin{align}
	a_0&=-N_0^2\sum_{n=1}^{\mathrm{N}}\frac{a_n}{N_n^2}\mathcal{P}_n(-1)=-\sum_{n=1}^\mathrm{N}\frac{(-1)^n(2n+\mathrm{D}-2)\;(n+\mathrm{D}-3)!}{n!}\,a_n.\label{eq:maxflat_a0}
\end{align}
%
%An alternative attempt
%\begin{align}
%	a_0&=\int_{-1}^1\left(\int_{-1}^{x}(1-t)^l(1+t)^m\mathrm{d}t\right)w\,\mathrm{d}x\nonumber\\
%	&=\int_{-1}^1\left\{\int_{0}^{1}\{1-[u(1+x)-1]\}^l\{1+[u(1+x)-1]\}^m(1+x)\mathrm{d}u\right\}w\,\mathrm{d}x\nonumber\\
%	&=\int_{-1}^1\left[\int_{0}^{1}2^l\left(1-\textstyle\frac{1+x}{2}u\right)^l(1+x)^mu^m\,(1+x)\mathrm{d}u\right]w\,\mathrm{d}x\nonumber\\
%	&=2^l\int_{-1}^1\underbrace{\left[\int_{0}^{1}\left(1-\textstyle\frac{1+x}{2}u\right)^lu^m\,\mathrm{d}u\right]}_{\frac{1}{m+1}F(-l,m+1;m+2;\frac{1+x}{2})}(1+x)^{m+\frac{\mathrm{D}-3}{2}+1}(1-x)^\frac{\mathrm{D}-3}{2}\,\mathrm{d}x\nonumber\\
%	&=\frac{2^{l+m+\mathrm{D}-2}}{m+1}\int_{0}^1F(-l,m+1;m+2;t)\,t^{m+\frac{\mathrm{D}-1}{2}}\,(1-t)^\frac{\mathrm{D}-3}{2}\,\mathrm{d}t
%\end{align}

Scaling can be fixed to unity by dividing all the $a_n\leftarrow a_n/b$ by $b=g_\mathrm{N}(1)-g_\mathrm{N}(-1)$,
\begin{align}
	b&=\sum_{n=1}^\mathrm{N}\frac{[1-(-1)^n](2n+\mathrm{D}-2)\;(n+\mathrm{D}-3)!}{n!}\,a_n\nonumber\\
	&=
	\sum_{k=0}^{\lfloor\frac{\mathrm{N}-1}{2}\rfloor}\frac{(4k+\mathrm{D})\;(2k+\mathrm{D}-2)!}{(2k+1)!}\,a_{2k+1}.\label{eq:maxflat_b}
\end{align}
%The constant part $a_0$ is obtained separately by integrating, with $N_0^2=\frac{\sqrt{\pi}\Gamma(\frac{\mathrm{D-2}}{2})}{\Gamma(\frac{\mathrm{D-1}}{2})}$, $C_0^{(\alpha)}=1$, $w=(1-x^2)^\frac{\mathrm{D}-3}{2}$, $\Delta N=\mathrm{M}-\mathrm{L}$,
%\begin{align}
%	a_0&=\frac{1}{N_0^2}\frac{(\mathrm{L+M}+1)!}{2^{\mathrm{L+M}+1}\,\mathrm{L}!\,\mathrm{M}!}\int_{-1}^1(1-x)^{\mathrm{L}+\frac{\mathrm{D}-3}{2}}\,(1+x)^{\mathrm{M}+\frac{\mathrm{D}-3}{2}}\mathrm{d}x\nonumber\\
%	&=
%	\frac{1}{N_0^2}\frac{(\mathrm{L+M}+1)!}{2^{\mathrm{L+M}+1}\,\mathrm{L}!\,\mathrm{M}!}\frac{2^{\mathrm{L+M+D}-2}\Gamma(\mathrm{L}+\frac{D}{2})\Gamma(\mathrm{M}+\frac{D}{2})}{(\mathrm{L+M+D}-1)!}
%\end{align}

For a non-recurrent, closed-form solution as for the non-generalized higher-order cardioid, only a simplified setting $\mathrm{M}=\mathrm{L}$ helps so that the $a_n$ term vanishes. We get
\begin{align}
\textstyle	(n-2)(\mathrm{N}-n+2)\,a_{n-2}+(n+2\alpha)(\mathrm{N}+n+2\alpha)\,a_{n}=0,\\
\textstyle a_n=\frac{(-1)^\frac{n}{2}(n-2)!!(2\alpha)!!(\mathrm{N}+2\alpha)!!\mathrm{N}!!}{(n+2\alpha)!!(\mathrm{N}+n+2\alpha)!!(\mathrm{N}-n)!!}\\
\textstyle
\frac{(n-2)(\mathrm{N}-n+2)}{(n+2\alpha)(\mathrm{N}+n+2\alpha)}\frac{(n-4)!!}{(n+2\alpha-2)!!(\mathrm{N}+n+2\alpha-2)!!(\mathrm{N}-n+2)!!}
-\frac{(n-2)!!}{(n+2\alpha)!!(\mathrm{N}+n+2\alpha)!!(\mathrm{N}-n)!!}
=0\quad\blacksquare
\end{align}

\newpage
\section{MATLAB Code examples}\label{apdx:matlab}
\subsection{gegenbauer{\_}me(x,N,D)}
Ultraspherical polynomials normalized to $\mathcal{P}_n(1)=1$ as given by the recurrence of eq.~\eqref{eq:ultraspherical_rec}.
\begin{lstlisting}[style=Matlab-editor,basicstyle=\small\mlttfamily,stepnumber=1,numbers=left, numbersep=5pt,numberstyle=\smaller, keepspaces, mlshowsectionrules]
function C=gegenbauer_me(x,N,D)
C=zeros(length(x),N+1);
C(:,1)=1;
if N>0
    C(:,2)=x(:);
end
for n=2:N
   C(:,n+1)=(2*n-4+D)/(n-3+D)*x(:).*C(:,n)-...
       (n-1)/(n-3+D)*C(:,n-1);
end
\end{lstlisting}

\subsection{chebyshev(x,N)}
Chebyshev polynomials $T_n(x)$ for $\mathrm{D}=2$ as given by the recurrence of eq.~\eqref{eq:chebyshev}.
\begin{lstlisting}[style=Matlab-editor,basicstyle=\small\mlttfamily,stepnumber=1,numbers=left, numbersep=5pt,numberstyle=\smaller, keepspaces, mlshowsectionrules]
function T=chebyshev(x,N)
x=x(:);
T=zeros(length(x),N+1);
T(:,1)=1;
if N>0
    T(:,2)=x;
end
for n=2:N
    T(:,n+1)=2*x.*T(:,n)-T(:,n-1);
end
\end{lstlisting}

\subsection{legendre{\_}u(x,N)}
Legendre polynomials $P_n$ as given by the recurrence of eq.~\eqref{eq:legendre}.
\begin{lstlisting}[style=Matlab-editor,basicstyle=\small\mlttfamily,stepnumber=1,numbers=left, numbersep=5pt,numberstyle=\smaller, keepspaces, mlshowsectionrules]
function P=legendre_u(x,N)
x=x(:);
P=zeros(length(x),N+1);
P(:,1)=1;
if N>0
    P(:,2)=x;
end
for n=1:N-1
    P(:,(n+2))=((2*n+1)*x.*P(:,n+1)-n*P(:,n))/(n+1);
end
\end{lstlisting}

\subsection{maxre(N,D)}
max-rE  as defined in the optimization problem of eq.~\eqref{eq:maxre}, and solved by Newton method, in order to get the maximum $r_\mathrm{E}$ used to evaluate the max-rE weights $a_n=\mathcal{P}_n(r_\mathrm{E})$.
\begin{lstlisting}[style=Matlab-editor,basicstyle=\small\mlttfamily,stepnumber=1,numbers=left, numbersep=5pt,numberstyle=\smaller, keepspaces, mlshowsectionrules]
function [a,re]=maxre(N,D)
re=cos(90*pi/180/(N+1)); % guessing by known D=2 max-rE
dx0=1/(N*10); % finite difference step
dx=dx0;
iterate=1;
it=0;
while iterate % solve for P_{N+1}(rE)=0 with rE->max
    reold=re; % store previous rE estimate
    p0=gegenbauer_me(re,N+1,D);
    p0=p0(end); % store previous P_{N+1}(rE)
    pp=gegenbauer_me(re+dx,N+1,D); % estimate change+
    pp=pp(end);
    pm=gegenbauer_me(re-dx,N+1,D); % estimate change-
    pm=pm(end);    
    dpdx=(pp-pm)/(2*dx); % central derivative estim.
    re=re-p0/dpdx; % Newton-Raphson improvement step
    dx=abs(p0)*dx0;
    if (abs(reold-re)<1e-9)||(abs(dx)<1e-11)||(abs(p0)<1e-9)
        iterate=0; % converged/good enough
    end    
    it=it+1;
end
a=gegenbauer_me(re,N,D)';
\end{lstlisting}

\subsection{one{\_}over{\_}gegenbauer{\_}norm{\_}square(N,D)}
Reciprocal of normalization squared $\frac{N_0^2}{N_n^2}$ as given in eq.~\eqref{eq:norm}
\begin{lstlisting}[style=Matlab-editor,basicstyle=\small\mlttfamily,stepnumber=1,numbers=left, numbersep=5pt,numberstyle=\smaller, keepspaces, mlshowsectionrules]
function nn=one_over_gegenbauer_norm_square(N,D);
nn=zeros(N+1,1);
nn(1)=1;
a=(D-2)/2;
for n=1:N
    if (n==1)&&(a==0)
        aaa=1; % dummy a->0 that cancels
        nn(n+1)=(a+n)*(2*aaa+0)/(n*(aaa+0))*nn(n);
    else
        nn(n+1)=(a+n)*(2*a+n-1)/(n*(a+n-1))*nn(n);
    end
end
\end{lstlisting}

\subsection{supercard(N)}
Supercardioid as defined in the optimization problem of eq.~\eqref{eq:supercard} and \autoref{apdx:gram}.
\begin{lstlisting}[style=Matlab-editor,basicstyle=\small\mlttfamily,stepnumber=1,numbers=left, numbersep=5pt,numberstyle=\smaller, keepspaces, mlshowsectionrules]
function a=supercard(N,D)
G=gram_matrix_supercard(N,D);
[Q,V]=eig(G);
a=Q(:,1);
a=a/a(1);
\end{lstlisting}

\begin{lstlisting}[style=Matlab-editor,basicstyle=\small\mlttfamily,stepnumber=1,numbers=left, numbersep=5pt,numberstyle=\smaller, keepspaces, mlshowsectionrules]
function G=gram_matrix_supercard(N,D)
G=zeros(N+1,N+1);
a=(D-2)/2;
n=0:N;
N0=sqrt(pi)*gamma((D-1)/2)/gamma((D)/2);
G(1:N+2:end)=one_over_gegenbauer_norm_square(N,D)/2;
for k=0:N/2
    for l=0:(N-1)/2
        terml=(-1)^(l+1)*prod(a+(1:2*l+1))/...
            (factorial(l)*prod(a+1+(l:2*l-1)));
        termk=(-1)^(k)*prod(a+(1:2*k))/...
            (factorial(k)*prod(a+(k:2*k-1)));
        termlk=2/((2*l+1)*(2*l+2*a+1)-(2*k)*(2*k+2*a));
        g=terml*termk*termlk/N0;
        G(2*l+2,2*k+1)=g;
        G(2*k+1,2*l+2)=g;
    end
end
\end{lstlisting}

\subsection{inphase(N)}
Higher-order cardioid or inphase weights $a_n$ as defined in eq.~\eqref{eq:inphase1} or derived around \eqref{eq:inphase}.
\begin{lstlisting}[style=Matlab-editor,basicstyle=\small\mlttfamily,stepnumber=1,numbers=left, numbersep=5pt,numberstyle=\smaller, keepspaces, mlshowsectionrules]
function a=inphase(N,D)
n=0:N;
a=factorial(N)./factorial(N-n);
b=factorial(N+D-2)./factorial(N+n+D-2);
a=a.*b;
a=a(:);
\end{lstlisting}

\newpage
\subsection{inphase{\_}maxflat(N,L,D)}
Inphase weights $a_n$ with $\mathrm{L}$ flatness degrees on axis, as defined in eqs.~\eqref{eq:maxflat_iteration} \eqref{eq:maxflat_a0} \eqref{eq:maxflat_b}.
\begin{lstlisting}[style=Matlab-editor,basicstyle=\small\mlttfamily,stepnumber=1,numbers=left, numbersep=5pt,numberstyle=\smaller, keepspaces, mlshowsectionrules]
function a=inphase_maxflat(N,L,D)
a=zeros(N+1,1);
M=N-L-1;
dN=L-M;
a(2)=1;
al=(D-2)/2;
for n=1:N-1
    a(n+2)=-2*dN*(n+al)*a(n+1);
    if n>1
        a(n+2)=a(n+2) - ...
            (N-n+1)*(n-1)*a(n);
    end
    a(n+2)=a(n+2)/((N+n+2*al+1)*(n+2*al+1));
end
nn=one_over_gegenbauer_norm_square(N,D);
g1=a'*nn;
nn(2:2:end)=-nn(2:2:end);
g2=a'*nn;
a(1)=-g2;
a=a/(g1-g2);

\end{lstlisting}

\vspace{-4mm}
\subsection{spherecap(N,D)}
\vspace{-2mm}
Spherical cap weights $a_n$ for $\mathrm{D}=2,3$ as defined in eq.~\eqref{eq:spherecap1} or derived in eqs.~\eqref{eq:spherecap},  \eqref{eq:spherecap0}, which would also contain a more arbitrary definition for other values of $\mathrm{D}$.
\begin{lstlisting}[style=Matlab-editor,basicstyle=\small\mlttfamily,stepnumber=1,numbers=left, numbersep=5pt,numberstyle=\smaller, keepspaces, mlshowsectionrules]
function a=spherecap(N,D,x0)
if D==2
    a0=acos(x0);
elseif D==3
    a0=1-x0;
else
    error("D that is not 2 or 3 is not implemented");
end
a=(D-2)/2;
w0=(1-x0.^2).^(a-1/2);
C0=gegenbauer_me(x0,N+1,D);
n=1:N;
a=w0./(2*n+2*a).*(C0(1:N)-C0(3:N+2));
a=[a0; a(:)];
return
\end{lstlisting}

\vspace{-6mm}
\subsection{eval{\_}PEQrVrEFBR(a,D)\label{apdx:metrics}}
\vspace{-3mm}
This function evaluates the level metrics $P$ and $E$, the directivity factor $Q$, the vector metrics $r_\mathrm{V}$ and $r_\mathrm{V}$ and the front-to-back ratio $FBR$ as given in eqs.~\eqref{eq:P1}\eqref{eq:E1}\eqref{eq:Q1}\eqref{eq:rV1}\eqref{eq:rE1}\eqref{eq:FBR1} and derived in eqs.~\eqref{eq:P}\eqref{eq:E}\eqref{eq:QFBR}\eqref{eq:rV}\eqref{eq:rE} of \autoref{apdx:axisymmetrics}.
\begin{lstlisting}[style=Matlab-editor,basicstyle=\small\mlttfamily,stepnumber=1,numbers=left, numbersep=5pt,numberstyle=\smaller, keepspaces, mlshowsectionrules]
function PEQrVrEFBR=eval_PEQrVrEFBR(a,D)
a=a(:);
N=length(a)-1;
n=0:N;
Nn=one_over_gegenbauer_norm_square(N,D)';
SD1=2*sqrt(pi)^D/gamma(D/2); % unit sphere
% analytic estimates of P,E,Q,rV,rE
P=a(1)*SD1;
E=SD1*Nn*(a(:).^2);
Q=SD1*(Nn*a(:))^2/E;
rV=a(2)/a(1);
if D==2 % recurrence weight
    b=(1+(n==0))/2;
else
    b=(n+D-2)./(2*n+D-2);
end
rE=SD1*(Nn(1:N).*b(1:N))*(2*a(1:N).*a(2:N+1))/E;
% numerical estimate of FBR
z=linspace(-1+1e-4,1-1e-4,300);
g=gegenbauer_me(z,N,D)*(Nn(:).*a(:));
w=sqrt((1-z.^2)).^(D-3);
FBR=sum(g(151:300).^2)/sum(g(1:150).^2);
PEQrVrEFBR=[P E Q rV rE FBR];
\end{lstlisting}

\vspace{-6mm}
\subsection{test{\_}PEQrVrEFBR(a,D)\label{apdx:metrics_verify}}
\vspace{-3mm}
Script for numerical verification:
\begin{lstlisting}[style=Matlab-editor,basicstyle=\small\mlttfamily,columns=fixed,stepnumber=1,numbers=left, numbersep=5pt,numberstyle=\smaller, keepspaces, mlshowsectionrules]
% numerical verification of analytic metrics
% P, E, Q, rV, rE, 
% and FBR is just numeric; with eval_PEQrVrEFBR.m
wtype={'basic','maxre','supercard','inphase'};
clc,
Npt=900;
for N=[1,2,3,4,5]
    n=0:N;
    for D=[2,3]
        gn=one_over_gegenbauer_norm_square(N,D);
        if D==2
            phi=2*pi*(0:Npt-1)/Npt;
            z=cos(phi);
        else
            z=linspace(-1+1/Npt,1-1/Npt,Npt);
        end
        for tt=1:length(wtype)
            switch wtype{tt}
                case 'basic'
                    a=ones(N+1,1);
                case 'maxre'
                    [a,rEm]=maxre(D,N);
                case 'supercard'
                    a=supercard(N,D);
                case 'inphase'
                    a=inphase(N,D);
            end
            % numerical verification estimates
            g=gegenbauer_me(z,N,D)*(a(:).*gn(:));
            if D==2
                P=sum(g)*2*pi/Npt;
                E=sum(g.^2)*2*pi/Npt;
                Q=g(1).^2*2*pi/E;
                rV=sum(g(:).*cos(phi(:)))/sum(g);
                rE=sum(g(:).^2.*cos(phi(:)))/sum(g.^2);
            else
                P=sum(g)*4*pi/Npt;
                E=sum(g.^2)*4*pi/Npt;
                Q=g(end).^2*4*pi/E;
                rV=sum(g(:).*z(:))/sum(g);
                rE=sum(g(:).^2.*z(:))/sum(g.^2);
            end
            rV=2*acos(rV)*180/pi;
            rE=2*acos(rE)*180/pi;
            vvnum=[P,E,Q,rV,rE];
            % analytic estimates:
            vv=eval_PEQrVrEFBR(a,D);
            vv(4:5)=2*acos(vv(4:5))*180/pi;
            vv(6)=10*log10(vv(6));
            % value output
            ee=real(vvnum-vv(1:length(vvnum)));
 disp(['D=' num2str(D) ', N=' num2str(N) ', ' wtype{tt}])
   disp('   P       E       Q       rV      rE      FBR')
            fprintf('%7.1f ',vv)
            fprintf('\n')
            fprintf('%7.1f ',vvnum)
            fprintf('\n')
            if max(abs(ee))>1
                fprintf('%7.1f ',abs(ee))
                fprintf('\n')
                error('maximum error larger than 1')
            end
        end
    end
end
\end{lstlisting}

\newpage
\section{2D Ambisonic panning as Fourier series}\label{apdx:circle_fourier}
Appendices~\ref{apdx:axiallaplacian} to \ref{apdx:matlab} derived the ultraspherical polynomials that uniformly satisfy arbitrary dimensions $\mathrm{D}$. Appendices \ref{apdx:circle_fourier}.1 until \ref{apdx:circle_fourier}.4 here are an optional, reassuring meditation/verification exercise that prove the results for $\mathrm{D}=2$ based on the Fourier series.
\vspace*{-6mm}
\subsection{Panning function as Fourier series}
\vspace*{-2mm}
Assuming that we first try to find a continuous panning function, we can regard the Fourier series as suitable means to express this function. The Fourier coefficient $a_{|m|}\,e^{-\mathrm{i}m\varphi_\mathrm{s}}$ is able to control the direction and length of the $\bm r_\mathrm{E}$- and $\bm r_\mathrm{V}$ vector perfectly:
\begin{align}
	g(\varphi)&=\frac{1}{2\pi}\sum_{m=-\infty}^\infty a_{|m|}\,e^{-\mathrm{i}m\varphi_\mathrm{s}}\,e^{\mathrm{i}m\varphi}.
\end{align}
While $a_{|m|}=1$, yields a directional Dirac delta $g(\varphi)=\delta(\varphi-\varphi_\mathrm{s})$, values for $a_{|m|}$ tending to zero for large $|m|$ yield a smoother curve $g(\varphi)$ that is easier to sample. 

Before looking into sampling the panning function at the loudspeaker positions, we seek whether the definitions of the linear and quadratic loudness and vector models are invariant with regard to the panning direction $\varphi_\mathrm{s}$, and what their relation is to $a_{|m|}$. 

For the loudness models, this is easily shown by replacing the integration variable $\varphi$ by $\phi=\varphi-\varphi_\mathrm{s}$, and for the quadratic measure by considering orthogonality:
\begin{align}
	P&=\int_{-\pi}^{\pi} g(\varphi)\,\mathrm{d}\varphi=\sum_{m=-\infty}^\infty\frac{a_{m}}{2\pi}\underbrace{\int_{-\pi-\varphi_\mathrm{s}}^{\pi-\varphi_\mathrm{s}}e^{\mathrm{i}m\phi}\,\mathrm{d}\phi}_{:=2\pi\,\delta_{m0}}=a_0,\\
	E&=\int_{-\pi}^\pi |g(\varphi)|^2\,\mathrm{d}\varphi=\sum_{m'=-\infty}^\infty\sum_{m=-\infty}^\infty\frac{a_{m}a_{m'}^*}{(2\pi)^2}\underbrace{\int_{-\pi-\varphi_\mathrm{s}}^{\pi-\varphi_\mathrm{s}}e^{\mathrm{i}(m-m')\phi}\,\mathrm{d}\phi}_{:=2\pi\,\delta_{mm'}}=\sum_{m=-\infty}^\infty\frac{|a_{m}|^2}{2\pi}.
\end{align}
Obviously the quadratic loudness measure involves all of the Fourier terms $a_m$ (Parseval's theorem) while the linear one only considers $a_0$.

The notation as complex phasors ($x$ as real and $y$ as imaginary part) simplifies the calculation of the vector models, i.e.\ using $\bm\theta=e^{\mathrm{i}\varphi}$. Orthogonality yields further simplifications:
\begin{align}	
	{\bm{r}_\mathrm{V}}&=\frac{\int_{-\pi}^\pi g(\varphi)\,e^{\mathrm{i}\varphi}\,\mathrm{d}\varphi}{ V}=\frac{\sum_{m} \frac{a_{m}}{2\pi}\,e^{-\mathrm{i}m\varphi_\mathrm{s}}\overbrace{\int_{-\pi}^\pi e^{\mathrm{i}(m+1)\varphi}\mathrm{d}\varphi}^{2\pi\delta_{m+1,0}}}{a_0}=\frac{a_1}{a_0}e^{\mathrm{i}\varphi_\mathrm{s}},\\	
	{\bm{r}_\mathrm{E}}&=
	\frac{\int_{-\pi}^\pi |g(\varphi)|^2\,e^{\mathrm{i}\varphi}\,\mathrm{d}\varphi}{ E}=\frac{\sum_{m,m'} \frac{a_{m}a_{m'}^*}{(2\pi)^2}\,e^{-\mathrm{i}(m-m')\varphi_\mathrm{s}}\overbrace{\int_{-\pi}^\pi e^{\mathrm{i}(m-m'+1)\varphi}\mathrm{d}\varphi}^{2\pi\delta_{m-m'+1,0}}}{\sum_{m} \frac{|a_{m}|^2}{2\pi}}\nonumber\\
	&=\frac{\sum_{m=-\infty}^\infty a_{m}a_{m+1}^*}{\sum_{m=-\infty}^\infty |a_{m}|^2}\,e^{\mathrm{i}\varphi_\mathrm{s}}.
\end{align}
In all of the above results, the vectors point exactly into the desired direction. The vector component $e^{-\mathrm{i}m\varphi_\mathrm{s}}$ just undergoes a sign change but remains untouched otherwise. Note that this is largely fulfilled independently of what the weights $a_{m}$ exactly look like.

The lengths of both vectors is independent of the panning direction $\varphi_\mathrm{s}$. This can also be logically concluded from two facts, see \autoref{sec:finite-order Fourier series}: (1) $g(\varphi)$ is symmetric around $\varphi_\mathrm{s}$, (2) and its shape is preserved under rotation. The length $\|\mathrm{r}_\mathrm{V}\|$ only depends on the weights $a_0$ and $a_1$, while the length $\|\bm{r}_\mathrm{E}\|$ depends on all the Fourier coefficient weights $a_{m}$ involved, which seems to be important as the $\bm r_\mathrm{E}$ vector was an excellent predictor in the listening experiments about direction and width.

Surely, both the quadratic criteria, $E$ and $\bm r_\mathrm{E}$ are the more critical ones when considering the continuous panning function as they involve all of the weights $a_{m}$.

\subsection{Finite-order Fourier series}\label{sec:finite-order Fourier series}
\paragraph{Rotation invariant shape.} The represented shape of a $2\pi$-periodic function in a finite-order Fourier series is rotation-invariant. For $2\pi\,g(\varphi)=\sum_{m=-\mathrm{N}}^{\mathrm{N}}a_{|m|}\,e^{\mathrm{i}m(\varphi-\varphi_\mathrm{s})}$ with arbitrary aiming $\varphi_\mathrm{s}$, rotation by $-\varphi_\mathrm{s}$ is losslessly expressed by modified coefficients $a_{|m|}\,e^{+\mathrm{i}m\varphi_\mathrm{s}}$ (\emph{shift theorem}), proved by $\sum_{m=-\mathrm{N}}^{\mathrm{N}}a_{|m|}\,e^{\mathrm{i}m\varphi_\mathrm{s}}\,e^{\mathrm{i}m(\varphi-\varphi_\mathrm{s})}=\sum_{m=-\mathrm{N}}^{\mathrm{N}}a_{|m|}\,e^{\mathrm{i}m\varphi}$, which obviously is $2\pi\,g(\varphi+\varphi_\mathrm{s})$, the shape of the original function rotated to aim at $0$.

\paragraph{Symmetry.}
A real-valued and symmetric Fourier coefficient $a_{|m|}=\Re\{a_{\pm m}\}$ yields a symmetric real shape
$\sum_{m=-\mathrm{N}}^\mathrm{N} a_m\,e^{\mathrm{i}m\phi}=a_0\,e^{\mathrm{i}0\phi}+\sum_{m=1}^\mathrm{N}\left(a_{-m}\,e^{-\mathrm{i}m\phi}+a_{m}\,e^{\mathrm{i}m\phi}\right)$. Symmetry yields for the second term $\sum_{m=1}^\mathrm{N}2a_m\Re\{e^{-\mathrm{i}m\phi}\}=\sum_{m=1}^\mathrm{N}2a_m\cos(m\phi)$. Such a cosine series is symmetric around $\phi=\varphi-\varphi_\mathrm{s}$, i.e.\ around the panning direction.

Is $\gamma_m=a_{|m|}$ a symmetric rectangular function, the finite geometric sum lets us obtain
\begin{align}
	g(\varphi)&=\frac{1}{2\pi}\sum_{m=-N}^Ne^{\mathrm{i}m\phi}=\frac{e^{-\mathrm{i}m\phi}-e^{\mathrm{i}(m+1)\phi}}{1-e^{\mathrm{i}\phi}}=\frac{\sin[(m+\frac{1}{2})\phi]}{2\pi\sin\frac{\phi}{2}},
\end{align}
the periodic sinc function shifted to $\varphi_\mathrm{s}$. However, there are better ways to define $a_{|m|}$.

\subsection{Finite-order series with maximum \texorpdfstring{$\bm{r}_\mathrm{E}$}{rE} vector length}
The maximization of the $\bm r_\mathrm{E}$ vector length by optimal Fourier weights $a_{m}$ yields a narrower perceived width and beneficial sidelobe suppression~\cite{Daniel99}. 
For this purpose, the length $r_\mathrm{E}=\frac{\sum_{m=-\infty}^\infty a_{m}a_{m+1}}{\sum_{m=-\infty}^\infty a_{m}^2}$ is differentiated with regard to the weight $a_m$ and equated to zero. 

The numerator can be re-written to $\sum_{m=-\infty}^\infty a_{m}a_{m+1}=\sum_{m'=0}^\infty a_{m'}a_{m'+1}+a_{m'}a_{m'-1}$ and the denominator to
$\sum_{m=-\infty}^\infty a_{m}^2=\sum_{m'=0}^\infty (2-\delta_{m'})a_{m'}^2$.
Formally, the differentiated fraction equated to zero $r_\mathrm{E}=\frac{A}{B}$ can be simplified to $r_\mathrm{E}'=\frac{A'}{B}-\frac{AB'}{B^2}=\frac{A'}{B}-\frac{r_\mathrm{E}\,B'}{B}$, and we may require $A'-r_\mathrm{E}\,B'=0$ 
\begin{align}
	\frac{\partial}{\partial a_m}\Biggl(\sum_{m'=0}^\infty a_{m'}a_{m'+1}+a_{m'}a_{m'-1}\Biggr)-r_\mathrm{E}\;\frac{\partial}{\partial a_m}\Biggl(\sum_{m'=0}^\infty(2-\delta_{m'})\, a_{m'}^2\Biggr)&=0\nonumber\\
	a_{m+1}+a_{m-1}-(2-\delta_m)\,r_\mathrm{E}\,a_{m}&=0.
\end{align}
As a result to our optimization problem we obviously got a recurrence relation for both $a_m$ and the minimum $r_\mathrm{E}$. We find a recurrence in \cite[eq.~18.9.1]{nist}, Tab.~\ref{tab:chebpoly}, as a recurrence defining the Chebyshev polynomials ($T_{m}(\cos\varphi)=\cos m\varphi$) from lower-degree ones,
\begin{align}
	T_{m+1}(x)+T_{m-1}(x)-(2-\delta_m)x\,T_{m}(x)=0,\nonumber
\end{align}
and by comparison we find for $a_m$ and $r_\mathrm{E}$: $r_\mathrm{E}=x$, $a_m=T_m(r_\mathrm{E})$. 
While $r_\mathrm{E}=1$ is only achievable with $N=\infty$, we want to get a maximum value for $r_\mathrm{E}$ with finite degree $N$.
We insert the solution into the equation for $r_\mathrm{E}$ with summation restricted to only contain  the existing finite-degree $a_m$ with $0\leq m\leq \mathrm{N}$ and find that the denominator multiplied with $r_\mathrm{E}$, on which the recurrence can be applied, yields a criterion for $a_{\mathrm{N}+1}$:
\begin{align}
	\frac{\sum_{n=0}^{\mathrm{N}-1}[T_m(r_\mathrm{E})T_{m+1}(r_\mathrm{E}) + T_m(r_\mathrm{E})T_{m-1}(r_\mathrm{E})]}{
		\sum_{n=0}^{\mathrm{N}}(2-\delta_m)[T_m(r_\mathrm{E})]^2}&=r_\mathrm{E}\\
	\sum_{n=0}^{\mathrm{N}-1}[T_m(r_\mathrm{E})T_{m+1}(r_\mathrm{E}) + T_m(r_\mathrm{E})T_{m-1}(r_\mathrm{E})]&=r_\mathrm{E}	\sum_{n=0}^{\mathrm{N}}(2-\delta_m)T_m(r_\mathrm{E})T_m(r_\mathrm{E})\nonumber\\
	&=
	\sum_{n=0}^{\mathrm{N}}[(2-\delta_m)r_\mathrm{E}T_m(r_\mathrm{E})]T_m(r_\mathrm{E})\nonumber\\
	&=
	\sum_{n=0}^{\mathrm{N}}[T_{m+1}(r_\mathrm{E})+T_{m-1}(r_\mathrm{E})]T_m(r_\mathrm{E})\nonumber\\
	0&=T_{\mathrm{N}+1}(r_\mathrm{E})T_{\mathrm{N}}(r_\mathrm{E}).\nonumber
\end{align}
While we want to require $a_0=1$ as a boundary condition, which is automatically fulfilled by $T_0(x)=1$, we seem to need to require that $a_{\mathrm{N}+1}=T_{\mathrm{N}+1}(r_\mathrm{E})=0$ must vanish. 
A suitable length for $r_\mathrm{E}$ must lie within the interval $[-1;1]$, so that we may substitute $r_\mathrm{E}$ by the cosine of a dispersion angle $\sigma_\mathrm{E}$ that should correspondingly be as small as possible and be a zero crossing of the $\mathrm{N}+1$ degree term $T_{\mathrm{N}+1}(\cos\sigma_\mathrm{E})=\cos[(\mathrm{N}+1)\sigma_\mathrm{E}]=0$, hence $r_\mathrm{E}$ must be defined by the angle $\sigma_\mathrm{E}=\frac{\pi}{2}(\mathrm{N}+1)$. The required $\mathrm{max}-\bm r_\mathrm{E}$ weight consequently is a $\cos$ half-wave window
\begin{align}
	a_m&=\cos(\textstyle\frac{\pi}{2}\frac{m}{\mathrm{N}+1}).
\end{align}

\newpage
\subsection{Optimally arranged surround loudspeaker ring}
We are looking for a discretization with $\mathrm{L}$ an optimal loudspeaker setup, characterized by the angles $\upvarphi_l$. For such a setup, the panning-invariant properties $P$, $E$, $r_\mathrm{V}$, $r_\mathrm{E}$ and the accurate directions of $\bm r_\mathrm{V}$ and $\bm r_\mathrm{E}$ should stay preserved
\begin{align}
	P&=\int_{-\pi}^{\pi}g(\varphi)\,\mathrm{d}\varphi={\textstyle\frac{2\pi}{\mathrm{L}}}\sum_{l=1}^\mathrm{L}g(\upvarphi_l),
	&\bm{r}_\mathrm{V}&=\frac{\int_{-\pi}^{\pi}g(\varphi)\,e^{\mathrm{i}\varphi}\,\mathrm{d}\varphi}{V}=\frac{\frac{2\pi}{\mathrm{L}}\sum_{l=1}^\mathrm{L}g(\upvarphi_l)\,e^{\mathrm{i}\upvarphi_l}}{V},\nonumber\\
	E&=\int_{-\pi}^{\pi}|g(\varphi)|^2\,\mathrm{d}\varphi={\textstyle\frac{2\pi}{\mathrm{L}}}\sum_{l=1}^\mathrm{L}|g(\upvarphi_l)|^2\,,
	&\bm{r}_\mathrm{E}&=\frac{\int_{-\pi}^{\pi}|g(\varphi)|^2\,e^{\mathrm{i}\varphi}\,\mathrm{d}\varphi}{E}=\frac{\frac{2\pi}{\mathrm{L}}\sum_{l=1}^\mathrm{L}|g(\upvarphi_l)|^2\,e^{\mathrm{i}\upvarphi_l}}{E}
	\nonumber.
\end{align}
What would be an optimal arrangement that can be defined without needing to prove all of the four equivalences? (\emph{see t designs, above}).

\end{appendices}

\bibliographystyle{IEEEtranSA}
\bibliography{bib_database}

% Generated by IEEEtranSA.bst, version: 1.14 (2015/08/26)
\providecommand{\etalchar}[1]{$^{#1}$}
\begin{thebibliography}{ZZFK17}
\providecommand{\url}[1]{#1}
\csname url@samestyle\endcsname
\providecommand{\newblock}{\relax}
\providecommand{\bibinfo}[2]{#2}
\providecommand{\BIBentrySTDinterwordspacing}{\spaceskip=0pt\relax}
\providecommand{\BIBentryALTinterwordstretchfactor}{4}
\providecommand{\BIBentryALTinterwordspacing}{\spaceskip=\fontdimen2\font plus
\BIBentryALTinterwordstretchfactor\fontdimen3\font minus
  \fontdimen4\font\relax}
\providecommand{\BIBforeignlanguage}[2]{{%
\expandafter\ifx\csname l@#1\endcsname\relax
\typeout{** WARNING: IEEEtranSA.bst: No hyphenation pattern has been}%
\typeout{** loaded for the language `#1'. Using the pattern for}%
\typeout{** the default language instead.}%
\else
\language=\csname l@#1\endcsname
\fi
#2}}
\providecommand{\BIBdecl}{\relax}
\BIBdecl

\bibitem[CG77]{Craven77}
P.~Craven and M.~A. Gerzon, ``Coincident microphone simulation covering three
  dimensional space and yielding various directional outputs,'' \emph{U.S.
  Patent}, no. 4,042,779, 1977.

\bibitem[CS72]{Cooper72}
D.~H. Cooper and T.~Shiga, ``Discrete-matrix multichannel stereo,'' \emph{J.
  Audio Eng. Soc.}, vol.~20, no.~5, pp. 346--360, 1972.

\bibitem[Dan01]{Daniel01}
J.~Daniel, ``Repr\'{e}sentation des champs acoustiques, application \`{a} la
  transmission et \`{a} la reproduction de sc\`{e}nes sonores complexes dans un
  contexte multim\'{e}dia,'' Ph.D. dissertation, Universit\'e Paris 6, 2001.

\bibitem[Dev15]{Devaraju}
\BIBentryALTinterwordspacing
B.~Devaraju, ``Understanding filtering on the sphere,'' Ph.D. dissertation,
  Universit{\"a}t Stuttgart, 2015. [Online]. Available:
  \url{https://elib.uni-stuttgart.de/bitstream/11682/4002/1/BDevarajuPhDThesis.pdf}
\BIBentrySTDinterwordspacing

\bibitem[DGS77]{DeslarteGoethalsSeidel77}
P.~Delsarte, J.-M. Goethals, and J.~J. Seidel, ``Spherical codes and designs,''
  \emph{Geometriae Dedicata}, vol.~6, no.~3, 1977.

\bibitem[DRP99]{Daniel99}
J.~Daniel, J.-B. Rault, and J.-D. Polack, ``Acoustic properties and perceptive
  implications of stereophonic phenomena,'' in \emph{AES 6th Int. Conf.:
  Spatial Sound Reproduction}, 1999.

\bibitem[Elk00]{Elko00}
G.~W. Elko, ``Superdirectional microphone arrays,'' in \emph{Acoustic Signal
  Processing for Telecommunication}, J.~Benesty and S.~L. Gay, Eds.\hskip 1em
  plus 0.5em minus 0.4em\relax Kluwer Academic Publishers, 2000.

\bibitem[Elk04]{Elko04}
------, ``Differential microphone arrays,'' in \emph{Audio Signal Processing
  for Next-Generation Multimedia Communication Systems}, Y.~Huang and
  J.~Benesty, Eds.\hskip 1em plus 0.5em minus 0.4em\relax Springer, 2004.

\bibitem[Fel74]{Felgett}
P.~Felgett, ``Ambisonic reproduction of directionality in surround-sound
  systems,'' \emph{Nature}, vol. 252, pp. 534--538, 1974.

\bibitem[Geg77]{Gegenbauer77}
\BIBentryALTinterwordspacing
L.~Gegenbauer, ``{\"U}ber die {F}unctionen {$C_n^\nu(x)$},''
  \emph{Sitzungsberichte der Kaiserlichen Akademie der Wissenschaften.
  Mathematisch-Naturwissenschaftliche Classe}, vol.~57, no.~8, 1877. [Online].
  Available:
  \url{https://viewer.acdh.oeaw.ac.at/viewer/image/MN_2Abt_75_1877/904/LOG_0074/}
\BIBentrySTDinterwordspacing

\bibitem[Ger75]{Gerzon75}
M.~A. Gerzon, ``The design of precisely coincident microphone arrays for stereo
  and surround sound,'' in \emph{prepr.\ L-20 of 50th Audio Eng. Soc. Conv.},
  1975.

\bibitem[GP11]{Graef11}
\BIBentryALTinterwordspacing
M.~Gr\"af and D.~Potts, ``On the computation of spherical designs by a new
  optimization approach based on fast spherical fourier transforms,''
  \emph{Numer. Math.}, vol. 119, 2011. [Online]. Available:
  \url{http://homepage.univie.ac.at/manuel.graef/quadrature.php}
\BIBentrySTDinterwordspacing

\bibitem[HC37]{HilbertCourant}
D.~Hilbert and R.~Courant, \emph{Methoden der mathematischen Physik II}.\hskip
  1em plus 0.5em minus 0.4em\relax Springer Berlin, 1937.

\bibitem[HS96]{Hardin96}
\BIBentryALTinterwordspacing
R.~H. Hardin and N.~J.~A. Sloane, ``Mclaren's improved snub cube and other new
  spherical designs in three dimensions,'' \emph{Discrete and Computational
  Geometry}, vol.~15, pp. 429--441, 1996. [Online]. Available:
  \url{http://neilsloane.com/sphdesigns/dim3/}
\BIBentrySTDinterwordspacing

\bibitem[LS75]{LawSledd75}
A.~G. Law and M.~B. Sledd, ``Normalizing orthogonal polynomials by using their
  recurrence coefficients,'' \emph{Proc. American Mathematical Society},
  vol.~48, no.~2, pp. 505--507, 1975.

\bibitem[OBC00]{nist}
F.~W.~J. Olver, R.~F. Boisvert, and C.~W. Clark, Eds., \emph{NIST Handbook of
  Mathematical Functions}.\hskip 1em plus 0.5em minus 0.4em\relax Cambridge,
  UK: Cambridge University Press, 2000, \url{http://dlmf.nist.gov}, accessed
  June 2012.

\bibitem[Raf19]{Rafaely19}
B.~Rafaely, \emph{Fundamentals of Spherical Array Processing}, 2nd~ed.\hskip
  1em plus 0.5em minus 0.4em\relax Springer Berlin Heidelberg, 2019.

\bibitem[Sch10]{Schuessler}
H.~W. Sch{\"u}{\ss}ler, \emph{Digitale Signalverarbeitung 2: Entwurf diskreter
  Systeme}, G.~Dehner, R.~Rabenstein, and P.~Steffen, Eds.\hskip 1em plus 0.5em
  minus 0.4em\relax Springer Heidelberg, 2010.

\bibitem[ZF12]{Zotter12b}
F.~Zotter and M.~Frank, ``All-round ambisonic panning and decoding,''
  \emph{Journal of the Audio Eng. Soc.}, 2012.

\bibitem[ZRF22]{Zotter22}
\BIBentryALTinterwordspacing
F.~Zotter, S.~Riedel, and M.~Frank, ``All-round ambisonic decoding: Spread and
  correlation,'' in \emph{Fortschritte der Akusti, DAGA}, Stuttgart, 2022.
  [Online]. Available:
  \url{https://pub.dega-akustik.de/DAGA_2022/data/articles/000344.pdf}
\BIBentrySTDinterwordspacing

\bibitem[ZZFK17]{Zotter17}
F.~Zotter, M.~Zaunschirm, M.~Frank, and M.~Kronlachner, ``A beamformer to play
  with wall reflections: The icosahedral loudspeaker,'' \emph{Computer Music
  Journal}, vol.~41, no.~3, 2017.

\end{thebibliography}
%\printbibliography[heading=bibintoc]

\end{document}